\documentclass[12pt]{article}
\RequirePackage{amsthm,amsmath,amsfonts,amssymb}
\RequirePackage{natbib}
\RequirePackage{graphicx}
\RequirePackage{xcolor}
\RequirePackage{soul}
\usepackage{float}
\usepackage[normalem]{ulem}

\theoremstyle{remark}
\newtheorem {Prop}{Proposition} [section]
 \newtheorem {Lemm}[Prop] {Lemma}
 \newtheorem {Theo}[Prop]{Theorem}
 
 \newtheorem {Nota}[Prop]{Remark}
 \newtheorem {Ejem}[Prop] {Example}

\newcommand{\blind}{0}

\begin{document}
\pagestyle{empty}

\definecolor{verde}{rgb}{0, 0.8, 0}
\definecolor{darkgris}{rgb}{0.4, 0.4, 0.4}
\definecolor{gris}{rgb}{0.6, 0.6, 0.6}
\definecolor{grisDos}{rgb}{0.8, 0.8, 0.8}
\definecolor{darkred}{rgb}{0.55,0,0}
\definecolor{cian}{rgb}{0.0, 0.9, 0.9}
\definecolor{morado}{rgb}{0.5, 0.0, 0.5}
 \definecolor{redo}{rgb}{0.7, 0.1, 0.6}
 \definecolor{azulito}{rgb}{0.45, 0, 1}
\definecolor{grisTres}{rgb}{0.45, 0.45, 0.45}
\definecolor{naranja}{rgb}{0.85, 0.35, 0}

\newcommand{\colred}{\color{red}}
\newcommand{\colblue}{\color{blue}}
\newcommand{\coloran}{\color{orange}}
\newcommand{\colverde}{\color{verde}}

\newcommand{\al}{\mbox{$\alpha$}}
\newcommand{\be}{\mbox{$\beta$}}
\newcommand{\ep}{\mbox{$\epsilon$}}
\newcommand{\gam}{\mbox{$\gamma$}}
\newcommand{\sig}{\mbox{$\sigma$}}

\DeclareRobustCommand{\FIN}{%
  \ifmmode 
  \else \leavevmode\unskip\penalty9999 \hbox{}\nobreak\hfill
  \fi
   \vspace{1mm}}

\newcommand{\calA}{\mbox{${\cal A}$}}
\newcommand{\calB}{\mbox{${\cal B}$}}
\newcommand{\calC}{\mbox{${\cal C}$}}

\newcommand{\muas}{\mbox{$\mu$-a.s.}}
\newcommand{\Nat}{\mbox{$\mathbb{N}$}}
\newcommand{\Rea}{\mbox{$\mathbb{R}$}}
\newcommand{\Prob}{\mbox{$\mathbb{P}$}}

\newcommand{\nin}{\mbox{$n \in \mathbb{N}$}}
\newcommand{\suc}{\mbox{$\{X_{n}\}$}}
\newcommand{\sucY}{\mbox{$\{Y_{m}\}$}}
\newcommand{\sucP}{\mbox{$\mathbb{P}_{n}\}$}}

\newcommand{\conv}{\rightarrow}
\newcommand{\convn}{\rightarrow_{n\rightarrow \infty}}
\newcommand{\convp}{\rightarrow_{p}}
\newcommand{\convs}{\rightarrow_{\mbox{a.s.}}}
\newcommand{\convw}{\rightarrow_w}
\newcommand{\convd}{\stackrel{\cal D}{\rightarrow}}

\def\spacingset#1{\renewcommand{\baselinestretch}%
{#1}\small\normalsize} \spacingset{1}


\if0\blind
{
  \title{\bf Invariant measures of disagreement with stochastic dominance\thanks{Research partially supported by grants No PID2021-128314NB-I00 and PID2022-139237NB-I00
funded by MCIN/AEI/10.13039/501100011033/FEDER.
}}
  \author{E. del Barrio\hspace{.2cm}\\
    Departamento de Estad\'{\i}stica e Investigaci\'on Operativa and IMUVA, 
    \\
    Universidad de Valladolid, Spain
    \\
    J.A. Cuesta-Albertos
     \\
   Departamento de
Matem\'{a}ticas, Estad\'{\i}stica y Computaci\'{o}n, 
\\
Universidad de Cantabria, Spain
\\
    C. Matr\'{a}n\\
    Departamento de Estad\'{\i}stica e Investigaci\'on Operativa and IMUVA, 
    \\
    Universidad de Valladolid, Spain}
  \maketitle
} \fi

\if1\blind
{
  \bigskip
  \bigskip
  \bigskip
  \begin{center}
Invariant measures of disagreement with stochastic dominance
\end{center}
  \medskip
} \fi

\begin{abstract}

Stochastic dominance has not been employed too often in practice due to its important limitations.
To increase its versatility, the concept has recently been adapted by introducing various indices that measure the degree to which one probability distribution stochastically dominates another. 
In this paper, starting from the fundamentals and using very simple examples, we present and discuss some of these indices when one intends to maintain invariance through increasing functions. This naturally leads to consideration of the appealing common representation, $\theta(F,G)=P(X>Y)$, where $(X, Y)$ is a random vector with marginal distributions $F$ and $G$. The indices considered here arise from different dependencies between X and Y. This includes the case of independent marginals, as well as other indices related to a contamination model or to a joint quantile representation.
We emphasize the complementary role of some of these indices, which, in addition to measuring disagreement with respect to stochastic dominance, enable us to describe the maximum possible difference in the status of a value $x\in \Rea$ under $F$ or $G$. We apply these indices to simulated and real-world datasets, exploring their practical advantages and limitations.
 The tour includes lesser-known facets of well-known statistics
such as Mann-Whitney, one-tailed Kolmogorov-Smirnov and Galton's rank statistics, even providing additional theory for the latter. 
\end{abstract}

\noindent%
{\it Keywords:}  relaxed stochastic dominance, asymptotics, inferential procedures, Galton's rank statistic
\vfill

\newpage
\setcounter{page}{1}
\pagestyle{plain}

\spacingset{1.45} 

\section{Introduction}
Comparison is a  common task in any kind of research or activity. Although it is trivial when  it involves  measurements on just two individuals, it is far from being obvious when  it involves  populations. 
Often the approach  just consists  in the comparison of {some   summary value (mean}, median, Gini index,\ldots), but the real meaning of such comparison is not so obvious and, sometimes, it is even misinterpreted  by the practitioners. This situation was addressed in \cite{Alv2017}, emphasizing on  location-scale models, defending  stochastic dominance (s.d.) as the natural gold standard in two-sample comparison problems. To pursue that direction, we  return here to the very principles of comparison, by considering the  meaning of s.d. between probability distributions and analyzing some natural relaxations  intended to measure the degree of disagreement with  s.d.

{Very often, a general statement like ``People are wealthier now than they were four years ago"  { is just supported by the fact} that the  per capita income is higher now than it was four years ago.}  However, such a comparison is compatible with very different shapes of the parent distributions, and could lead to a false picture that is highly inappropriate if, for example, skewed, very heavy-tailed distributions are involved. An order between populations or distributions should indicate a comprehensive relationship between them, improving those based on comparisons of single indices or features of the distributions.

Suppose that we are given two probability distributions on the real line, $\Rea$, defined through their  distribution functions (d.f.'s) $F$ and  $G$. Denoting by $F^{-1}$ to the quantile function of $F$ (defined by $F^{-1}(t):= \inf\{x : \ t\leq F(x)\}$, 
for $t \in (0,1)$), we say that 
{\it $F$ is stochastically dominated by $G$, denoted $F \leq_{st}G$, if 
\begin{equation}\label{storder2}
F^{-1}(t)\leq G^{-1}(t) \mbox{ for every } t\in (0,1).
\end{equation}}
This definition, as noted in \cite{Lehmann}, 
is more intuitive than the usual  one:
\begin{equation}\label{storder}
F \leq_{st}G \mbox{ \it whenever } F(x)\geq G(x) \mbox{ \it for every } x\in\Rea.
\end{equation}

Given $x\in \Rea$,  the value $F(x)$ (resp. $G(x)$) can be considered as the ``status" of  $x$ in the first (resp. second) population. In this sense, the meaning of (\ref{storder}) is that the status of any individual value  would be higher when considered in the first population than when considered in the second. 
 
{We stress the fact that $F^{-1}$ is a particular inverse function of $F$, which allows a representation of the probability $P_F$ associated to $F$ through a random variable and the length measure $\ell$, which is a probability on the space $(0,1)$: Since $\ell \left(t\in (0,1):  a<F^{-1}(t)\leq b\right)=F(b)-F(a)=P_F\left((a,b]\right)$, the quantile function $F^{-1}$ has probability law $P_F.$}  This fact and (\ref{storder2}) lead to the following characterization of s.d., which is  the basis of this paper:
\begin{equation}
\label{storder3}
\begin{array}{c}
F\leq_{st}G \ \mbox{\it  if and only if  there exist r.v.'s } X, Y 
\mbox{\it defined on some }\\ \mbox{\it probability space } (\Omega,\sigma,P), 
\mbox{ \it with d.f.'s } F,G,   \mbox{\it  such that } {P(X\leq Y )=1.}
\end{array}
\end{equation}

\begin{Nota}\label{Nota.Invarianza}
{This representation shows that  s.d. is invariant through increasing functions because if $X,Y$ are two  r.v.'s verifying $P(X\leq Y )=1$, and $\psi:\Rea \to \Rea$ is increasing, then  $P(\psi(X)\leq \psi(Y) )=1$, so the  d.f. of $\psi(Y)$ also stochastically dominates that of $\psi(X)$.} 
\end{Nota}

 {Notice that, whenever $F\leq_{st}G$, (\ref{storder3}) guarantees that   we can obtain random samples $(x_1,y_1),\ldots,(x_n,y_n)$ such that $x_1,\ldots,x_n$ (resp. $y_1,\ldots,y_n$) is a random sample from $F$ (resp. $G$) and $x_i\leq y_i, i=1,\ldots,n$}. The quantile functions provide one such representation  but (\ref{storder3}) allows for other representations of s.d. in terms of almost sure dominance.

\begin{figure}
\begin{center}\includegraphics[width=8cm,height= 5cm]{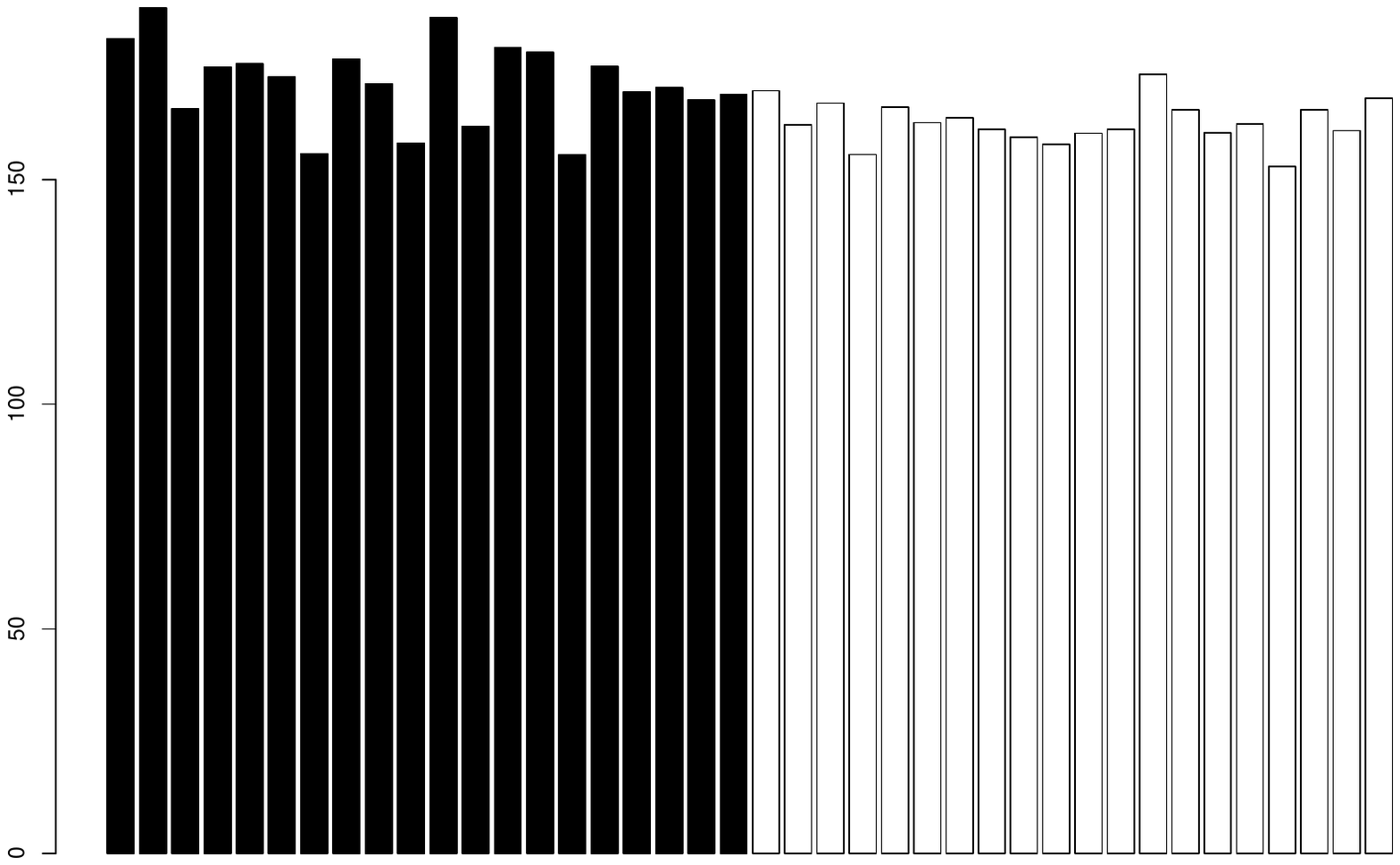}
\includegraphics[width=8cm,height= 5cm]{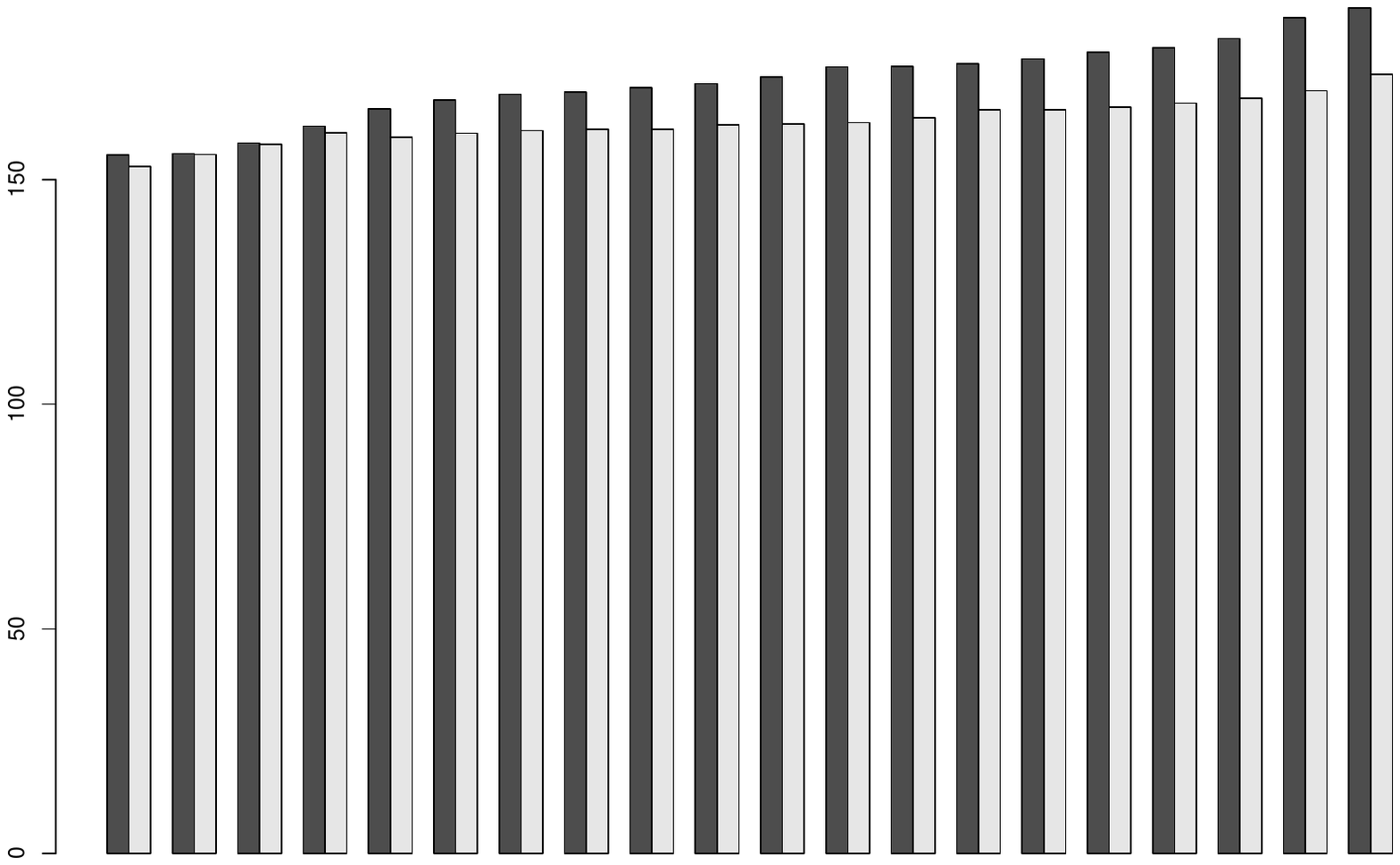}
\end{center}
\vspace{-5mm}
\caption{{Incomes of   two samples of 20  individuals: today (black bars) and  four years ago (white bars).  {
Left: Unsorted bars of both samples. Right: Paired bars of  sorted samples}.} 
\label{primer_ejemplo}}
\end{figure}

Let us begin by comparing the  incomes of two equal-sized   samples of 20  individuals today and four years ago, represented in Figure \ref{primer_ejemplo} through black and white bars. In the first graph, we can see that there is some tendency for the black bars to be higher than the white ones, although there are white bars that are larger than some of the black ones. To check whether the black sample stochastically dominates the white sample, the second graph shows the same bars, but now the smallest black bar is matched with the smallest white one, the second with the {second,} 
and so on.   We  see that every  black bar is larger than   its paired white one. 
 According to \eqref{storder3}, we have  s.d. of the incomes of the first  sample over the second one. In fact,  the comparison between the individuals with  same rank is just the comparison between the quantile functions given in (\ref{storder2}).
 
While {the interest of  s.d. is clear}, it is often observed
that such a relation is too strong as to be satisfied in practice. 
{For instance,} \cite{Arcones} notes that  s.d. (\ref{storder}) may well 
 hold over {most  of the domain but it may fail over a small part of it, or may 
 simply be unknown or unknowable over the entire range {(a fact also implicit in \cite{Lesh2002})}. 
 Even worse, if we want to conclude that s.d. holds we should gather statistical evidence to reject the null 
 in the testing problem of 
$
  H_0: F \nleq_{st}G \mbox{ vs } H_a: F \leq_{st}G.
$
 But, {
 this is an impossible task}, because, even if $F \leq_{st}G$, there exist $F^*$ and $G^*$ as close as desired to $F$ and $G$ with $F^* \nleq_{st}G^* $ (see \cite{Berger}).

 Following ideas that go back to \cite{Hodges}, {we can find in the literature contributions whose} goal is to enlarge the null by including ``similar" distributions leading to reject it just if there are ``relevant { differences" between $F$ and $G$.
\cite{Munk1998}, \cite{Liu},  \cite{Alv2011b}, \cite{Dette1} and \cite{Dette2} share this point of view.

 There also exist   indices 
 to measure the extent to which s.d. is violated. In this line, we are aware of the   {\it almost s.d.} defined in
 \cite{Lesh2002} {
  through}:
\begin{equation}\label{Levyindex}
 \epsilon(F,G):=\frac{\int(G(x)-F(x))^+dx}{\int|G(x)-F(x)|dx},
 \end{equation}
where $x^+:=\max(x,0)$,
{and the variant introduced in  \cite{PedroGil}}:
 \begin{equation}\label{TransportIndex}
 \epsilon_{2}(F,G):=\frac{\int((F^{-1}(x)-G^{-1}(x))^+)^2dx}{\int(F^{-1}(x)-G^{-1}(x))^2dx}.
 \end{equation}
 
Also, \cite{Alv2015} proposed another index which, although  based on a contamination model, admits the simple expression (see Subsection \ref{piapproach}):
\begin{equation}\label{pilevel0}
 \pi(F,G)=\sup_{x \in \mathbb{R}} (G(x)-F(x)).
 \end{equation}

{In this paper we focus  on some indices which are invariant w.r.t. increasing maps and, more specifically, based on Remark  \ref{Nota.Invarianza}, on indices which can be represented as $P(X>Y)$ by  suitable pairs of r.v.'s $(X,Y)$ with marginal d.f.'s $F$ and $G$. The difference between them coming from  different dependences between $X$ and $Y$ including the independence, and two cases of maximal dependence. They are introduced in Section \ref{presentacionindices}.

The statistical analysis of $P(X>Y)$  began
in \cite{Birnbaum}, and under the suggestive name of Stress-Strength Model is widely recognized by its multiple 
applications (\cite{Kotz} gives general account). Here, we  analyse their behaviour in inference 
using the  plug-in estimators of the indices 
(the indices computed on the 
 empirical d.f.'s). We show that   the
indices considered  involve well known statistics:  Mann-Whitney,
 one-sided Kolmogorov-Smirnov, and Galton's rank order.  
 While the available literature on the asymptotics  for the  first two statistics suffices for their use, 
 only recent  results justify the use of Galton's statistic in this setting {and even the first results allowing to employ bootstrap are presented here (see 
 Theorem 4.2 in the supplementary material).}}

{The indices (\ref{Levyindex}) and (\ref{TransportIndex})
are only invariant with respect to increasing linear functions, but not in general, so  they  do not admit a $P(X>Y)$-representation. However, we will show in Subsection \ref{piapproach} that the index $\pi(F,G)$ admits that representation. }

The paper is organized as follows.   In Subsections 2.1 to 2.3, we give an overview of the  {indices considered here}. In Subsection \ref{fram}, we present the framework for analyzing these indices from a common perspective. Section \ref{testing} contains the relevant comments and theory for the statistical use of the indices.
 In Section \ref{SCS} we   analyse several  simulated and real data sets
which show the performance of these indices in applied settings.  
The paper ends with a section devoted to  conclusions.  The supplementary material, at the end of this paper,  (``the supplement" in the sequel), contains some theoretical results as well as additional information on  the analysed real and simulated datasets.

{
Concerning the notation, the symbols $X,Y$, with or without sub/super-indices, will be real r.v.'s with respective d.f.'s $F$ and $G$.   $\mathcal L(Z)$ will denote the law  of the 
random vector or variable $Z$ and   $\mathcal L(Z/A)$  refers to the conditional law given the event $A$. {$\ell$ will denote the length measure } on the  unit interval $(0,1)$. Convergences in the almost sure or in law senses will be respectively denoted by $\convs$ and $\convw$.   We use the notation  $x^+$  for $\max(x,0)$. The term increasing must be interpreted in the strict sense.}

 \section{Measuring departures from stochastic dominance}\label{presentacionindices}

  To give an intuitive idea of how the different approaches focus on measuring the lack of s.d., 
 we will consider the following example.
\begin{Ejem}\label{Ej.finiteSamples}
{\rm

Assume that  ${\mathcal  X}=\{x_1,\dots ,x_n\}$ and ${\mathcal  Y}=\{y_1,\dots,y_n\}$ are the heights of some individuals belonging to different populations. 
To analyse a possible s.d. of the {data in} 
${\mathcal Y}$ over { those in} 
${\mathcal X}$, we can rearrange the  heights, resulting in ${\mathcal X}=\{x^*_1 \leq \dots \leq x^*_n\}$ and ${\mathcal Y}=\{y^*_1 \leq \dots \leq y^*_n\}$. In this way, s.d. is 
equivalent to $x^*_i \leq y^*_i$ for each $i=1,\dots,n$.
}
\end{Ejem}

To illustrate the behavior of the different indices, we  include comparisons between fixed normal distributions. For a  visual illustration of the relationship between these indices,
we refer to the contour plots in \cite{Alv2017} 
and \cite{PedroGil}.

Our goal is to analyse several indices to measure how far are two d.f.'s $F,G$ to satisfy $F \leq_{st}G$ using the value of $P(X>Y)$ where  $X,Y$ are two r.v.'s with respective d.f.'s $F$ and $G$. More precisely, in each of Subsections \ref{quantileapproach} to \ref{piapproach} we analyse an index with this property, and where the dependence between $X$ and $Y$ varies from an index to another.

  \noindent
 {\it Example  \ref{Ej.finiteSamples} (Cont.)}
The values of the indices $\epsilon$ and $\epsilon_2
$ (defined in \eqref{Levyindex} 
and \eqref{TransportIndex}) are
$$\epsilon(F,G)=\frac{\sum_{i=1}^n({ x^*_i-y^*_i})^+}{\sum_{i=1}^n|{ x^*_i-y^*_i}|},
\ \mbox{ and }
\epsilon_2
(F,G)=
\frac{\sum_{i=1}^n(({ x^*_i-y^*_i})^+)^2}{\sum_{i=1}^n{ (x^*_i-y^*_i)^2}}.
$$
Therefore, they depend not only on the number of pairs satisfying $x^*_i > y^*_i$, but also  on the quantities ${x^*_i-y^*_i}$; {so they do not accept the $P(X>Y)$-representation}.
\FIN

 \subsection{The quantile approach}\label{quantileapproach}
The quantile representation $X^*=F^{-1}$ and $Y^*=G^{-1},$ defined on   {$(0,1)$}, translates the  more abstract concept of s.d. 
 to the more familiar pointwise ordering between  $X^*$ and  $Y^*$.   
 From this point of view, \cite{Alv2017}  introduced an index  measuring the size of the set where  $X^*$ and $Y^*$  
do not satisfy the  point-wise order:
 \begin{equation}\label{gamma}
\gamma(F,G):=\ell(t \in (0,1) : F^{-1}(t)>G^{-1}(t)).
 \end{equation}
 
We  present this index 
through a  modification in the example
in Figure \ref{primer_ejemplo}. The barplot on the left of Figure \ref{segundo_ejemplo} has been produced in the same way as the one on the right  of Figure \ref{primer_ejemplo},  but now
 the third and fourth 
 black bars are shorter than the corresponding white ones;
thus, 
there is no s.d. of the  inputs of the second sample  over those in the earlier.
However we can give a precise measure of the extent to which  s.d. is not satisfied: 2/20. More precisely, for 90\% of the individuals associated with the white bars, their status according to their income level would be worse if they were considered in the sample associated with the black bars. 
The worsening of  status of a  person would mean that he/she would be considered less rich  today than four years ago if his/her income did not change.
  

\begin{figure}
\begin{center}\includegraphics[width=8cm,height= 5cm]{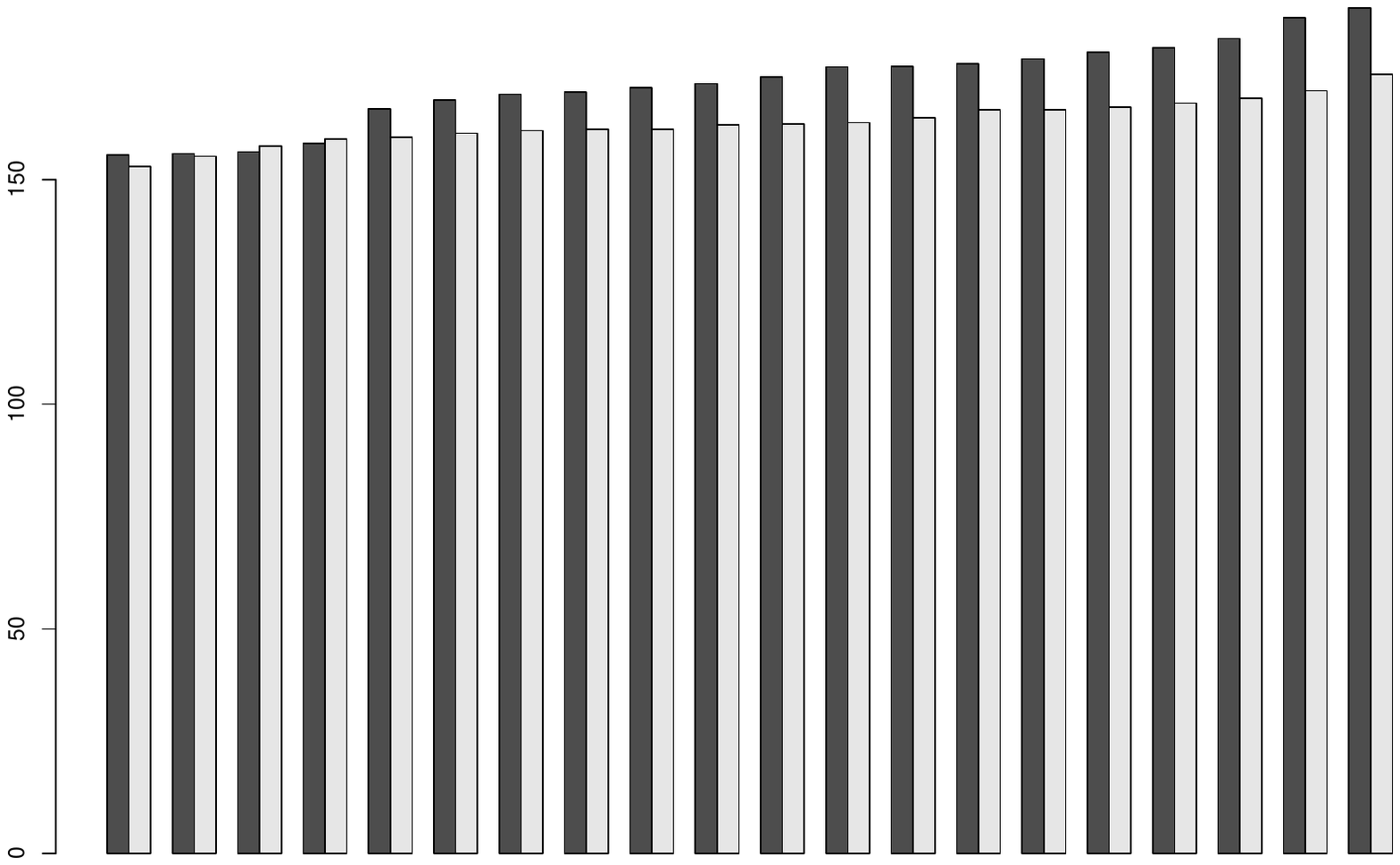}
\includegraphics[width=8cm,height= 5cm]{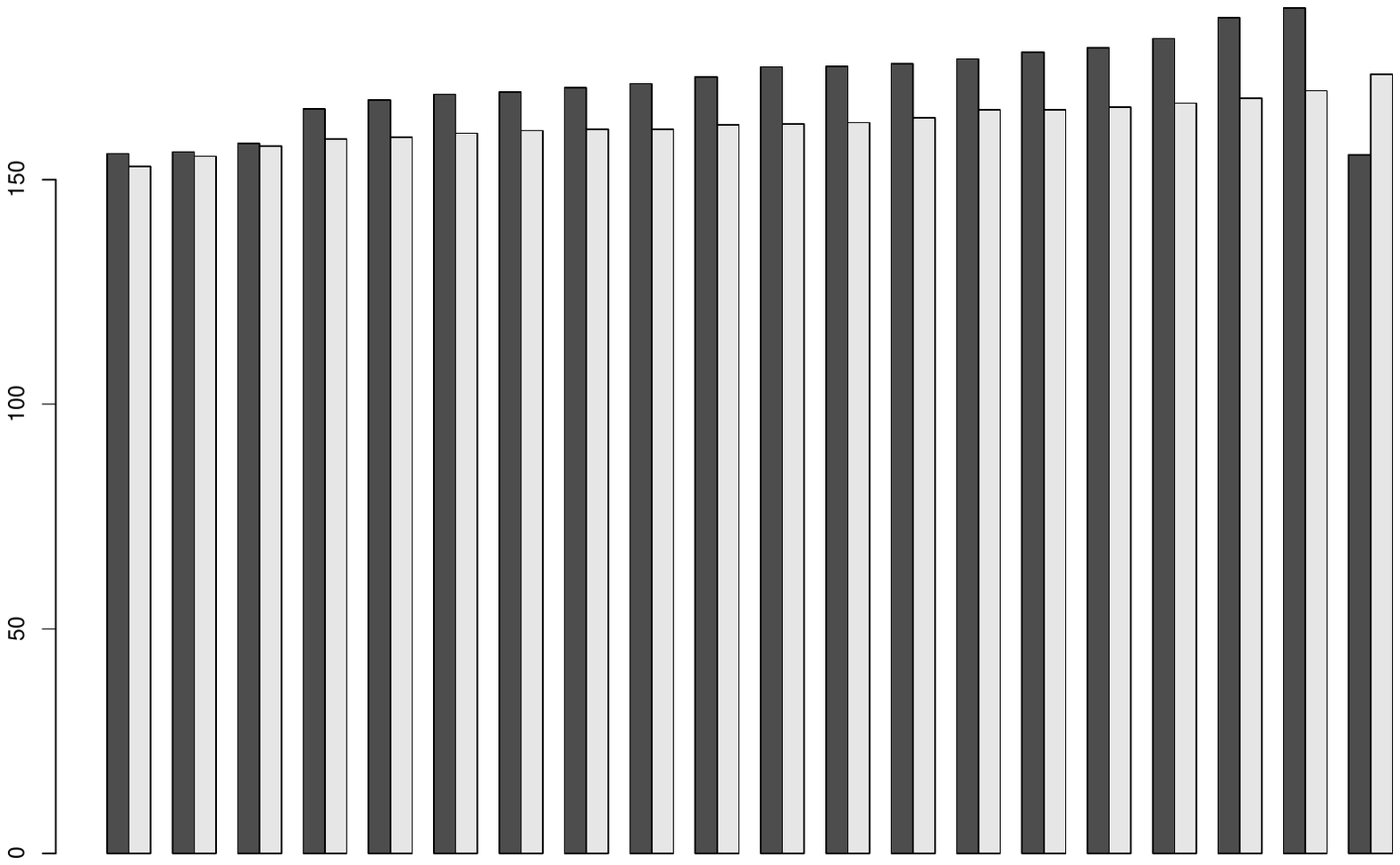}
\end{center}
\vspace{-5mm}
\caption{Left: 
Pairing of 20 individuals  with equal rank in their respective sample.    Right: White bars  coincide with the ones in the left; black bars are moved one position to the left, excepting the first one that is sent 
to the last position. 
\label{segundo_ejemplo}}
\end{figure}

  From  definition \eqref{storder2},  it  easily follows that if $F$ and $G$ are d.f.'s of normal distributions 
 $N(\mu_1,\sigma_1^2)$ and $ N(\mu_2,\sigma_2^2)$, respectively, $F\leq_{st}G$ holds 
 if and only if 
 $\sigma_1=\sigma_2$ and $\mu_1\leq \mu_2$. 
Therefore,  if e.g.  $\sigma_1=1, \sigma_2=2,$  the s.d. relation is 
impossible. For instance, 
if
 $\mu_1=0,\mu_2=2$ the set where (\ref{storder2}) fails  is the interval $(0,0.023)$.
 In other words, 
$\gamma(N(0,1),N(2,2^2)) = 0.023$, 
  while e.g. $\gamma(N(0,1),N(1,2^2)) = 0.16$.

  \vspace{2mm}
  
  \noindent
 {\it Example  \ref{Ej.finiteSamples} (Cont.)}
If there exist  exactly $m$ pairs $(x_i^*,y_i^*)$   such that $x^*_i> y^*_i$, then the $\gamma$-index approach  would report the value $m/n$ as the measure of lack of s.d. 
\FIN 
 
The index $\gamma$ allows one to quantify the importance of the set where  a (restricted) s.d. holds,  a concept already considered in \cite{Berger},  \cite{Lehmann92} and \cite{Davidson2013}. 
 A statement like $\gamma(F,G)\leq\gamma_0$,  for some fixed (small) $\gamma_0$ would give a quantified 
 approach to an approximate s.d.  Let us also note the  trivial facts that  $\gamma(F,G)=0$ is equivalent to $F\leq_{st}G$
 and that for any pair of continuous d.f.'s which coincide at most  on a denumerable set of points, the relation $ \gamma(F,G)+ \gamma(G,F)=1$ holds.

\subsection{The independent sampling approach}  \label{Subs.Indep}
{If we take $X,Y$ independent r.v.'s with d.f.'s $F$ and $G$, }
we can consider the index
  \begin{equation}\label{rho}
    \rho(F,G):=P(X> Y).
    \end{equation}
    
    Although not explicitly considered as an index, the value $\rho(F,G)$ was used in \cite{Arcones} to introduce 
  the concept of `stochastic precedence' of $F$ to $G$ ($F\leq_{sp}G$) which corresponds to the 
  case $\rho(F,G)\leq 1/2$, leading to a weaker relation than  s.d. In fact, if $F\leq_{st}G$, without any additional regularity requirement we see that
  \begin{eqnarray*}
  \rho(F,G)&=&\int_{-\infty}^\infty G(x-)dF(x)
  \leq \int_{-\infty}^\infty F(x-)dF(x)=P(X> X^{*})\leq \frac 1 2, 
  \end{eqnarray*}
 where $X^{*}$ is an independent copy of $X$. Of course the value 1/2 can be considered as a maximal value of 
 $\rho(F,G)$ to guarantee some advantage of $Y$ over $X$ in the sense considered in (\ref{rho}),  but lower values 
 of $\rho$ would confirm a larger guarantee of improvement.  
 
 In \cite{Arcones} it is mentioned, as a convenient feature of stochastic precedence, that it holds for normal 
 distributions whenever their means satisfy the corresponding order (thus being an alternative to shift testing). 
 {
 In Section  2 
 in the supplement, we show that this extends to a wide class of location-scale families}.

The relation $\rho(F,G)=P(X\leq Y)=1$ is too extreme,
  because it requires that  the support of $\mathcal L(X)$ 
 (resp. $\mathcal L(Y)$) must be contained in $(-\infty,z]$ (resp. $[z,-\infty)$) for some value $z\in \Rea$. However, in order to compare e.g. treatments applied to different populations,
  it would also be very informative to know that 
$\rho(F,G)$ is very small, because this would mean that there is a high probability that treatment $X$ will not give better results than treatment $Y$ when applied to independent samples of patients. For the  parameters  considered in Subsection \ref{quantileapproach}, $\rho(N(0,1),N(2,2^2))= 0.19$ 
while $\rho(N(0,1),N(1,2^2))= 0.33.$

Notice that $\rho(F,G)+\rho(G,F)\leq1$ with equality  if $F$ or $G$ is continuous. Furthermore, $\rho(F,G)=0$ implies 
that $F\leq_{st}G$, but the opposite fails.

  \vspace{2mm}
  
  \noindent
{\it Example  \ref{Ej.finiteSamples} (Cont.)}
 The value of the  $\rho$ index   in this situation is $\#\{ (i,j): x_i>y_j\}/ n^2$.
 \FIN


\subsection{The contamination approach} \label{piapproach}
This  approach was proposed in \cite{Alv2015}. It is based on the fact that always exist  
 $ \alpha \in (0,1)$, and d.f.'s  $ \tilde F, \tilde G, R , S \mbox{ such that } \tilde F\leq_{st} \tilde G$   which satisfy 
\begin{equation}\label{modelocontaminadotwosample1}
F  =  (1-\alpha)\tilde F+\alpha R
\hspace{5mm} \mbox{ and } \hspace{5mm}
G  =  (1-\alpha)\tilde G+\alpha S.
\end{equation}

The decomposition \eqref{modelocontaminadotwosample1} can be interpreted  in terms of a two stage  random procedure 
that,  when generating values from $F$ (resp. $G$), with  probability equal to $1-\alpha$, chooses the  distribution $\tilde F$  (resp.  $\tilde G$)  that    satisfy  s.d. Therefore, if   such a $\alpha$ is small enough, we could say that the greater part of the distribution $G$ dominates that of $F$.  Thus, a level of disagreement with  s.d.  is the lowest $\alpha$ compatible with  \eqref{modelocontaminadotwosample1}:
 \begin{equation}\label{pilevel}
 \alpha(F,G):=\inf \{\alpha \in (0,1) \mbox{: decomposition (\ref{modelocontaminadotwosample1}) holds}\}.
 \end{equation}

The theory developed in \cite{Alv2015} guarantees  (see Subsection 3.1 
in the supplement) that $\pi(F,G)=\alpha(F,G)$,
which gives a new 
characterization of $\pi(F,G)$ as defined in (\ref{pilevel0}). 
 This characterization allows us to define a quantified approximation to  s.d.:  given $\pi \in (0,1)$, we say that  $F\leq_{st}^{\pi}G$, whenever 
 $\pi \geq \alpha(F,G)$.


 The next proposition characterizes the relation $F\leq_{st}^{\pi}G$. It will be used later. Its short proof is deferred to Subsection 3.2 
  in the supplement.

\begin{Prop}\label{trim_quantile}
For d.f.'s $F$ and $G$ we have $F\leq_{st}^{\pi}G$ if and only if
$F^{-1}(y)\leq G^{-1}(\pi+y), \mbox{ \ for every $y$ such that }  0<y<1-\pi$.
\end{Prop}

Obviously, $\pi(F,G)=0$ if and only if $F\leq_{st}G$. Also  $\pi(F,G)+ \pi(G,F)\leq 1$ holds, although 
strict inequality is the typical situation, and  quite often $\pi(F,G)$ and $\pi(G,F)$ are both small. Notice that $\pi(F,G)=1$  requires $F(x_0)=0$ and $G(x_0)=1$ for some $x_0\in \Rea$.

\begin{Nota}
\label{Pi_Cosas}
Relation \eqref{pilevel}  allows us to interpret the index $\pi$ as {\it ``there exist subpopulations containing $\pi(F,G)\times100\%$ of their respective original populations for which s.d. holds''}. 
Regrettably, 
 these subpopulations are not unique (see Remark \ref{Pi_Cosas_2}) and the way in which they are obtained is  somewhat artificial, which makes their interpretation difficult. Therefore, we  often retain  the main interpretation of
 $\pi(F,G)$ as the maximum possible difference in status for an individual with the value $x$ when moving from population $F$ to $G$.
 \end{Nota}

 Returning to the example already considered in the preceding subsections, 
 we have 
 $\pi(N(0,12),N(2,2^2))= 0.006$, while $\pi(N(0,1),N(1,2^2))= 0.045.$ Additional comments about this index  appear after Proposition \ref{relations} and Remark \ref{NotaUpperBound}. 

 \vspace{2mm}
  
  \noindent
 {\it Example  \ref{Ej.finiteSamples} (Cont.)}
The $\pi$-index  is the infimum value, say $k/n$, such that deleting the $k$ greatest ranked individuals in $\mathcal X$ and the $k$ lowest ranked in $\mathcal Y$, the remaining subsets $ \mathcal X_{k}=\{x_1^*,\dots,x_{n-k}^*\}$ and ${\mathcal Y}_{k}=\{y_{k+1}^*,\dots,y_{n}^*\}$ verify  s.d. (thus $x_i^*\leq y_{k+i}^*$ for  $i=1,\dots,n-k$).
\FIN

\subsection{A unifying framework}\label{fram}

 In this section we provide a common framework for the proposed indices. It is based on the $P(X>Y)$ representation and is  one of the keystones of the paper.

Despite their different meanings, $\gamma(F,G)$ and $\rho(F,G)$ share a common principle that can be generalized as follows.  
If $(X,Y)$ is an arbitrary bivariate random vector,
then  the distribution of $X$  can be decomposed as the mixture
\begin{eqnarray}\label{relation7}
\mathcal L(X)&=&\mathcal L(X / X\leq Y)P(X\leq Y)
+\mathcal L(X / X > Y)P(X > Y).
\end{eqnarray}
 and similarly for ${\mathcal L}(Y)$. 
Trivially, 
regardless of the joint distribution of $(X,Y),$  
the conditional laws satisfy the  s.d. relations 
 \begin{eqnarray*}
  \mathcal L(X / X\leq Y) \leq_{st} \mathcal L(Y / X\leq Y)
 \quad \mbox{ and }   \quad 
\mathcal L(X / X> Y) \geq_{st} \mathcal L(Y / X> Y),
 \end{eqnarray*}
  that embedded in (\ref{relation7}), and taking $\lambda= P(X>Y)$, gives decompositions of $F$ and $G$  as 
\begin{equation}\label{relation8}
F=(1-\lambda)F_1+\lambda F_2 \ \mbox{ and } \ \ G=(1-\lambda)G_1+\lambda G_2, 
\end{equation}
 which depends on the joint law $\mathcal L(X,Y)$, but that
 satisfies $F_1\leq_{st}G_1 \mbox{ and } F_2\geq_{st}G_2$. 

As a first byproduct,  from (\ref{pilevel}) we conclude that $\lambda \geq \pi(F,G)$  regardless of  the chosen representation.
Particularizing \eqref{relation7} for the pair given by the quantile functions $(F^{-1},G^{-1})$, $\lambda$ takes the value $\gamma(F,G)$, and the d.f.'s $F_1$ and $F_2$ (resp. $G_1$ and $G_2$) are the conditional d.f.'s of the quantile function $F^{-1}$ (resp. $G^{-1}$) given the subsets $\{F^{-1}\leq G^{-1}\}$ and $\{F^{-1}>G^{-1}\}$ of $(0,1).$
 Therefore, 
$
\pi(F,G)\leq \gamma(F,G).
$

Recall that $\rho(F,G) =P(X> Y)$ for independent r.v.'s $X$ and $Y$, 
leading to (\ref{relation8}) with $\lambda=\rho(F,G)$,  and $F_1$ and $F_2$  being the conditional d.f.'s of the first coordinate given the half-spaces $\{(x,y)\in \Rea^2: x \leq y \}$ and  $\{(x,y)\in \Rea^2: x > y \}$ of $\Rea^2$ equipped with the product probability associated to the d.f. $F(x)G(y)$ on $\Rea^2$
 and similarly for $G_1$ and $G_2$.

Other decompositions based on different dependence structures may be  of some interest, 
but instead let us focus on the  problem of searching for a pair $(X,Y)$, if it exists, 
that minimizes $\lambda$ in the decompositions (\ref{relation8}). This would result in the 
new suggestive index 
  \begin{equation}\label{newindex}
\upsilon(F,G):=\inf \{P(X>Y), (X,Y) \mbox{ with marginals d.f.'s } F \mbox{ and } G\}.
\end{equation}

Now,  since $\pi(F,G)$ is a lower bound for any $\lambda$ satisfying (\ref{relation8}),  we have that
\begin{equation}\label{relaciones}
\pi(F,G)\leq\upsilon(F,G)\leq \gamma(F,G).
\end{equation}

The next tasks are to show that the first inequality in \eqref{relaciones} is an equality, and find the expression of a pair $(X,Y)$  yielding the minimum in \eqref{newindex}. Before that, 
let us return to the example in Figure \ref{segundo_ejemplo}. If we couple the highest white bar to the lowest black one and couple the rest by order (see the right graph in Figure \ref{segundo_ejemplo}), we 
 would obtain  the lower possible value of $P(X>Y) =1/20$.  
The answer to the proposed tasks relies  on this construction.

 Now, let us begin by noticing that 
  under  s.d.  we  have $\ell(F^{-1}>G^{-1})=0$ thus both inequalities in 
(\ref{relaciones}) are equalities. Additionally, if  $0<\pi(F,G)=\pi_0$, 
then $F\leq_{st}^{\pi_0}G$. Thus, by Proposition \ref{trim_quantile}, $F^{-1}(y)\leq G^{-1}(\pi_0+y)$ 
holds for every $y \in (0,1-\pi_0)$.  Let us define
\begin{equation}\label{modifquantile}
\overline{G^{-1}}(y)=
\left\{
\begin{matrix}
G^{-1}(\pi_0+y), \quad &\mbox{ if }& y \in (0,1-\pi_0)
\\[0mm]
G^{-1}(y-(1-\pi_0)), \quad &\mbox{ if }& y \in [1-\pi_0,1).
\end{matrix}
\right.
\end{equation}

It is easy to  check that, seen as a r.v. defined on  $(0,1)$,
the d.f. of $\overline{G^{-1}}$ is also $G$. Our construction guarantees that $F^{-1}(y)\leq \overline{G^{-1}}(y)$ 
for every $y \in (0,1-\pi_0)$. Therefore $\ell(F^{-1}>\overline{G^{-1}})\leq \pi_0$; hence by definition of $\upsilon(F,G)$ 
and the first inequality in (\ref{relaciones}), 
$$\pi_0\leq\upsilon(F,G)\leq \ell(F^{-1}>\overline{G^{-1}})\leq \pi_0.$$ 

 {We summarize all these facts in the following proposition.

\begin{Prop}\label{relations}
Let $F,G$ arbitrary d.f.'s and $\gamma(F,G),$ $\rho(F,G), \pi(F,G)$,  and $\upsilon(F,G)$ be the indices defined in 
(\ref{gamma}), (\ref{rho}),  (\ref{pilevel0}) and (\ref{newindex}), respectively.  
 We have that
\begin{eqnarray*}
\pi(F,G)=\ell(F^{-1}>\overline{G^{-1}})
=
\upsilon(F,G)  \leq \min ( \gamma(F,G),   \rho(F,G)).
\end{eqnarray*}
\end{Prop}

 Proposition \ref{relations} shows that the indices $\pi$ and $\upsilon$ coincide, {
 thus in the sequel we will only use $\pi$. As a by-product,  we also know} that all the indices considered here can be expressed as $P(X>Y)$  for suitable pairs 
$(X,Y)$:
  the quantile functions $(F^{-1},G^{-1})$ in the case of $\gamma(F,G)$,   independent r.v.'s for $\rho(F,G)$, and  $(F^{-1},\overline{G^{-1}})$ in the case of  $\pi(F,G)$.


\begin{Nota}\label{Pi_Cosas_2}
Given $F$ and $G$, it is often possible to  consider  ``less extreme" couplings   than $(F^{-1},\overline{G^{-1}})$   to get the same result.  As an example, let us consider the situation in Figure \ref{segundo_ejemplo}. The pairing   shown in the right graph there is the one   given  by $(F^{-1},\overline{G^{-1}})$,   which associates  the 
 higher white income   with  the  lower black income, while the rest of the individuals are paired according to their (updated) ranks. The same  result is 
 obtained by transposing the third and fourth white bars, leaving the rest of the bars  unchanged.
\end{Nota}

 \begin{Nota}\label{generalidades}
 To get a proper coupling for particular d.f.'s $F,G$ we need to know the value $\pi(F,G)$. This was not necessary for  $\gamma(F,G)$ neither for $\rho(F,G)$. 

 Since we are mainly interested in  indices that are able to evaluate small deviations from s.d., a required property of any index  is to be 
 zero whenever s.d. holds. This property is shared by $\pi$ and $\gamma$ (or even $\epsilon$ or $ \epsilon_{2}$ as defined in (\ref{Levyindex}) and \eqref{TransportIndex}), but not by $\rho$.   
 \end{Nota}

\begin{Nota}\label{NotaUpperBound}

  It is also worth noting that the indices $\gamma$ and $\pi$ are loosely  related by the inequality $\pi(F,G)\leq \gamma(F,G)$. Thus,  if $\gamma(F,G)$ is small, both indices have similar values. 
  However,  it is enough to consider $G$ uniform on $(0,1)$ and $F$ uniform on $(\delta,1+\delta)$ for $\delta >0$ as small as desired, to obtain examples with $\gamma(F,G)=1$ and $\pi(F,G)=\delta$.
 A real example  with $\gamma$ large and $\pi$ small occurs in the INE data starting in 2008 (see Section \ref{INE_Data}).

To 
 clarify the role of the indices, let us assume that  $F$ and $G$ 
  are the assets 
 of each individual in a population four years ago and today.
Let us also assume that   the 
wealth has increased 
over this period.   This 
would be 
described by  $F\leq_{sd} G$ 
or, instead of strict s.d., by confidence intervals for $\gamma$ and $\pi$ with the upper extreme close to 0.
 Conversely, in a period of economic contraction, we should get a large  value for $\gamma$, while     
a small value of $\pi$ might be possible (meaning that the situation has worsened, but not too much for anyone). Alternatively, a large $\pi$ would indicate that  the economy of some 
people 
worsened a lot.

The $\rho$ index refers to the comparison of independently selected individuals from both populations. If samples of size $n$ from both populations are randomly paired, the number of pairs that do not agree with the improvement has a binomial distribution with parameters $(n, \rho(F,G))$. Obviously, small values would be associated with economic prosperity, while large values would be associated with contraction time.

 \end{Nota}

\begin{Nota}\label{NotaUpperBound_2}
Taking $\vartheta(F,G):=1-\pi(G,F),$ we obtain  
 \begin{eqnarray*}
\vartheta(F,G)=
 \sup\{P(X\geq Y): \ (X,Y) \mbox{ has marginal d.f.'s } F \mbox{ and } G\},
 \end{eqnarray*}
  giving an upper bound for the probability $P(X>Y)$ and the less favorable decomposition, if we are looking for s.d. of $G$ over $F$, in the way considered in (\ref{relation8}).   
 \end{Nota}

\section{Testing the levels of stochastic dominance}\label{testing}
In this section $X_1,\dots,X_n$ and $Y_1,\dots,Y_m$ are independent samples of i.i.d. r.v.'s with  d.f.'s $F$ and $G$.   $F_n$ and $G_m$  denote the respective  empirical d.f.'s based on the  $X's$  and $Y's$. 

 The Mann-Whitney version of Wilcoxon statistic (\cite{Mann1947}): $$U_{n,m}:=\#\{(i,j)\in\{1,\dots,n\}\times \{1,\dots,m\}  : (X_i >Y_j)\}$$ allows us to obtain a  natural estimator for $\rho(F,G)$ through
$
\hat \rho_{n,m}:=\rho(F_n,G_m)=\frac{U_{n,m}}{nm}.
$

This estimator has been widely analysed in the   literature  from the beginning 1950's (see \cite{Birnbaum} and references therein, \cite{Govindarajulu}, \cite{Yu}). Chapter 5 in \cite{Kotz}  describes the asymptotic properties of  $\hat \rho_{n,m}$ and   provides  asymptotic confidence intervals and bounds for $\rho(F,G)$ based on   the asymptotic normality of $\hat \rho_{n,m}$. Therefore we do not pursue on this topic here.

{We also use 
the  plug-in estimator
$\hat \pi_{n,m}:=\pi (F_n,G_m)  = \sup_{x\in \mathbb{R}}\left(G_m(x)-F_n(x)\right)$}
of the index $\pi(F,G)$. This is widely known as the one sided Kolmogorov-Smirnov statistic, with an important role in the framework of nonparametric goodness of fit and   to testing s.d. (see e.g. \cite{McFadden}, \cite{Barrett}, \cite{Linton}), although mainly in the context of testing  $H_0: F\leq_{st} G$ vs $H_a: F \nleq_{st} G$.
 
 The asymptotic distribution of $\hat \pi_{n,m}$ under the hypothesis $F=G$ was  obtained by Smirnov in the late 1930's, while \cite{Ragh73}   solved    the general   case:  
 
 \begin{Theo}[\cite{Ragh73}]
Let $F$ and $G$ be continuous, and   $n,m \to \infty$ 
in such a way that $\frac n {n+m} \to 
\lambda \in (0,1)$. If we denote $\Gamma(F,G):=\{x\in\mathbb{R}: 
G(x)-F(x)=\pi(F,G) \}$ and  $B_1(t)$ and $ B_2(t)$  are independent Brownian 
Bridges on $(0,1)$, then
\begin{eqnarray} \label{asympt}
{\textstyle \sqrt \frac {mn}{m+n} \left(\hat \pi_{n,m}-\pi(F,G)\right) \convw}
 \sup_{x \in 
\Gamma(F,G)}\left(\sqrt{\lambda} \ B_1(G(x))-\sqrt{1- \lambda} \  B_2(F(x))\right).
\end{eqnarray}
\end{Theo}

A more general result including a bootstrap version appears in \cite{Alv2014}. These results allow to develop consistent procedures for statistical assessment of almost s.d. like $F\leq_{st}^{\pi_0} G$ when rejecting the null at the fixed level. We refer to \cite{Alv2015} and 
 \cite{Alv2014}, for details and  
simulations showing the sample performance of the tests and confidence bounds relative to this index. 
A semiparametric approach, under  a density ratio model, and just a crossing point between the d.f.'s, has been developed in \cite{Zhuang}.}

  We estimate $\gamma(F,G)$ with the plug-in statistic $\hat \gamma_{n,m}:=\gamma(F_n,G_m)$.
  This  statistic  has been considered in the literature {
  mainly} to reject, for small enough values of  $\hat \gamma_{n,m}$, the null hypothesis that the treatment has no effect ($F=G$), in favour of the alternative that the treatment 
  increases the  values  ({
  $F>_{st}G$}). 
  
 {Galton's rank-order statistic    coincides with  {
 $n \hat \gamma_{n,n}$. As reported in \cite{Hodges-Galton}, and included in  Feller's celebrated book (\cite{Feller1}), Galton used it 
 to answer a question by  Darwin about 
 two samples of size 15 in which the order was reversed only twice.  Galton considered this as a  rare event but, as showed by Hodges   
 if $F=G$, under continuity, the $p$-value associated with 
 Darwin's data is 
 3/16, which is not as rare as Galton thought. 
 }
 

The available results on $\hat \gamma_{n,m}$ include \cite{Gross}, which treated the  
 case where 
 $\ell(F^{-1}=G^{-1})>0$. Then, \cite{Alv2017} analysed the case of d.f.'s with a unique crossing point (as is usually the case in location-scattering families).   
 The theory developed in that paper 
 was adapted in \cite{ZhuangEtAl2019}   under the additional assumption of  an exponential density ratio model and using semiparametric estimates of the quantile functions,  to cover 
 a finite number of crossings.  Finally, \cite{del Barrio2021} shows the complex behaviour that 
  $\hat \gamma_{n,m}$ 
 exhibits in the case of a finite number of  contact points between $F$ and $G$ (explained below) 
 in the non-parametric setting.
 In Section 4 
 in the supplement, there is 
 some 
 new  theory  on this index. 

For d.f.'s that coincide on some interval,  $\gamma(F,G)$ cannot be consistently estimated by its  plug-in version 
(\cite{Gross}  and 
\cite{del Barrio2021}). That last paper 
shows}  that the key of the problem   involves the {\it set of contact points} between the  composed function $F(G^{-1})$ (or $G(F^{-1})$) with the identity. I.e., the set  $(t\in [0,1]: F(G^{-1})(t)=t)$,  where we define  $F(G^{-1})$ at 0 and 1 by continuity. In particular, Theorem 1.1 in \cite{del Barrio2021} shows that as $m,n\to \infty$,  
\begin{eqnarray*}  
\hat \gamma_{n,m} \convs \gamma(F,G), 
\mbox{ 
if and only if } \ell(t\in (0,1): F(G^{-1}(t))=t)=0.
\end{eqnarray*}



 
 The analysis on the asymptotic distribution of $\hat \gamma_{n,m}$ relies on a careful analysis of the situation of the contact points and their {\it orders of  contact}.  Let us say that a contact point $t_0\in[0,1]$ is  {\it regular} if $\Delta(h):=F(G^{-1}(t_0+h)) - (t_0 + h)$   is Lipschitz in a neighborhood of $t_0$ and there exist $r_L,r_R \geq 1$ and $C_L,C_R\neq 0$, depending on $t_0$, such that 
 \begin{equation}\label{EqOrden}
\Delta(h) = 
\left\{
\begin{matrix} 
C_L |h|^{r_L}  + o(|h|^{r_L}),& & \mbox{ if } h <0,
\\[0mm]
C_R |h|^{r_R}  + o(|h|^{r_R}), & &\mbox{ if } h >0.
\end{matrix}
\right.
\end{equation}
In the extreme cases $t_0=0 \mbox{ or }1$ only positive (resp. negative) values of $h$ are considered.

\begin{figure}
\begin{center}
\includegraphics[width=7.8cm,height= 5.1cm]{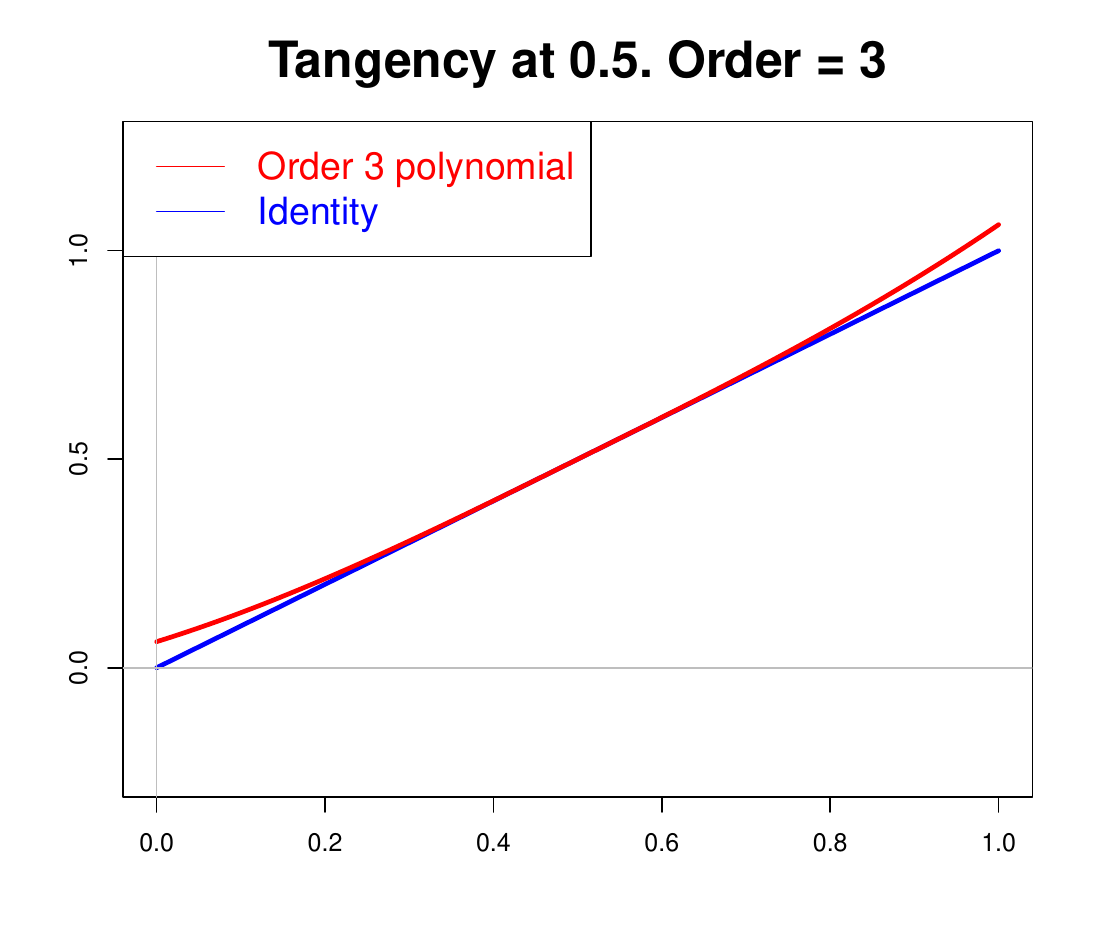}
\hspace{3mm}
\includegraphics[width=7.8cm,height= 5.1cm]{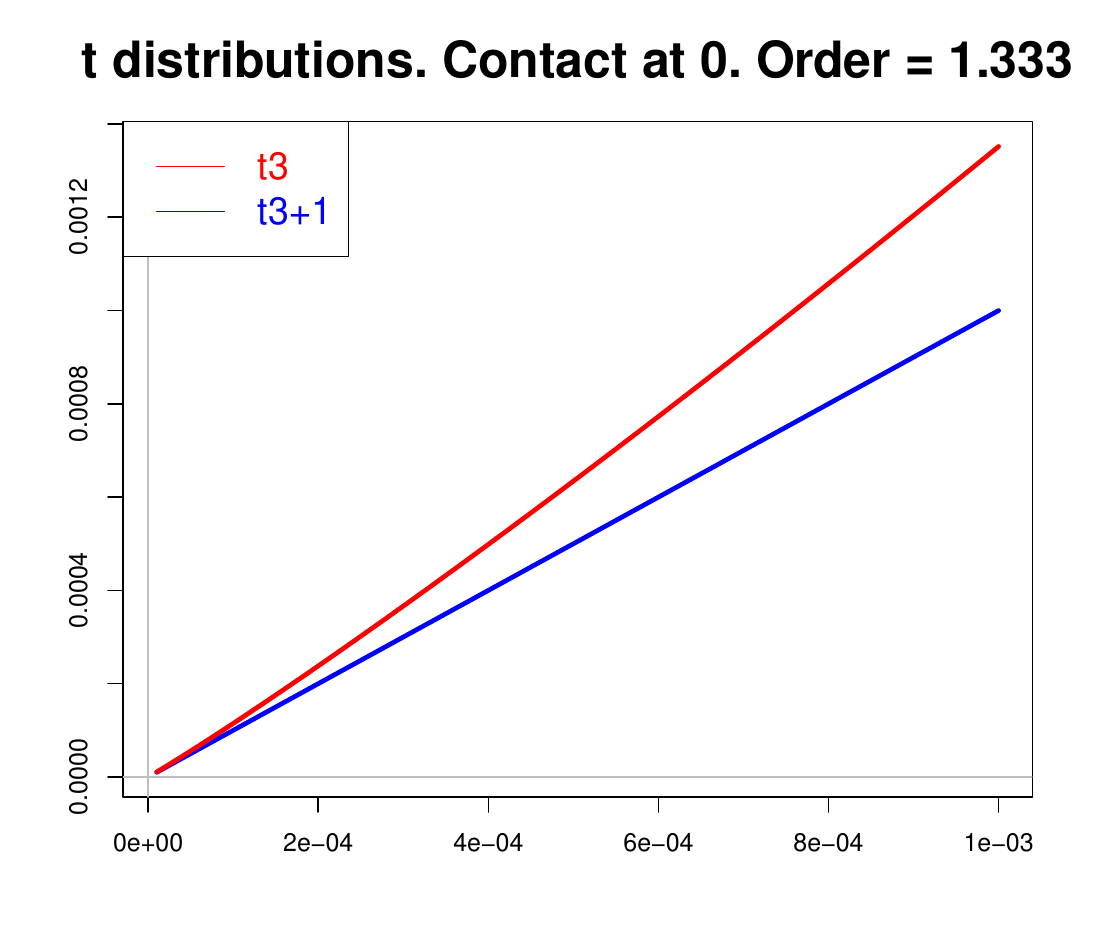}
\end{center}

\vspace{-5mm}
\caption{Two examples of contact. In the left a tangency that is  an order 3 contact. In the right a contact at 0 with a non-integer order. 
\label{Fig.Contactos}}
\end{figure}

\begin{Nota}
Expression \eqref{EqOrden} is closely related to the Taylor expansions. Main differences being that $r_L$ and $r_R$ (resp. $C_L$ and $C_R$) may be different and that  $r_L$ and $r_R$ are not necessarily integers. An additional minor difference is shown in the left hand graph in Figure \ref{Fig.Contactos}: there the contact order at 0.5  is 3 but it does not correspond to a crossing.
The other graph in Figure \ref{Fig.Contactos} shows a case  in which,  for the contact at 0, $r_R$ is not integer (here $F$ is the d.f. of a r.v., $X$, with Student distribution $t_3$ and $G$ is the d.f. of $X+1$).

The main point of interest in Figure \ref{Fig.Contactos} is that it shows that the difficulty of estimating $\gamma(F,G)$ increases considerably with the intensity of the contact. It is easiest to see that $\gamma(F,G)= 0$ in the second graph than in the first. Additionally, if the order-3 contact, shown in the graph on the left, had occurred at $t=0$, estimation would have been easier, and this would  have increased the value of $r_0$  in Theorem \ref{TCL.principal}  by a half (see Theorem 4.1 in the supplement).

\end{Nota}

The next theorem  
covers   all the situations involving  finite sets of
contact points 
between smooth enough d.f.'s  (for a more precise statement, see Theorem 4.1  
in the supplement).
\begin{Theo}\label{TCL.principal}
Assume that    $t_1,\ldots,t_k$ are the contact points between $F(G^{-1})$  and the identity and that  they are regular.
Then, there exist $r_0\geq 0.5$ and a non-degenerate  law $\mu$ such that
$$(n+m)^{\frac 1 {2r_0}}(\hat \gamma_{n,m}-\gamma(F,G)) \convw \mu.$$

Here $\mu$ and $r_0$ depend on the values $r_L(t_i),r_R(t_i)$ and, also on  when 0 or 1 is a contact point;  $\mu$ also depends on when the contacts are tangencies or crossings.
\end{Theo}


 The optimal convergence rate  $2r_0=1$ is reached when there exists an order one contact on 0 or 1. Any value $r_0\in [1/2,\infty)$ can be reached, see \cite{del Barrio2021}.

   The only case of a Gaussian $\mu$ happens, with $r_0=1$, when  $F$ and $G$ have  densities, $f,g$, positive on possibly unbounded intervals where they are   continuously differentiable with $F^{-1}(t_i)=G^{-1}(t_i)$ and $f(F^{-1}(t_i))\neq g(G^{-1}(t_i)), i=1,\ldots,k$ (see Theorem 4.9 in \cite{del Barrio2021}). Such 
   Gaussian  law  is also  the one  obtained  in \cite{ZhuangEtAl2019}.

Given the different rates and possible limit laws, the bootstrap is 
a good candidate to produce asymptotic 
confidence bounds for $\gamma(F,G)$. Unfortunately,  our simulations when resorting to the naif bootstrap (with sizes $\tilde n =n,\tilde m=m$) are dissapointing, even for sample sizes as high as $n=m=50000$, because the simulations often show a notable dependence of the bootstrap law on the initial samples, making such law
 useless (for additional details see Remark 4.5 in the supplement)}. Instead,  if $\{t_1,\ldots, t_k\}\subset (0,1)$ 
a bootstrap 
with sizes $\tilde n=o(\frac n {\log \log n}), \ \tilde m=o(\frac m {\log \log m})$ 
 works (see Theorem 4.2 in Section 4 
 in the supplement).

\begin{Nota}\label{NotaSobreGamma}
%
The value $r_0$ {
in Theorem \ref{TCL.principal}} is unknown in the applications. However, we can use several 
  resampling sizes with different orders, say $\tilde {\tilde n},\tilde {\tilde m}$ and $\tilde n, \tilde m$ (see Section 4 
  in the supplement) to obtain different  values, $\tilde z_\alpha, \tilde {\tilde z}_\alpha$, of the $\alpha-$quantile  bootstrap estimates 
  to obtain that the exponent  must verify
\begin{equation}\label{res_sizes}
\frac 1 {2r_0}\approx  \frac {\log \left((\tilde z_{\beta}-\tilde z_\alpha)/(\tilde {\tilde z}_{\beta}-\tilde {\tilde z}_\alpha)\right)}  {\log \left(({\tilde {\tilde n}+\tilde {\tilde m})/(\tilde n +\tilde m})\right)}\mbox{ , for any } \alpha, \beta\in (0,1).
\end{equation} 

For applications  we 
compute a confidence lower bound for $r_0$ from several bootstrap replicates of (\ref{res_sizes}), handling different resampling sizes and values $\alpha, \beta$  (see 
Subsection 5.1 
in the supplement). 
 \cite{Bertail} includes a 
discussion on this kind of problem. 
%

 Last challenge is the choice of 
the resampling sizes $\tilde n, \tilde m$. Too small or unbalanced rates of $\tilde n/n$ and $\tilde m/m$  would produce inadequate bootstrap samples,   being a poor representation of the original ones and giving unstable results.
However, we can take $\tilde n, \tilde m$ as large as desired as long as the relations  $\tilde n=o(\frac n {\log \log n}), \ \tilde m=o(\frac m {\log \log m})$ hold. In the applications, our choice has been to  take  (rounded values) $\tilde n=n/(n+m)^{0.05}$ and $\tilde m=m/(n+m)^{0.05}$.

\end{Nota}

\section{Simulated and real data examples}\label{SCS}

In this section we analyse  some simulated and  real data sets. They cover situations with relatively large, medium and small sample sizes. We describe and analyse them in Subsections  \ref{simulations} and \ref{realDataSets}.  
The computational technical details appear in  Section 5 
 in the supplement.

\subsection{Simulations}\label{simulations}
 We have assumed that in practice we compare distributions belonging to the same family, and 
 we have simulated 
 pairs of normal or uniform distributions (to cover the cases of bounded and unbounded support)  with 
  one distribution dominating the other, except in a small part of the left tail. {
 We considered sample sizes $n=m \in \{250,1000,5000,15000\}$, but here we only present the cases $n=m=250$ (resp. 15000) for the indices $\pi$ and $\rho$ (resp. $\gamma$)  in Tables \ref{Table.Normal_15000} to \ref{TablaSimulUnif_rho_250}. The  results for $n=m \in \{250,1000,5000\}$ appear  
in Tables 2 
to 7 
in Section 6 
in the supplement. The results are quite good for the indices $\pi$ and $\rho$,  with good covering rates even for $n=m=250$ (with $\pi$ giving  over-covering for all sample sizes). In contrast, the results for $\gamma$ are rather poorer, even for high sizes, such as 15000.
 }

We have taken $\gamma =0.01, 0.05, 0.10$ and we have fixed the parameters to get these exact values of   $\gamma$. 
 In the normal case, we always take  $F=N(0,1)$. Then, for instance,  we  fix $\sigma =1.1$, take $\mu= 0.233
 $, and handle  $G=N(\mu,\sigma^2)$; this choice 
 gives $\gamma(F,G) =0.01.$ 
 In the uniform family, we always take  $F=U(0,1)$. Then  we  take  pairs $(a,H), a<0, H>1$,  and handle  $G=U(a,H)$. 
 The same pairs of distributions are used for the remaining indices. The specific values of the parameters are given in the  tables.

For each data set, we have  computed   $\hat \gamma_{n,m},\hat  \pi_{n,m}$ and $\hat \rho_{n,m}$,  
the extremes of the associated confidence intervals and whether or not these intervals cover the true value of the index. We have also  estimated the value of $2r_0$  involved in the asymptotic distribution of $\hat \gamma_{n,m}$.

For each combination of parameters, our tables show the mean values of the estimated quantities along  500 repetitions. Below each mean, between parenthesis, we include the standard deviation of the obtained values.

\subsubsection{Gaussian simulations}\label{SimulGauss}
We analyse the results obtained for  $F=N(0,1)$ vs $G=N(\mu,\sigma^2)$ for the pairs $(\mu,\sigma)$ shown in the tables. As noted,  we present here only the results for $n=15000$ (resp. $n=250$) for $\gamma$ (resp. $\pi$ and $\rho$), although the comments include all cases.  The situations for  all the  considered sample sizes  are similar for $\pi$ and $\rho$ but not for $\gamma$ {
which seems to require larger sample sizes to work properly. In particular, the coverages obtained suggest that the use of the $\gamma$  index 
should not be considered for sample sizes below 1000 and to be extremely cautious for sample sizes below  5000}.

{
\noindent{\bf Simulations for $\gamma$.} The means of $\widehat{2r_0}$ are quite stable, taking values between 1 and 2 (remember that    
 $2r_0=2$ if  the contact point is   in (0,1)). With size 250, we obtained some negative estimates (always less than 5\% of the time, except in the comparison with $N(2.326,2^2)$, where  it happened 15\% of times; in the uniform case these proportions are much higher). This did not happen for sizes 1000 and above. 

}
{ 
}

\begin{table}[ht]
\centering
{\tiny
\begin{tabular}{cccccccc}
  \hline
$\gamma$ &  $\mu$ &  $\sigma$  & $\widehat{2r_0}$ & $\hat \gamma_L$ & $\hat \gamma$ & $\hat \gamma_U$ & Coverage  \\ 
  \hline
 0.01 &0.233 & 1.10 & 1.658 &0.0010 &0.0118 &0.0200 &0.6400 \\[-1mm] 
 &  &  &(0.474) &(0.0038) &(0.0096) &(0.0163) &(0.4805) \\ 
  & 1.163 & 1.50 & 1.927 &0.0046 &0.0100 &0.0152 &0.8540 \\ [-1mm] 
 &  &  &(0.420) &(0.0031) &(0.0029) &(0.0041) &(0.3535) \\ 
  & 2.326 & 2.00 & 1.757 &0.0066 &0.0100 &0.0134 &0.8920 \\ [-1mm] 
 &  &  &(0.260) &(0.0020) &(0.0018) &(0.0023) &(0.3107) \\ 
\hline
0.05 &0.164 & 1.10 & 1.899 &0.0110 &0.0494 &0.0805 &0.7200 \\ [-1mm] 
 &  &  &(0.461) &(0.0168) &(0.0237) &(0.0384) &(0.4494) \\ 
  &0.822 & 1.50 & 1.713 &0.0381 &0.0501 &0.0618 &0.8980 \\ [-1mm] 
 &  &  &(0.262) &(0.0070) &(0.0065) &(0.0079) &(0.3030) \\ 
 & 1.645 & 2.00 & 1.733 &0.0426 &0.0501 &0.0575 &0.9340 \\ [-1mm] 
 &  &  &(0.185) &(0.0042) &(0.0039) &(0.0045) &(0.2485) \\ 
 \hline
0.10 &0.128 & 1.10 & 1.970 &0.0386 &0.1029 &0.1597 &0.7760 \\ [-1mm] 
 &  &  &(0.529) &(0.0369) &(0.0375) &(0.0566) &(0.4173) \\ 
  &0.641 & 1.50 & 1.706 &0.0829 &0.0995 &0.1156 &0.8900 \\ [-1mm] 
 &  &  &(0.217) &(0.0099) &(0.0089) &(0.0104) &(0.3132) \\ 
 & 1.282 & 2.00 & 1.759 &0.0896 &0.1000 &0.1103 &0.9100 \\ [-1mm] 
 &  &  &(0.166) &(0.0059) &(0.0057) &(0.0063) &(0.2865) \\
   \hline
\end{tabular}
}
\caption{Means of the {
estimates of $2r_0$,  of $\gamma$ and of its confidence intervals at  level 0.05}, from 500  simulations of $N(0,1)$ vs  $N(\mu,\sigma)$. Sizes $n=m=15000$. Coverage is the proportion of times in which $\gamma\in (\hat \gamma_L,\hat\gamma_U)$. Between parenthesis,  standard deviations of the  estimations. } 
\label{Table.Normal_15000}
\end{table}

The mean  of $\hat \gamma$ is reasonably close to the target except for the case $\sigma=1.1$. In this case the  standard deviations are relatively large although they decrease with the sample size  and are reasonable for the size  15000.
However, the coverage is not satisfactory enough: it is quite low when $\sigma=1.1$,  improves when $\sigma=1.5,2$, but it never reaches the nominal 0.95.

This could be due to the difficulty of the problem. Figure \ref{Fig.QF_Gaussians_500} shows one example for each variance.   It seems that, when $\sigma=1.1$, we have two high-order contacts 
at   $t=0,1$. If $\sigma=2$ a not so high order   crossing appears  at $t= 0.1.$ 
  The case $\sigma=1.5$ is intermediate.

\begin{center}
\begin{figure}[h] 
\centering
\includegraphics[width=7.9cm,height= 5.5cm]{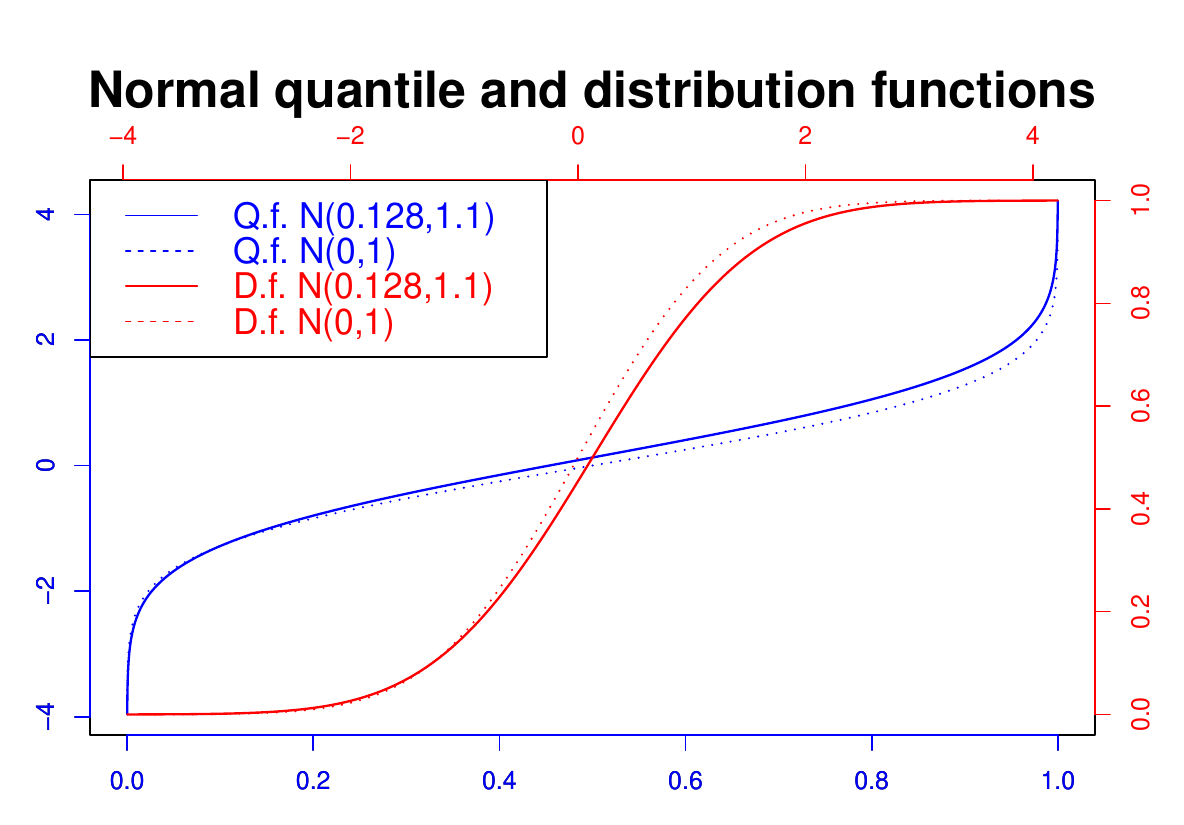}\hspace{3mm}
\includegraphics[width=7.9cm,height=5.5cm]{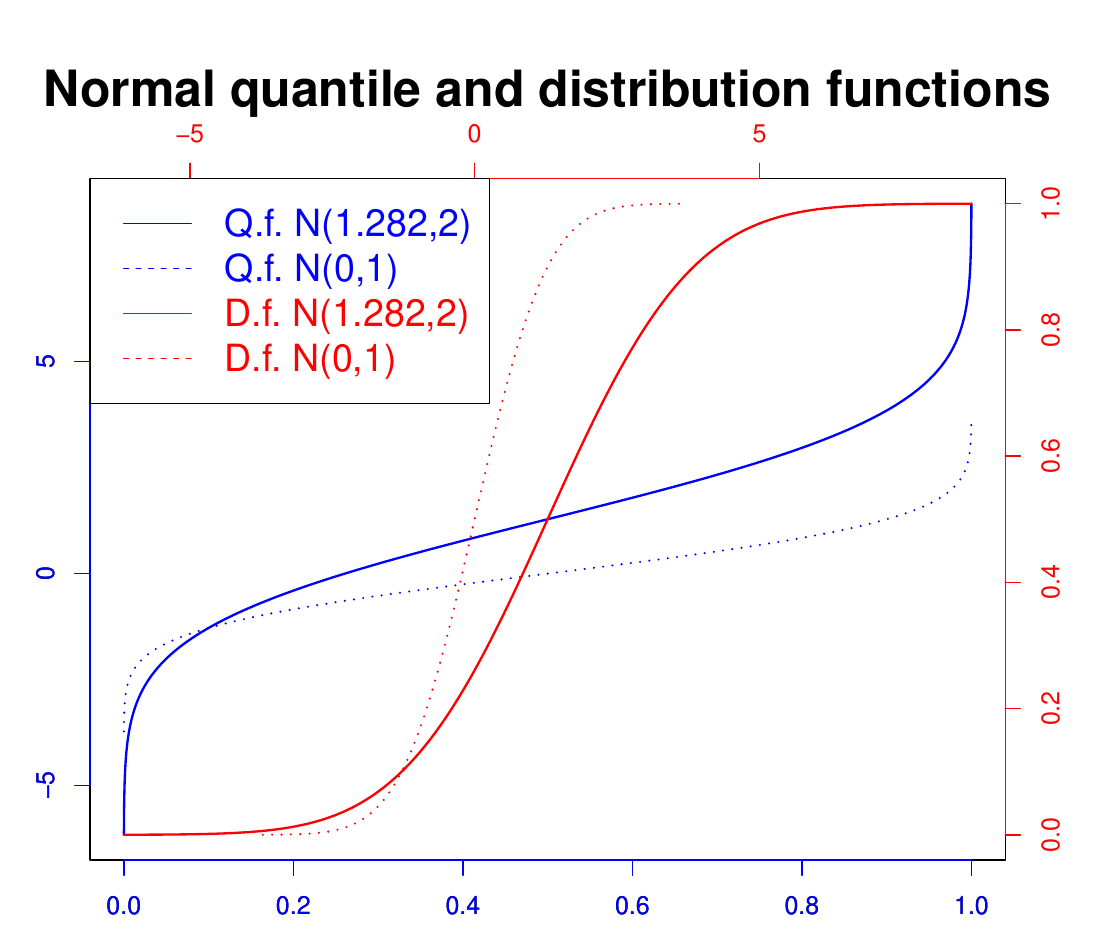}
\caption{Simultaneous quantile (in blue) and distribution functions (in red)  plots.}
\label{Fig.QF_Gaussians_500}
\end{figure}
\end{center}

\vspace{-1cm}
\noindent{{\bf Simulations for $\pi$.}
 The coverages in this case (including the size 250) are quite high, always above 0.98 and quite often 1. This could be related to the conservative upper bound handled for $\pi$, based on the worse possible variance (see   \cite{Alv2015}).

 The values of $\pi$ to be estimated are small (the largest is 0.0277) and the lower extremes of the confidence interval are quite often 0, while the upper ones are relatively large when $\sigma=1.1$ and decrease with $\sigma$ as well as with the sample size.} For size 5000, the upper values are 20 times $\pi$ when $\sigma=1.1$, decreasing to less than 3 times when $\sigma=2.0$. The ratio also decreases with $\mu$ giving a minimum of 1.29 times $\pi$ for $\pi=0.0277$.

Regarding the relation with $\sigma$, Figure \ref{Fig.QF_Gaussians_500} shows that in all cases,  
$\pi$ is reached slightly above 0.2; moreover, the  d.f. $F$  
becomes flatter in this region as $\sigma$ increases, making it easier to estimate $\pi$.

\begin{table}[ht]
\centering
{\tiny
\begin{tabular}{ccccccc}
  \hline
 $\pi$ & $\mu$ & $\sigma$  & $\hat \pi_L$ & $\hat \pi$ & $\hat \pi_U$ & Coverage \\ 
  \hline
0.0004    &0.233   &  1.10       &  0 &0.0172    & 0.0712    &  1 \\[-1mm] 
   &    &         &  (0) &(0.0149)    & (0.0231)     &  (0) \\ 
0.0019    & 0.164    &  1.10       &  0 &0.0270    & 0.0843    &  1 \\ [-1mm]
   &    &         &  (0) &(0.0202)    & (0.0268)     &  (0) \\ 
0.0039    & 0.128    &  1.10       & 0.0001    & 0.0321    & 0.0938    &  1 \\ [-1mm]
   &    &      & (0.0016)    & (0.0233)    & (0.0286)     &  (0) \\ 
   \hline 
0.0016    & 1.163   &  1.50       &  0 &0.0082    & 0.0310    & 0.992 \\ [-1mm]
   &    &         &  (0) &(0.0072)    & (0.0120)     & (0.089) \\ 
0.0081    & 0.822    &  1.50       &  0 &0.0199    & 0.0521    &  1 \\ [-1mm]
   &    &         &  (0) &(0.0122)    & (0.0162)     &  (0) \\ 
0.0165    & 0.641    &  1.50       &  0 &0.0319    & 0.0659    &  1 \\ [-1mm]
   &    &         &  (0) &(0.0160)    & (0.0192)     &  (0) \\ 
   \hline 
0.0026    & 2.326   &  2.00       &  0 &0.0081    & 0.0262    & 0.982 \\ [-1mm]
   &    &         &  (0) &(0.0066)    & (0.0103)     & (0.133) \\ 
0.0136    & 1.645   &  2.00       &  0 &0.0230    & 0.0483    &  1 \\ [-1mm]
   &    &         &  (0) &(0.0113)    & (0.0142)     &  (0) \\ 
0.0277    & 1.282   &  2.00       &  0 &0.0393    & 0.0683    & 0.994 \\ [-1mm]
   &    &         &  (0) &(0.0166)    & (0.0212)     & (0.077) \\ 
   \hline
\end{tabular}
  }
\caption{Means of the estimations of $\pi$ and  its confidence intervals at  level 0.05, from 500  simulations of $N(0,1)$ vs  $N(\mu,\sigma)$. Sizes $n=m=250$. Coverage is the proportion of times in which $\pi\in (\hat \pi_L,\hat\pi_U)$. Between parenthesis the standard deviations of the above estimates.} 
\label{TablaSimulNormal_pi_250}
\end{table}

\vspace{3mm}

\noindent{\bf Simulations for $\rho$.}
It seems that $\rho$ is easier to estimate than $\gamma$ and $\pi$, because here  the values of $\hat \rho$ and the coverages are quite close to the targets  for every combination of parameters and sample sizes, and the standard deviations are small compared to the $\rho$'s.

A noticeable problem with this parameter is its small variation between cases. Take, for example, the cases with $\sigma=1.1$. Here the highest values of $\gamma$ and $\pi$ in  Tables \ref{Table.Normal_15000} and \ref{TablaSimulNormal_pi_250} are about 10 times the lowest. However, the values of $\rho$ only go from 0.5344 to 0.5623, making it difficult to distinguish between situations.
\begin{table}[ht]
\centering
{\tiny
\begin{tabular}{cccccccc}
  \hline
  	 $\rho$  & 	  $\mu$  & 	  $\sigma$    & 	  $\hat \rho_L$  & 	  $\hat \rho$  & 	  $\hat \rho_U$  & 	  Coverage \\ 
  \hline
0.5623 &0.233 & 1.10 &0.5121 &0.5624 &0.6126 &0.9520 \\ [-1mm]
&&&(0.0257) &(0.0253) &(0.0249) &(0.2140) \\ 
0.5439 &0.164 & 1.10 &0.4927 &0.5432 &0.5937 &0.9540 \\ [-1mm]
&&&(0.0266) &(0.0263) &(0.0260) &(0.2097) \\ 
0.5343 &0.128 & 1.10 &0.4850 &0.5356 &0.5862 &0.9560 \\ [-1mm]
&&&(0.0255) &(0.0253) &(0.0251) &(0.2053) \\ 
\hline
0.7406 &1.163& 1.50 &0.6956 &0.7394 &0.7832 &0.9500 \\ [-1mm]
&&&(0.0237) &(0.0222) &(0.0207) &(0.2182) \\ 
0.6758 &0.822 & 1.50 &0.6284 &0.6758 &0.7232 &0.9480 \\ [-1mm]
&&&(0.0253) &(0.0241) &(0.0230) &(0.2222) \\ 
0.6389 &0.641 & 1.50 &0.5884 &0.6374 &0.6864 &0.9460 \\ [-1mm]
&&&(0.0255) &(0.0246) &(0.0238) &(0.2262) \\ 
\hline
0.8509 &2.326& 2.00 &0.8145 &0.8497 &0.8850 &0.9440 \\ [-1mm]
&&&(0.0204) &(0.0182) &(0.0160) &(0.2302) \\ 
0.7690 &1.645& 2.00 &0.7271 &0.7699 &0.8128 &0.9600 \\ [-1mm]
&&&(0.0222) &(0.0206) &(0.0189) &(0.1962) \\ 
0.7168 &1.282& 2.00 &0.6710 &0.7175 &0.7639 &0.9480 \\ [-1mm]
&&&(0.0246) &(0.0232) &(0.0218) &(0.2222) \\ 
   \hline
\end{tabular}
}
\caption{
Means of the estimates of $\rho$ and  of its confidence intervals at 0.05 level, from 500 simulations of $N(0,1)$ vs  $N(\mu,\sigma)$. Sizes $n=m=250$. Coverage is the proportion of times that $\rho\in (\hat \rho_L,\hat \rho_U)$. Between parenthesis are the standard deviations of the above estimates.} 
\label{TablaSimulNormal_rho_250}
\end{table}

 \subsubsection{Simulations from uniform distributions}\label{SimulUnif}
 We analyse the cases  $F=U(0,1)$ against $G=U(a,H)$ for the pairs $(a,H)$ shown in the tables. 
 We only present here the results  for  the size 15000 (resp. 250) for $\gamma$ (resp. $\pi,$ and $\rho$), while the comments apply to all sizes. The problems we noticed for $\gamma$  in the Gaussian case appear here too, although the variation in  size is 
smaller. The results here are generally better than there because (as can be seen in Figure \ref{Fig.QF_Uniform_500}) the contacts are  clearer  here

\noindent{\bf Simulations for $\gamma$.} The estimator  $\widehat{2r_0}$ is  quite stable with means  between 0.982 and 1.771 and  reasonable standard deviations. Remember that the right value is 1 or 2 depending on whether we assume that the contact point is 0 or greater than 0; this is coherent with the fact that the mean value of $\widehat{2r_0}$ increases with $\gamma$ and, also, with the sample size (the larger the size, the better the chance of detecting that the crossing is not at 0).

{

As for the negative values of $\widehat{2r_0}$, we only got them with size 250. We got 20\% of them when $\gamma=0.01$ and at most $2\%$ cases when $\gamma=0.05$ or $0.10$.

 \begin{table}[ht]
\centering
{\tiny
\begin{tabular}{ccccccccc}
  \hline
  $\gamma$ & $a$ & $H$  & $\widehat{2r_0}$ & $\hat \gamma_L$ & $\hat \gamma$ & $\hat \gamma_U$ & Coverage \\ 
  \hline
0.01 & -0.051 & 6.00  & 1.619 & 0.0081 &0.0100 &0.0117 &0.9140 \\ [-1mm]
  &  &  & (0.148) &(0.0010) &(0.0010) &(0.0012) &(0.2806) \\ 
& -0.101 & 11.00  & 1.595 & 0.0083 &0.0100 &0.0116 &0.8840 \\ [-1mm]
  &  &  & (0.136) &(0.0009) &(0.0010) &(0.0011) &(0.3205) \\ 
& -0.202 & 21.00  & 1.567 & 0.0083 &0.0099 &0.0114 &0.9120 \\ [-1mm]
  &  &  & (0.130) &(0.0008) &(0.0008) &(0.0009) &(0.2836) \\ 
  \hline
0.05 & -0.050 & 1.95  & 1.736 & 0.0423 &0.0504 &0.0574 &0.9280 \\ [-1mm]
  &  &  & (0.181) &(0.0038) &(0.0038) &(0.0047) &(0.2587) \\ 
  & -0.100 & 2.90  & 1.749 & 0.0445 &0.0500 &0.0551 &0.9060 \\ [-1mm]
  &  &  & (0.150) &(0.0029) &(0.0030) &(0.0033) &(0.2921) \\ 
  & -0.200 & 4.80  & 1.751 & 0.0456 &0.0500 &0.0542 &0.9440 \\ [-1mm]
  &  &  & (0.127) &(0.0021) &(0.0022) &(0.0024) &(0.2302) \\ 
  \hline
0.10 & -0.050 & 1.45  & 1.708 & 0.0823 &0.1004 &0.1154 &0.8980 \\ [-1mm]
  &  &  & (0.216) &(0.0091) &(0.0087) &(0.0106) &(0.3030) \\ 
  & -0.100 & 1.90  & 1.742 & 0.0895 &0.1003 &0.1101 &0.9160 \\ [-1mm]
  &  &  &  (0.167) &(0.0056) &(0.0056) &(0.0064) &(0.2777) \\ 
  & -0.200 & 2.80  & 1.771 & 0.0928 &0.1003 &0.1074 &0.9320 \\ [-1mm]
  &  &  & (0.134) &(0.0039) &(0.0040) &(0.0043) &(0.2520) \\    \hline
\end{tabular}
}
\caption{Means of the  estimates of $2r_0$,  of $\gamma$ and of its confidence intervals at  level 0.05, from 500 simulations of $U(0,1)$ vs  $U(a,H)$. Sizes $n=m=15000$. Coverage is the proportion of times in which $\gamma\in (\hat \gamma_L,\hat\gamma_U)$. Between parenthesis, standard deviations of the  estimations. } 
\label{TablaSimulUnif_gamma_15000}
\end{table}
}

{
The means of $\hat \gamma$ are quite close to the target, with reasonable standard deviations. These deviations decrease with $H$ (and of course with the sample size).

The coverage increases with the sample size  and  $\gamma$. The coverages for $\gamma=0.01$ are well below the target;  for $n=15000$ the minimum  is 0.884, with all but two values above 0.90. }
\begin{center}
\begin{figure}[h] 
\centering
\includegraphics[width=7.8cm,height=5.5cm]{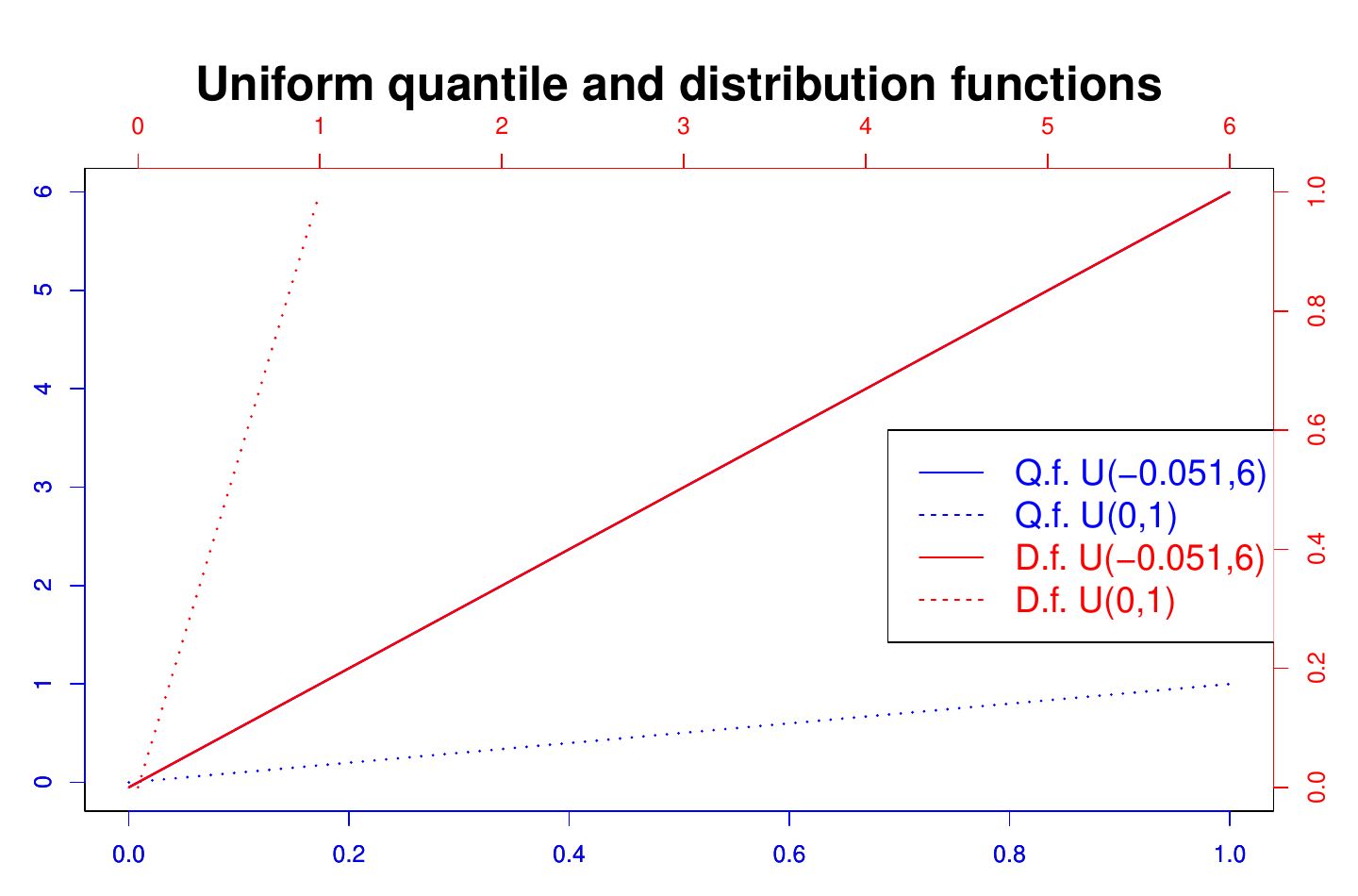}\hspace{3mm}
\includegraphics[width=7.8cm,height=5.5cm]{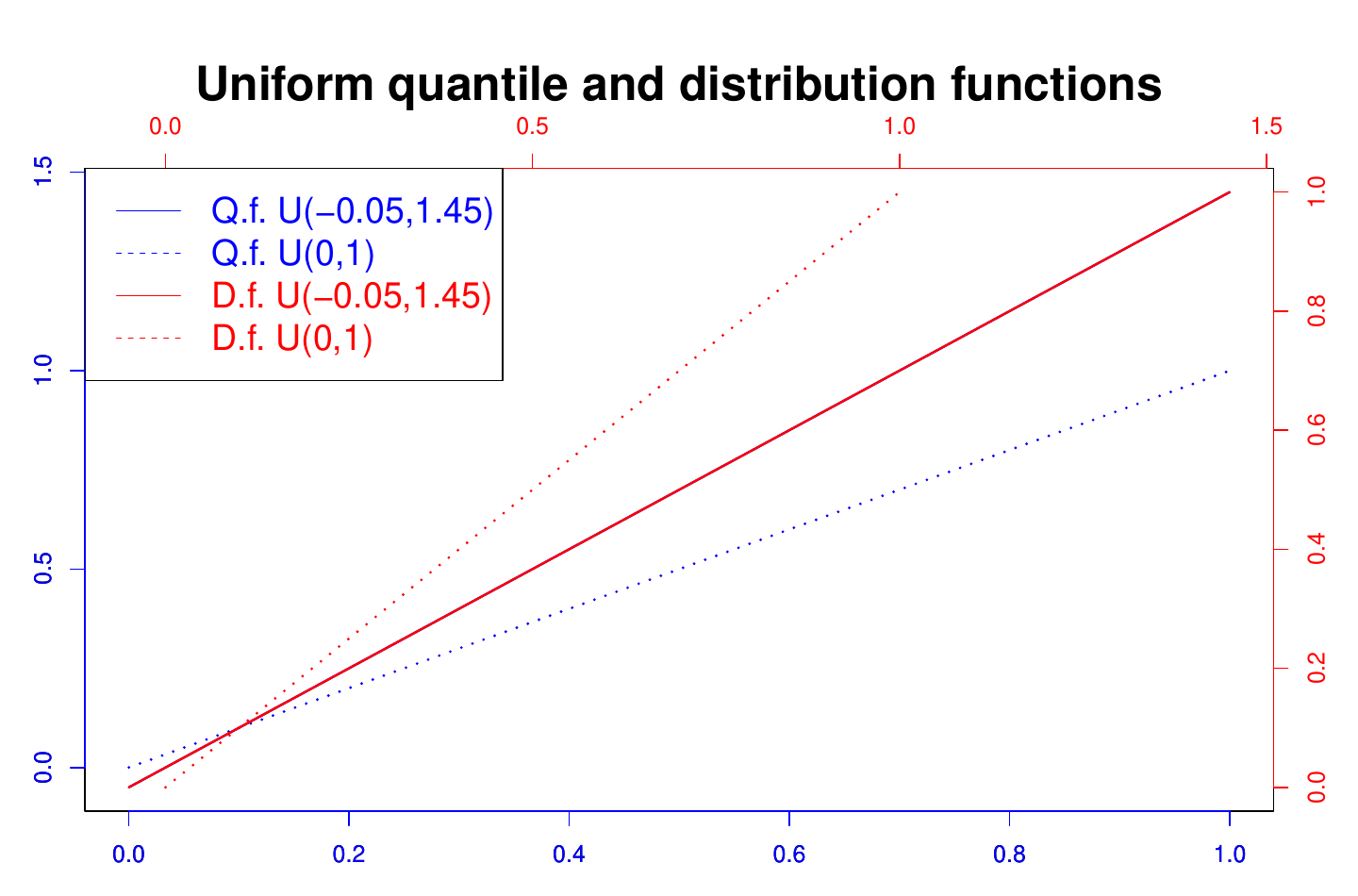}
\caption{ Simultaneous quantile (in blue) and distribution functions (in red) plots for several  Uniform distributions}
\label{Fig.QF_Uniform_500}
\end{figure}
\end{center}

\vspace{-1cm}
\noindent {\bf Simulations for $\pi$ and $\rho$.}
We do not pay special attention to these  indices 
because, in all cases  they are 
correctly estimated with reasonable coverings.

 \begin{table}[ht]
\centering
 {\tiny
\begin{tabular}{cccccccc}
  \hline
 $\pi$ & $a$ & $H$ & $\hat \pi_L$ & $\hat \pi$ & $\hat \pi_U$ & Coverage \\ 
  \hline
0.0083 & -0.051 & 6 & 0 &0.0088 &0.0221 &0.9300 \\ [-1mm]
   &  &  & (0) &(0.0058) &(0.0091) &(0.2554) \\ 
0.0091 & -0.101 & 11 & 0 &0.0093 &0.0214 &0.9200 \\ [-1mm]
   &  &  & (0) &(0.0060) &(0.0100) &(0.2716)  \\ 
0.0095 & -0.202 & 21  & 0 &0.0099 &0.0220 &0.9220 \\ [-1mm]
   &  &  & (0) &(0.0062) &(0.0102) &(0.2684)  \\ 
 \hline 
0.0250 & -0.050 & 1.95  & 0 &0.0302 &0.0514 &0.9840 \\ [-1mm]
 &  &    & (0) &(0.0113) &(0.0154) &(0.1256)  \\ 
0.0333 & -0.100 & 2.90 &  0 &0.0353 &0.0572 &0.9500 \\ [-1mm]
 &    &  & (0) &(0.0115) &(0.0154) &(0.2182)  \\ 
0.0400 & -0.200 & 4.80 & 0 &0.0405 &0.0640 &0.9540 \\ [-1mm]
   &  &  & (0) &(0.0124) &(0.0161) &(0.2097)  \\ 
 \hline 
0.0333 & -0.050 & 1.45  & 0 &0.0398 &0.0650 &0.9720 \\ [-1mm]
 &  &  &   (0) &(0.0138) &(0.0201) &(0.1651)  \\  
0.0500 & -0.100 & 1.90  & 0 &0.0530 &0.0796 &0.9520 \\ [-1mm]
 &  &    & (0) &(0.0147) &(0.0188) &(0.2140)  \\  
0.0667 & -0.200 & 2.80  &  0 &0.0695 &0.0996 &0.9560 \\ [-1mm]
 &    &  &(0.0001) &(0.0159) &(0.0194) &(0.2053)  \\  
   \hline
\end{tabular}
 }
\caption{Means of the estimations of $\pi$ and of its confidence intervals at  level 0.05, from 500  simulations of $U(0,1)$ vs  $U(a,H)$. Sizes $n=m=250$. Coverage is the proportion of times in which $\pi\in (\hat \pi_L,\hat\pi_U)$. Between parenthesis the standard deviations of the 
estimates.} 
\label{TablaSimulUnif_pi_250}
\end{table}

\vspace{3mm}

\begin{table}[ht]
\centering
{\tiny
\begin{tabular}{cccccccc}
  \hline
 $\rho$ & $a$ & $H$  & $\hat \rho_L$ & $\hat \rho$ & $\hat \rho_U$ & Coverage \\ 
  \hline
0.9090 & -0.051 & 6.00 &0.8797 &0.9094 &0.9390 &0.9520 \\ [-1mm]
&&&(0.0171) &(0.0146) &(0.0122) &(0.2140) \\ 
0.9459 & -0.101 & 11.00 &0.9216 &0.9453 &0.9691 &0.9480 \\ [-1mm]
&&&(0.0149) &(0.0121) &(0.0093) &(0.2222) \\ 
0.9669 & -0.202 & 21.00 &0.9470 &0.9663 &0.9856 &0.9160 \\ [-1mm]
&&&(0.0134) &(0.0101) &(0.0069) &(0.2777) \\  
\hline
0.7250 & -0.050 & 1.95 &0.6800 &0.7259 &0.7718 &0.9580 \\ [-1mm]
&&&(0.0246) &(0.0231) &(0.0216) &(0.2008) \\ 
0.8000 & -0.100 & 2.90 &0.7584 &0.8002 &0.8421 &0.9500 \\ [-1mm]
&&&(0.0228) &(0.0209) &(0.0189) &(0.2182) \\  
0.8600 & -0.200 & 4.80 &0.8225 &0.8598 &0.8970 &0.9560 \\ [-1mm]
&&&(0.0207) &(0.0183) &(0.0160) &(0.2053) \\  
\hline
0.6333 & -0.050 & 1.45 &0.5827 &0.6321 &0.6816 &0.9460 \\ [-1mm]
&&&(0.0260) &(0.0251) &(0.0241) &(0.2262) \\  
0.7000 & -0.100 & 1.90 & 0.6508 &0.6986 &0.7464 &0.9520 \\ [-1mm]
&&&(0.0254) &(0.0241) &(0.0228) &(0.2140) \\  
0.7667 & -0.200 & 2.80 & 0.7228 &0.7678 &0.8129 &0.9460 \\ [-1mm]
&&&(0.0239) &(0.0222) &(0.0205) &(0.2262) \\
    \hline
\end{tabular}
 }
\caption{Means of the estimations of $\rho$ and of its confidence intervals at  level 0.05, from 500  simulations of $U(0,1)$ vs  $U(a,H)$. Sizes $n=m=250$. Coverage is the proportion of times in which $\rho\in (\hat \rho_L,\hat\rho_U)$. Between parenthesis  the standard deviations of the 
estimates.} 
\label{TablaSimulUnif_rho_250}
\end{table}

\subsection{Analysis of two real data sets}\label{realDataSets}
For each data set, we  compute the 
 estimates $\hat \gamma_{n,m}, \hat \pi_{n,m}$ and $\hat \rho_{n,m}$, 
plus the extremes of a 0.05 confidence interval computed as described in Section \ref{testing}. We also provide estimates of the rate of convergence for the index $\hat \gamma_{n,m}$. 

\subsubsection{INE data} \label{INE_Data}

This dataset contains   data
adjusted for inflation from the Living Conditions Survey (ECV) of the Spanish Statistics Institute (INE, for its 
name: Instituto Nacional de Estad\'\i s\-tica) for the  period 2003-2011. 
It contains  the annual income of Spanish families. The period covers the economic crisis of 2008, and our aim is to analyse  s.d. of the ECV of a given year with respect to its counterpart four years before. 
{
A quarter of the sample is replaced each year, so that no individual remains in the sample for more than four years, justifying the independence assumption of the samples being compared. The data are available at } 
{\tt https://doi.org/10.7910/DVN/1A5FZU}. 
Plots of the quantile and distribution functions associated with some of these datasets can be seen in Figure 1 
in the supplement. 

The analysis of s.d. of data related to poverty or welfare of the populations has notably contributed to the renewed interest in  s.d. in the econometric literature (see e.g. \cite{McFadden}, \cite{Anderson}, \cite{Barrett}, \ldots). 
{
However,  as discussed 
in the introduction, most published  works 
have chosen, for example, $H_0: F\leq_{st} G$. Thus,
these studies were based only on the lack of rejection of the null hypothesis of s.d. Therefore their conclusions are at best that there is no evidence against  $H_0: F\leq_{st} G$. 
So their aim was quite different from ours making the comparison between techniques impossible.
}


In  periods 
of economic prosperity,  the 
household disposable incomes should show an 
 improvement  which 
would be mathematically described through  s.d. of the distribution at the beginning of  the 
period by  that one at 
its end; 
or, instead of strict s.d., by confidence intervals for $\gamma$ and $\pi$  with 
upper extremes close to 0 and contained in $[0,0.5]$  for 
$\rho$.

The sample sizes of the involved data, by years,   appear in Table \ref{INE_Sizes}. The results of the computations appear in Tables 8, 
9 
and 10 
in Subsection 7 
in the supplement. A graphical representation of the results appears in Figure \ref{Figure.INE}. 

\begin{table}[ht]
\begin{center}
{
\begin{tabular}{ccccccccc}
\hline
{2003}&
{2004}&
{2005}&
{2006}&
{2007}&
{2008}&
{2009} &
{2010}&
{2011}
\\
\hline
15355&12996&12205&12329&13014&
13360&13597&13109&12714
\\                             
\hline
\end{tabular}
}
 \caption{Data sizes by year in the INE data.}
\label{INE_Sizes}
 \end{center}
 \end{table}

The first graph in Figure \ref{Figure.INE} and 
Table 8 
in the supplement
show the values of the $\gamma$ index when comparing each year  between 
2003 and 2007 with the corresponding one four years later. We   see there 
that the status of all individual incomes at the starting date would be considered worse  if 
considered in the final date, except, at most, 6.8\% of them in the periods 2003-07  to 
2005-09.  Between 2006 and 2010, the situation  reversed: 
at least 86.9\% of individuals (possibly  all of them)   improved 
their status. This percentage increased to 99.6\%  in 2007-11.
The inverse of the rates of convergence are around 2, which corresponds to clean crosses in the interior of (0,1) but, also, with order two contacts on  $t=0$ and $1$.

\begin{center}
\begin{figure}[ht] 
\centering

\includegraphics[width=5.2cm,height=5.5cm]{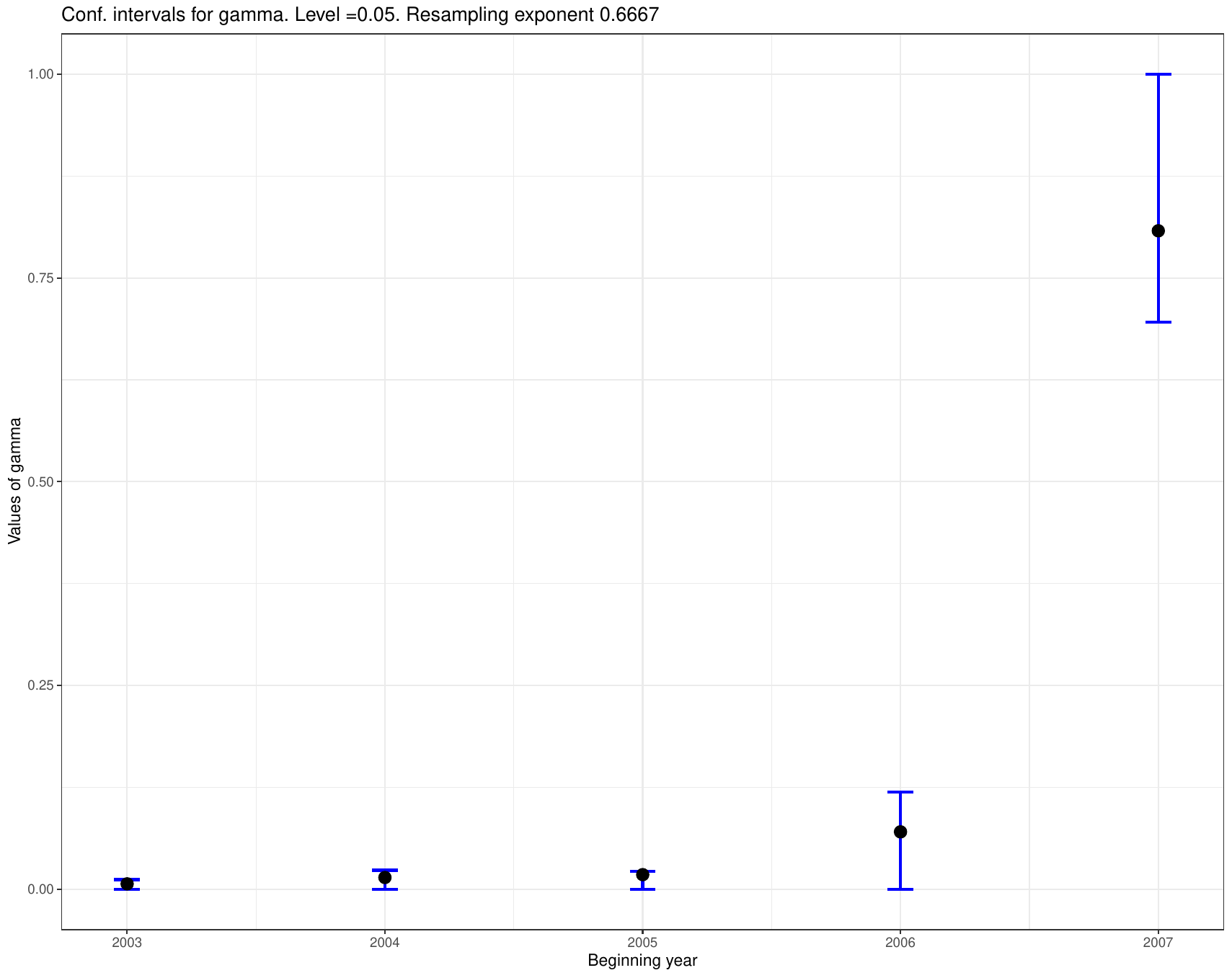}\hspace{3mm}
\includegraphics[width=5.2cm,height=5.5cm]{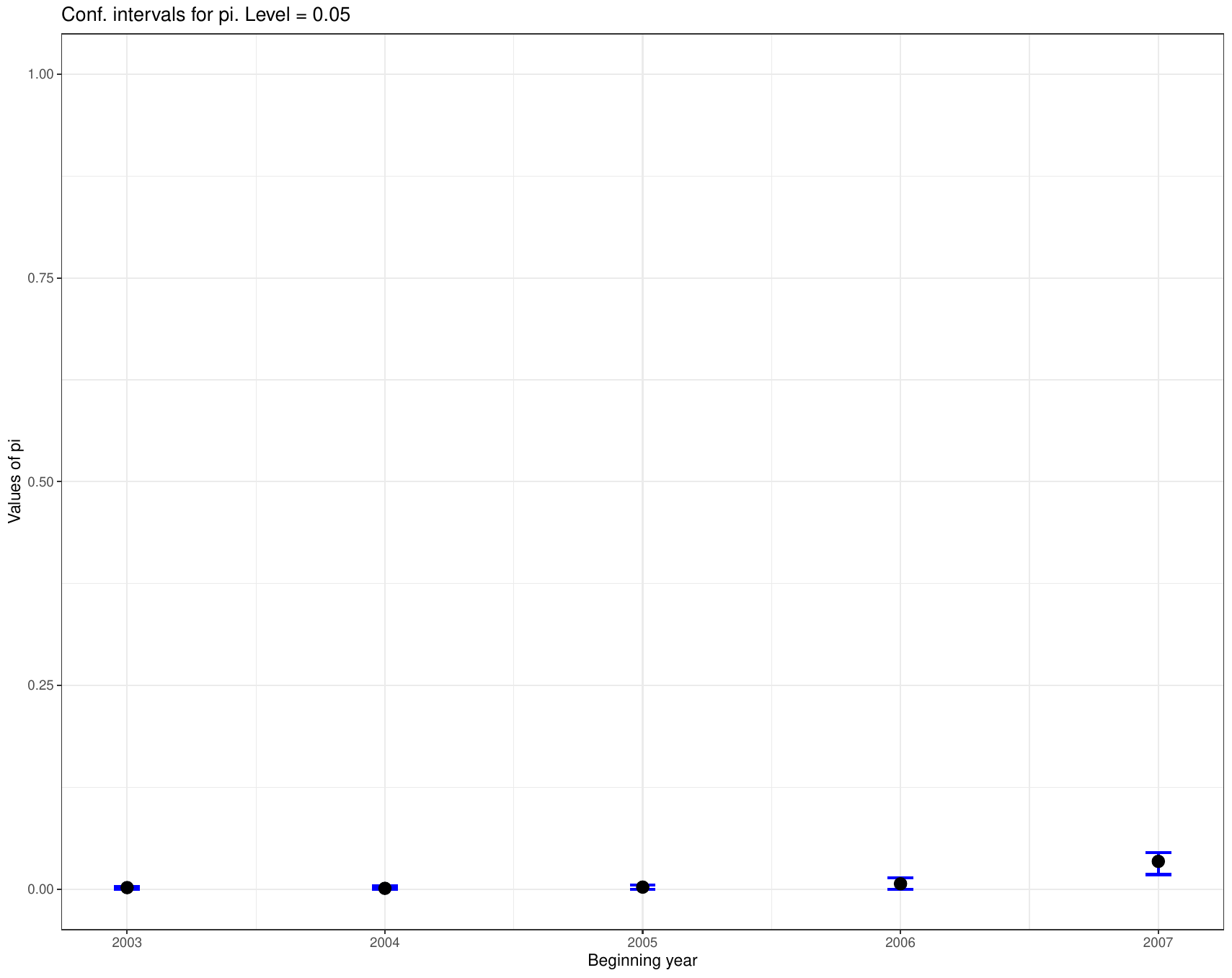}
%
\includegraphics[width=5.2cm,height=5.5cm]{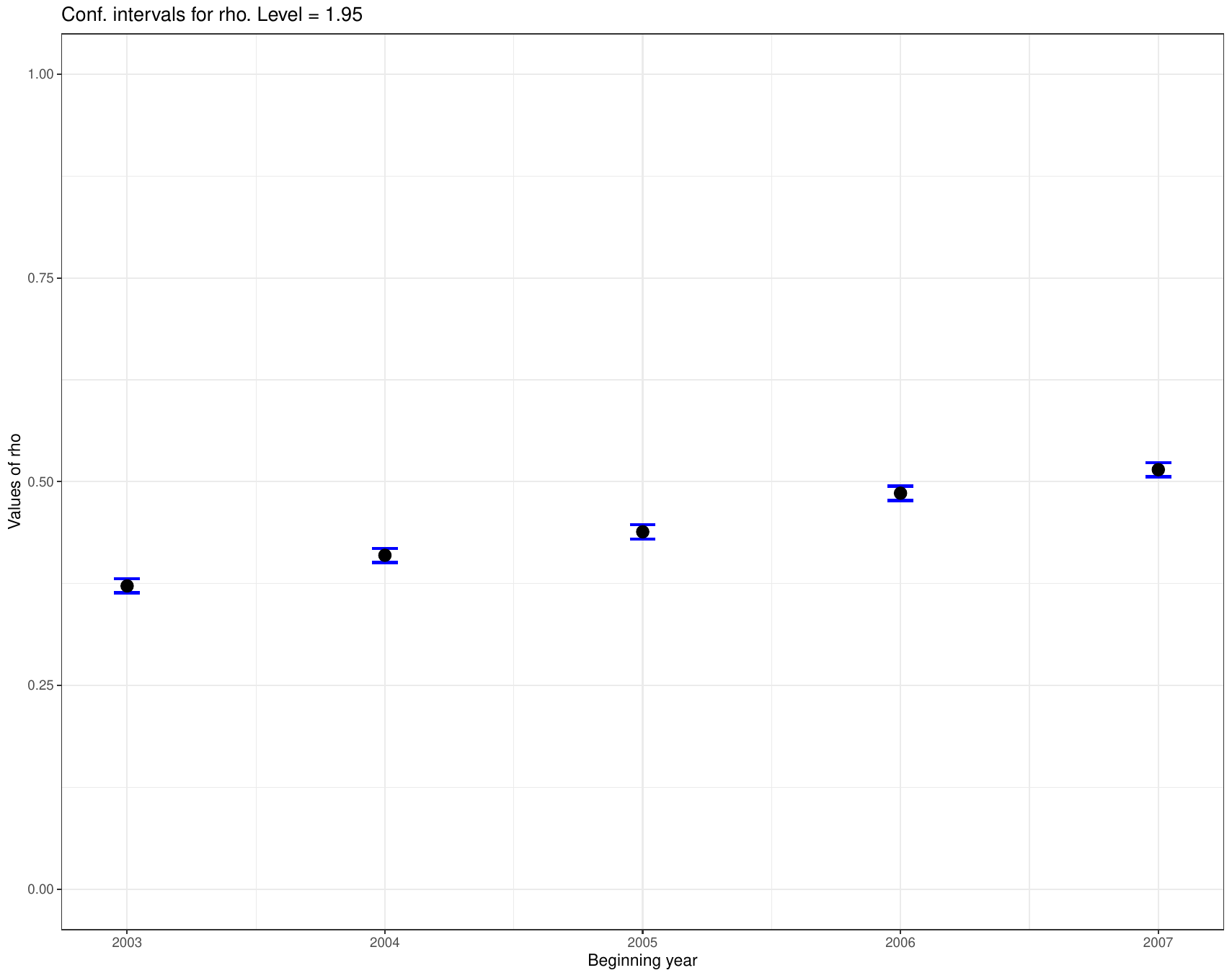}
\caption{INE data. Estimations and confidence intervals for $\gamma$, $\pi$ and $\rho$. Level = 0.05}
\label{Figure.INE}
\end{figure}
\end{center}

\vspace{-1.3cm}
Table 9 
in the supplement and the second graph in Figure \ref{Figure.INE} show the  estimations related to 
the $\pi$ index. 
  These estimations 
should be lower than those in Table 8 in the supplement 
(see Proposition \ref{relations})
which is 
true,  including 
the extremes of all the confidence intervals. As explained before, the $\pi$ coefficient tells, for instance, that if we  translate a given income in 2007 
to 2011 
then, its percentile in 2011 
would be at most 9.14\% 
larger than in 2007.

The  results 
for the $\rho$ index appear in Table 10 in the supplement 
and the last graph in Figure  \ref{Figure.INE}. 
Values of $\rho$ lower/higher than 0.5 indicate an improvement/worsening 
 in the  economy. 
Thus,  we 
conclude 
that 
 the living conditions in Spain 
improved in each of the four-year periods that began from 2003 to 2005, while they deteriorated 
 in the periods starting on 2006 and 2007, in line with the conclusions obtained with the $\gamma$ index.
However, the differences between the indices are smaller than to those of the  $\gamma$ index. This reinforces previous comments about the scarce information provided by variations of the index $\rho$.

\subsubsection{NHANES data} \label{NHANES_Data}

This dataset is from NHANES (U.S. National Health and Nutrition Examination Survey); it can be found at 
{\tt https://doi.org/10.7910/dvn/shbf2g}.   It contains heights from boys and girls with ages from 2 to 14 years  from the 1999, 2001,  \ldots  2009 surveys.    Table \ref{NHANES_Sizes} shows the sample sizes for each cohort by sex and age. We analyse  s.d. of the heights of boys over those of girls at each age.

 At each age we can consider independent the heights of boys and girls. Moreover, although the  dataset 
  has a longitudinal component,  once an age has been fixed, we can  also consider  the  heights 
  within each sex independent.  Plots of the quantile and distribution functions for both sexes, at ages 9, 11  and 14 appear in Figure 2 
 in the supplement. 

\begin{table}[h] 
\centering
\begin{tabular}{lccccccccccccc}
 \hline
 Age &2&3 &4& 5&6&7&8 &9&10&11&12&13 & 14
\\
\hline
$n$ (boys) &796 & 632 & 633 & 563 &557 & 582 &579 & 543 &556 & 556 & 735 & 728 & 704
\\
$m$ (girls) &776 & 563 & 620& 567 & 542& 564 &572 &579 &536& 587& 733& 757 & 764
\\
\hline

\end{tabular}
 \caption{Sample sizes by age (boys, top row; girls, bottom row)}

\label{NHANES_Sizes}
 \end{table}

The results of the analyses appear in Tables 11, 
12 
and 13 
 in Section 7 
 of the supplement. 
 A graphical representation of these results appears in Figure \ref{Fig.NHANES}.

Regarding the $\gamma$   index, 
Table 11 
and the left graph in Figure \ref{Fig.NHANES} show that boys are 
taller than girls at ages 2, 4 to 6, and 14  (more precisely: all male percentiles at these ages are higher than the corresponding female ones except for at most 2.44\% of them). 
The  opposite is true 
at 10 and 11 years (excepting, at most, for  8.22\% of the boys). For ages 7 to 9 and 12 and 13, when the transitions take place, the situation is 
doubtful, although we could accept that boys look taller than girls at age 8, since the proportion of percentiles that do not agree with this is at most 10.71\%. At age 3 this proportion is 6.44\%.

 The rates of convergence are, mostly, between 1 and 2, with three cases between 4 and 5 and the age 9 where the estimation is 13.7.   However, the difference between the distributions of the heights at age 9 is negligible (see Figure 2 in the supplement).   
 Therefore, the contacts between those curves have a very hight order,  supporting   that the distributions of the heights at this age could coincide
(recall that $2r_0$ should be $\infty$ in that case).

\begin{figure}[tb] 
\centering
\includegraphics[width=5.2cm,height=5.5cm]{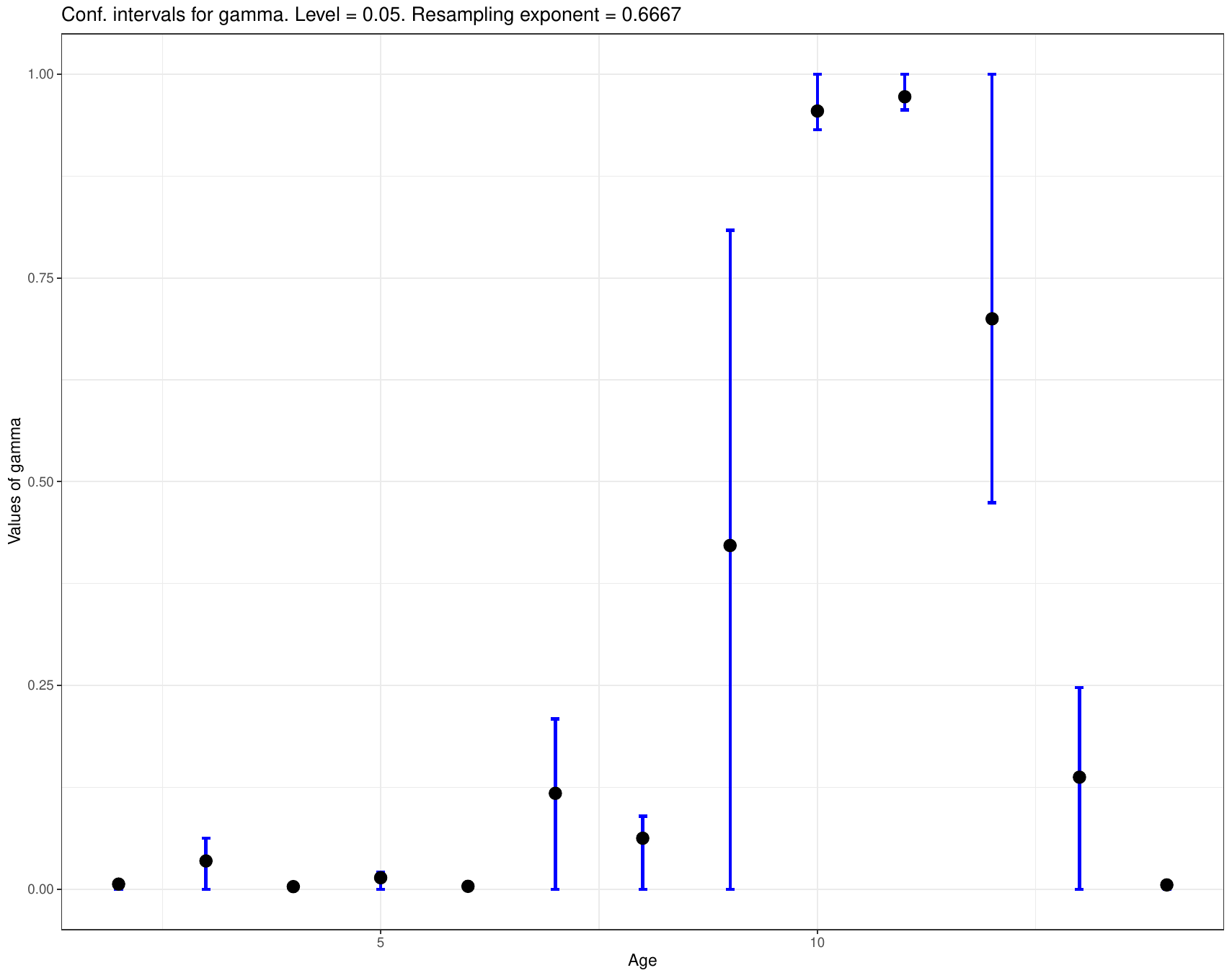}
\includegraphics[width=5.2cm,height=5.5cm]{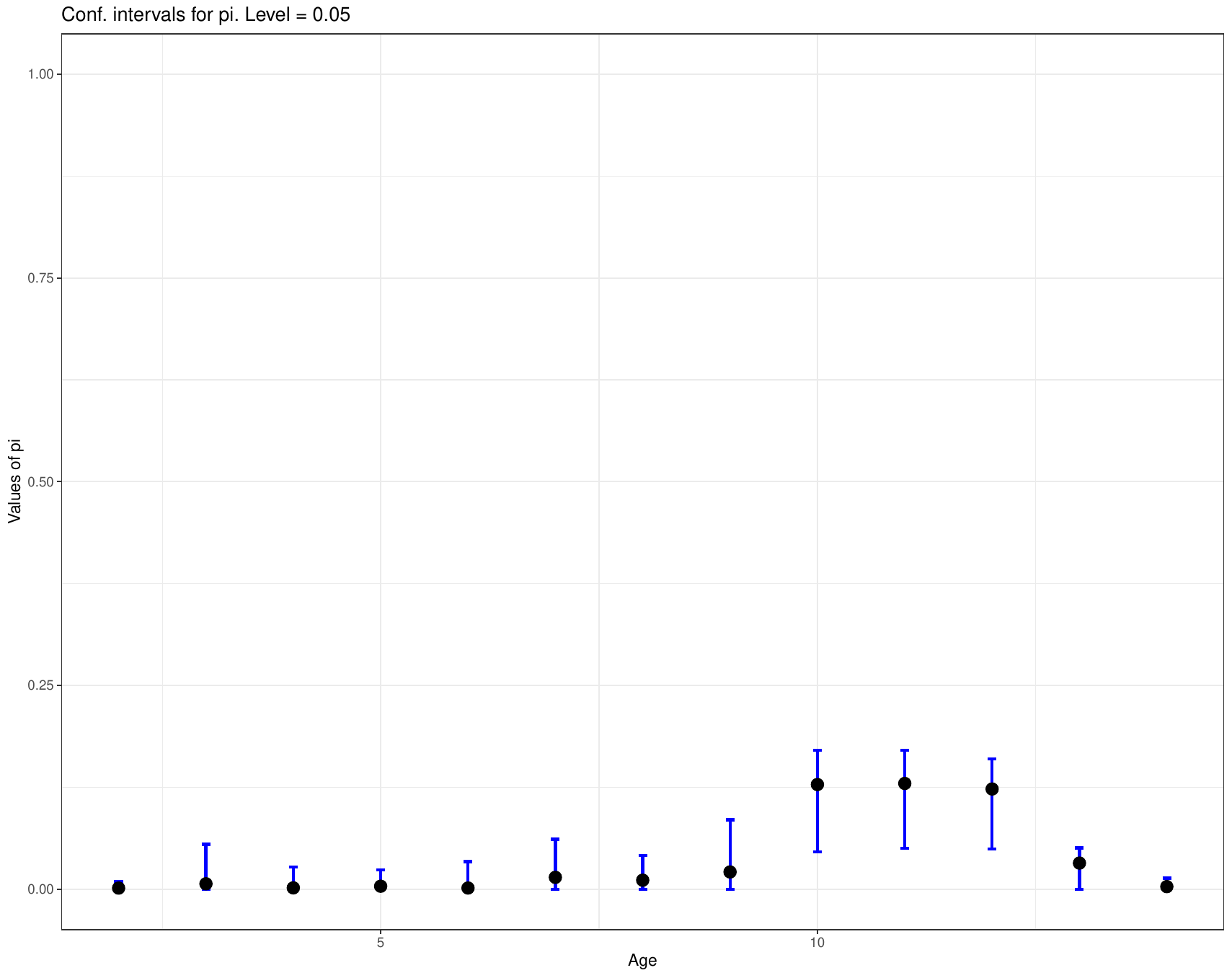}
\includegraphics[width=5.2cm,height=5.5cm]{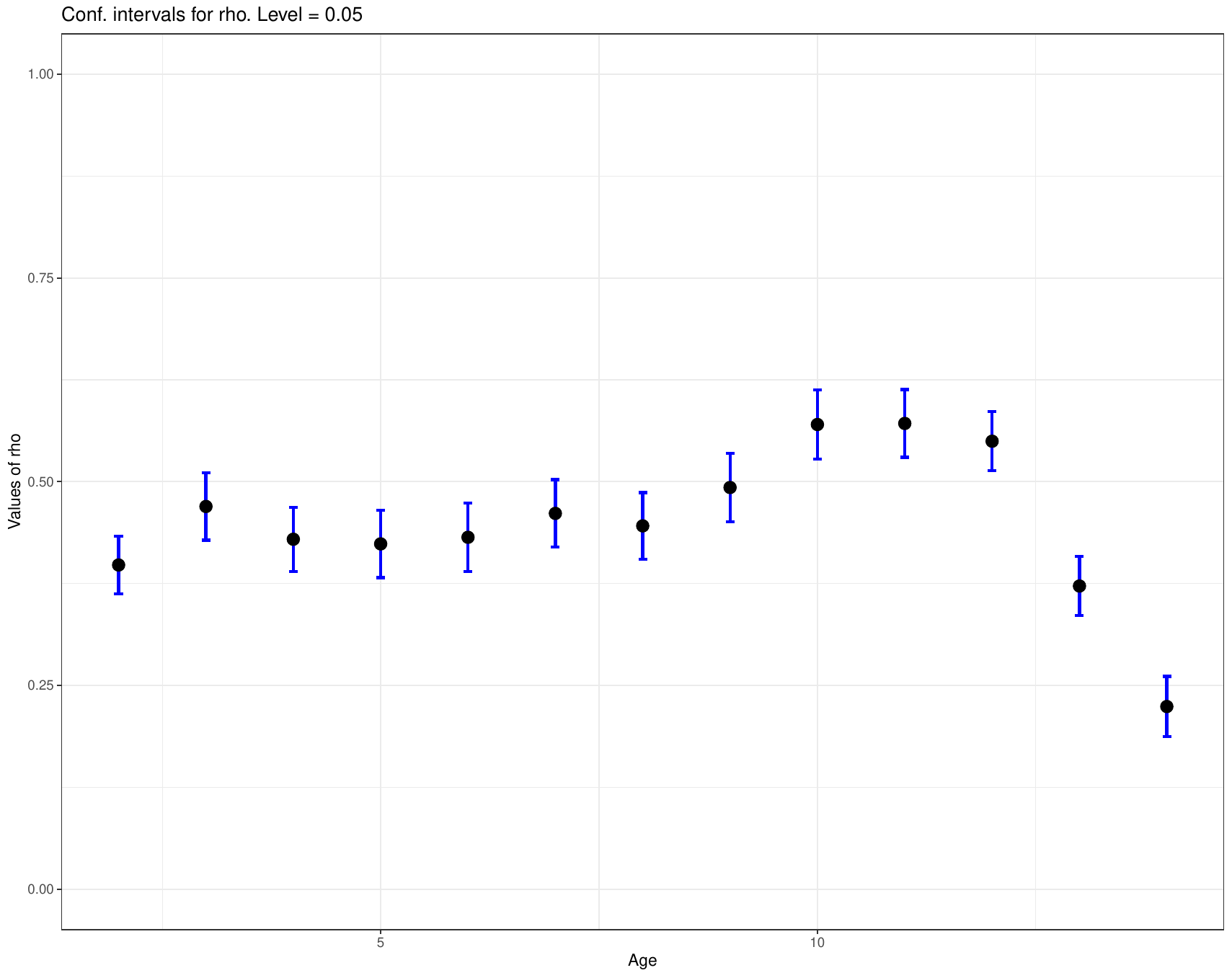}
\caption{NHANES data. Estimations and confidence intervals for $\gamma$, $\pi$ and $\rho$. Level = 0.05}
\label{Fig.NHANES}
\end{figure}

Table 12 
and  the central graph in Figure \ref{Fig.NHANES} show the results for the $\pi$ index, which are similar to those obtained for the $\gamma$ index, with
the proportions of people being replaced by {the maximum loss in status if a boy is included in the population of girls.}

 The comments for the economic situation in 2007-11 apply to the ages 10 to 12.  
The estimates of the $\pi$ indices   and the lower bounds  are always much lower than   those of  the $\gamma$ indices. The same happens with 
 the upper bounds, except for age 5 (where the upper bound of $\pi$ is  only 
slightly lower)  and for ages  6 and 14: in the last case the difference is not too large, while  the upper bound for 6 (= 0.0343)  is five times  larger. 
 This is due to the conservative estimation chosen for the upper bound of $\pi$ (see comment in Subsection  \ref{SimulGauss}
), but also to the different rates of convergence and  sample sizes. 

The $\rho$ index, with values below 0.5 indicating stochastic precedence in the sense of \cite{Arcones} (and the opposite for values greater than 0.5), further supports the conclusions drawn from the $\gamma$ and $\pi$ indices. 

{
\section{Conclusions}
{
 Looking for indices to measure how far $G$ is from reaching s.d. over $F$, we explore quantifiable  
 relaxations of the s.d.  definition 
 while maintaining  its invariance with respect to increasing transformations. We also introduce a simple decomposition of the marginal laws of any random vector $(X,Y)$, which allows us to consider such indices under the common representation $P[X>Y]$, where ${\mathcal L}(X)=F$, ${\mathcal L}(Y)=G$, for any given dependence structure between $X$ and $Y$. These representations guarantee that  these indices are invariant under increasing maps (we suspect that only the indices that allow such a representation could  have this invariance). We have particularized this representation for the indices $\rho, \gamma$ and $\pi$, highlighting some properties and alternative characterizations with emphasis on the  equality: 
$\pi(F,G)=\min \{P(X>Y): (X,Y) \mbox{ with marginal d.f.'s } F \mbox{ and } G\}.$

The indices considered here 
allow to give prominence to a particular choice of joint dependency structure, or 
to get a more complete description of the deviation from s.d. 
  The $\rho$ index, assuming  independent populations, is an 
example of the first goal, 
a point of view pursued in \cite{Montes} in the context of copulas, looking for conditions under which s.d. implies statistical preference. 
Usually, $\rho$  varies not so much, the only important point being the value 0.5 which is where the dominant variable changes.

The indices $\gamma$ and $\pi$ vary more than $\rho$. $\gamma$ gives a more accurate idea than $\pi$ of how far is $G$ from s.d. $F$: 
   $\gamma(F,G)=1$ and    $\gamma(F,G)=0$  are respectively equivalent to $F\geq_{st}G$ and $G\geq_{st}F$. However, $\pi(F,G)=0$ implies $G\geq_{st} F$, but it  is 
easy to imagine situations where 
may happen that $F\geq_{st}G$, while  $\pi (F,G)$ is as close to zero as desired (see Remark \ref{NotaUpperBound}). 

Our preferred indices are $\gamma$ and $\pi$ which are complementary and jointly give a good description of the situation.  As stated, the former is  the most natural  to measure   how far is $G$ to s.d. $F$. Additionally, if $z\in \Rea$ and we assume that the value $F(z)$ (resp. $G(z)$) measures the status of the value $z$ in the population $F$ (resp. $G$), it is clear that, if $F\leq_{st}G$, then the status of $z$ should be higher in $F$ than in $G$. Then, $\pi(F,G)$ measures the maximum loss of status of any individual when moving from $F$ to $G$.

The plug-in estimators, 
$\rho(F_n,G_m), \pi(F_n,G_m), \gamma(F_n,G_m)$, of the indices are well-known statistics widely used in nonparametric inference, but scarcely explored in the actual setting, especially in the case of Galton's statistic (whose theoretical behaviour has only recently been reported, although yet not discussed in applications).
We have recompiled the asymptotic statistical theory of all of them while noting that, contrary to what it was previously assumed, {the naif bootstrap showed in the simulations a poor behaviour  for $\gamma$}, the  arguably simpler of the  indices considered here, and we give new results showing how a low resampling bootstrap can fill the gap. However,   such new theory does not cover 
contact points at 0 or 1 between the theoretical quantile functions involved and, as our simulations with Gaussian distributions show, estimates of $\gamma$ can be poor  when the contact points are  on those points. 

The above mentioned limitations of $\gamma$  for inferential purposes, enhance the role   of  $\pi$ 
for such goal.
Its good performance shown in the analysed examples and simulations, and  the 
interpretation of $\pi$ associated to \eqref{pilevel0}, 
lead us to encourage its use for inferential purposes.

 The approach} taken is purely nonparametric, in a very complex scenario where small changes in the tails can have significant effects.  Nevertheless, as shown in \cite{ZhuangEtAl2019} and \cite{Zhuang}, if appropriate,  a semi-parametric approach under the density ratio model could significantly improve the behaviour of appropriately adjusted estimates of $\gamma(F,G)$ and $\pi(F,G)$ (although losing the invariance property).
\vspace{5mm}

\noindent
ACKNOWLEDGMENT.  The authors would like to express their sincere gratitude for the numerous and interesting suggestions they have received from the reviewers, the associate editor, and the editor responsible for the article, all of whom have contributed to significantly improving the article. Many thanks to all of them.

The authors report there are no competing interests to declare.

\begin{appendix}

\vspace*{3cm}

\begin{center}
{
  \title{\bf \Large Supplementary material for ``Invariant measures of disagreement with stochastic dominance"}
  \\
  
  \maketitle
} 
\end{center}

\bigskip

\section{Content of this supplementary material}
This paper contains some supplementary material of \cite{del Barrio2025}. We employ here the notation and abbreviations introduced in that paper. The supplementary
 is organised as follows: Section \ref{ProofOfProposition2.2} is devoted to prove an interesting relation between location-scale families and s.d. Section \ref{AppendixIndicepi} contains the proofs of the results in Subsection 2.3 in the main paper. The asymptotic distribution of $\gamma$ is obtained in Section \ref{Sec.gamma.asympt}. Section \ref{Subs.Technicalities} explains the computational details of the three parameters $\gamma, \pi$ and $\rho$. Finally, Sections \ref{Tablas_Simul} and \ref{Tablas_Datos} include the tables with the complete results of Subsections 4.1.1 
and 4.1.2 
and of Subsections 4.2.1 
and 4.2.2 
respectively. Section \ref{Tablas_Datos} also show the graphs of the quantile and empirical cumulative distribution functions of the INE and NHANES data sets.

\section{Stochastic precedence and location-scale families}\label{ProofOfProposition2.2}
{
We begin with the generalization to location-scale families of the property that the stochastic precedence coincides with the order of the means. It is not difficult to get counterexamples to Proposition \ref{Prop.rho_LocationScale} if the symmetry or the {
strict increasing} assumptions fail.}

 \begin{Prop} \label{Prop.rho_LocationScale}
 Let $X_0,Y_0$ be two { r.v.'s} whose distributions are symmetrical w.r.t. zero. Let $F_0, G_0$ be their d.f.'s which we assume  increasing on $(-\varepsilon,\varepsilon)$ for some $\varepsilon>0$. Let $\mu_1,\mu_2\in \Rea$ and $\sigma_1,\sigma_2>0$ and let $F,G$ be respectively the d.f.'s of $\sigma_1X_0+\mu_1$ and $\sigma_2Y_0+\mu_2$. Then,   $F\leq_{sp}G$ if and only if $\mu_1\leq\mu_2$.
 \end{Prop}
 \textbf{Proof:}
{
Let us take $X_0, Y_0$  independent}. We have $\mathcal L((X_0,Y_0)) =\mathcal L((-X_0,-Y_0))$, hence 
\begin{eqnarray*}
P[X_0<Y_0]=P[-X_0<-Y_0]=P[X_0>Y_0] 
\quad \mbox{and } P[X_0\leq Y_0]=P[X_0\geq Y_0]\geq 1/2.
\end{eqnarray*}
Thus, $F_0\leq_{sp}G_0$ is always true, a property that share the distribution functions of $\sigma_1X_0$ and $\sigma_2Y_0$, because they also verify the hypotheses. If we take $\mu_1 \leq \mu_2$, then 
obviously $P[\mu_1+X_0\leq \mu_2+Y_0]\geq P[X_0\leq Y_0]\geq 1/2.$ In fact,  if the reverse inequality $\mu_1 > \mu_2$
holds and $P[\mu_1+X_0\leq \mu_2+Y_0]\geq 1/2,$ under the symmetry hypothesis, it should happen
$P[|X_0-Y_0|<a]=0$ for any $0<a\leq \mu_2-\mu_1,$ what is impossible by assumption. 
\FIN

\section{Proofs related to Subsection 2.3
}\label{AppendixIndicepi}

 \subsection{Relation of $
 \pi(F,G)$ with contamination mixtures}
 The index   $\pi(F,G)$ defined through (6) in the main paper 
 can be successfully characterized by relying on the intrinsic relationship between trimmings and contamination mixtures, as pointed out e.g. in Proposition 2.1 in \cite{Alv2011b}. Trimming procedures are common in statistical practice as a way of downplaying the influence of outliers in the analysis. Trimming usually refers to the removal of extreme observations in the sample, usually the same number at both extremes.  A more flexible approach  allows partial trimming by reweighting the data. For a fixed $\pi\in [0,1)$,  trimming a data set $\chi=\{x_1,...,x_n\}$ { (at most)} at the level $\pi$
consists in replacing the uniform distribution on $\chi$ by  a distribution giving new weights to the data. This is carried through a trimming function $w$, giving weights $w(x_i)$,  that satisfy
$$0\leq w(x_i)\equiv w_i \leq 1
, \ i=1,...,n \ \mbox{ and } \sum_{i=1}^n w_i\geq n (1-\pi).$$
 
 {
 As an example, let us assume that we have two samples with the values  $\{1,3,6\}$ and $\{0,4,5\}$. If we remove the central observation in both cases (I.e. if we use the weight $w_1=w_3=1$ and $w_2=0$ in both cases), we would obtain that the trimmed first sample s.d. the trimmed second one.
 }
 
 The weight $w_i$   measures the relevance of $x_i$ in the sample, the extreme values $w_i=0 \mbox{ or } 1$  respectively meaning that the observation $x_i$ is deleted or kept. {
Intermediate values are also possible, but its interest is mainly theoretical or limited to situations in which the sample sizes are different}. Each trimming function has an associated trimmed probability  given by $\tilde P_w(x_i) =  \frac {w_i}{S_w}$, where $S_w=\sum_{i=1}^nw_i$.

 The extension to general probability measures  is simple. The probability $\tilde{P}$ is a 
$\pi$-trimming of the probability $P$, a fact denoted by $\tilde P \in \mathcal{R}_\pi(P),$ if there exists a function $w$ satisfying
$ 0\le w \le 1$,   $P$-a.s., and $S_w:=\int w\ dP\geq 1-\pi$ such that
\[
\tilde P(B)=\frac 1 {S_w}\int_Bw\ dP.
\]

 Proposition \ref{prop1} shows the  link between 
the contamination  model and  trimmings as well as some  relevant facts in our present setting.  It is included in  
 Propositions 2.3 and 2.4 in \cite{Alv2015} where more details and additional discussion can be found.

\begin{Prop}\label{prop1}
Let   $\pi \in [0,1)$ and $\tilde P, P$ be  probability distributions on $\mathbb{R}$ with d.f.'s 
$\tilde F$ and $F$, respectively. Also define the d.f.'s
\begin{equation*}
F^\pi(x) = \max({\textstyle \frac 1 {1-\pi}} (F(x) - \pi),0)
\quad \mbox{and}\quad
F_\pi(x) =
\min({\textstyle \frac 1 {1-\pi}} F(x),1). 
\end{equation*}

Then, the relation $\tilde P\in \mathcal{R}_\pi(P)$:
\begin{itemize}
\item[a)] implies the relations $F_{\pi}\leq_{st} \tilde F \leq_{st}F^{\pi}$, and

\item[b)] is equivalent to $F=(1-\pi)\tilde F+\pi R,$ for some  d.f. $R$.

\end{itemize}
\end{Prop}

Statement a) implies that the set of the trimmed versions of a given distribution on $\Rea$ has a minimum 
and a maximum with respect to  s.d., $F_{\pi}$ and $F^{\pi}$, and that  {the d.f.'s of the probabilities in } $\mathcal{R}_\pi(P)$ are included in the interval $[F_{\pi},F^{\pi}]$ (however, probabilities whose d.f.'s $H$ verify $F_{\pi}\leq_{st} H \leq_{st}F^{\pi}$ do not necessarily belong to $\mathcal{R}_\pi(P)$).

The  extreme d.f.'s, $F_{\pi}$ and $F^{\pi}$
are respectively obtained by trimming at level $\pi$ just on the right (resp. left) tail the probability with d.f. $F$. From this, 
it is easy to show that the decompositions (9) in the main paper 
hold for $\pi \in [0,1)$ if and only if 
$F_{\pi}\leq_{st}G^{\pi}$, and this holds if and only if $\pi \geq\sup_{x \in \mathbb{R}} (G(x)-F(x))$. 

\subsection{Proof of Proposition 2.2 
}\label{Prooftrim_quantile}
We use the d.f.'s $F_{\pi}$ and $F^{\pi}$  introduced in Proposition \ref{prop1}.
A simple computation shows that their associated quantile functions are
\begin{eqnarray*}
(F_{\pi})^{-1}(t)=F^{-1}((1-\pi)t),
 (F^{\pi})^{-1}(t)=F^{-1}(\pi+(1-\pi)t),\quad 0<t<1.
 \end{eqnarray*}

From Proposition \ref{prop1} we have that the quantile function of any d.f. $\tilde F$ in $\mathcal{R}_\pi(F)$
satisfies
\[
F^{-1}((1-\pi)t)\leq \tilde F^{-1}(t)\leq F^{-1}(\pi+(1-\pi)t),  0<t<1,
\]
thus the characterizations following Proposition \ref{prop1} lead to 
$F\leq_{st}^{\pi}G$ if and only if 
$$F^{-1}((1-\pi)t)\leq G^{-1}(\pi+(1-\pi)t), \quad 0<t<1$$
or, equivalently, if and only if the relation 
$$F^{-1}(y)\leq G^{-1}(\pi+y), \mbox{ \ for every $y$ such that }  0<y<1-\pi$$ holds.
\FIN

\section{Asymptotic distribution of $\gamma$}\label{Sec.gamma.asympt}

The first difficulty to deal with is the exponent that determines the asymptotic rate in Theorem 3.2. 
This rate depends on the greater ``intensity  (or order) of the contact'' between the curves $F^{-1}$ and $G^{-1}$ (see Theorem \ref{Bootstrap} below), where the contact points include  even some extensions of  crosses and tangencies, and the intensity of each contact point can be described in terms of generalized local expansions  of the composed functions $F(G^{-1})$ and $G(F^{-1})$: 
For a point $t_0 \in (0,1)$ 
such that $F(G^{-1}(t_0))=t_0$, let us  consider  the function 
\begin{equation*} \label{Eq.Delta}
\Delta (h) := F(G^{-1}(t_0+h)) - t_0 - h. 
\end{equation*}

Following \cite{del Barrio2021}, we say that $t_0$ is a {\it regular contact point} if $\Delta$ is locally Lipschitz at $0$ and 
there exist $\eta>0$,  $r_L, r_R\geq 1$  and $C_L, C_R\ne 0$,  depending on $t_0$, such that
\begin{equation*}\label{Eq.DesarrolloDelta}
\Delta(h)=\left\{
\begin{matrix} 
C_L |h|^{r_L}  + o(|h|^{r_L}),& & \mbox{ if } h \in (-\eta,0),
\\[0mm]
C_R |h|^{r_R}  + o(|h|^{r_R}), & &\mbox{ if } h \in (0,\eta).
\end{matrix}
\right.
\end{equation*}
 Here,  $r_L$ (resp. $r_R$) is the {\it  intensity 
}of the contact on the left (resp.  
right) between $F^{-1}$ and $G^{-1}$ at $t_0$. 
 Such assumptions cover  the case of  smooth enough d.f.'s with at most a finite number of  crossing or tangency points.   When considering contact points on the left (resp. on the right)  with $t_0=0$ (resp. $t_0=1$), with an abuse of notation we can take $r_L=C_L=0$ (resp. $r_R=C_R=0$) in the cases in which the expression on the right (resp. 
 left) has sense.

 When $\ell(F^{-1}=G^{-1})>0$ and $F(G^{-1})$ is Lipschitz, $\hat \gamma_{n,m}$ has a non-degenerate  limit law, while if the set of generalized contact points is finite, the rate will depend  on the 
 intensity of the contacts and on  their location.
 We refer to \cite{del Barrio2021} for the general treatment but, with statistical applications in mind, we give here a simplified version of  Theorem 1.7 there, which emphasizes on the rates of convergence without additional details on the non-degenerated limit laws.

\begin{Theo}\label{TCL.principal.App}
Assume that   {$t_1,\ldots,t_k$ are the contact points between $F(G^{-1})$  and the identity and that  all of them are regular}
 with the left and right intensities of the contact at $t_i$  being $r_L(t_i),r_R(t_i)$. 
Set $r_{i}=\max(r_L(t_i),r_R(t_i))$ if $t_i\in (0,1)$, and $r_{i}=\max(r_L(t_i),r_R(t_i))-\frac 1 2$ if $t_i\in\{0,1\}$, and let $r_0=\max_{1\leq i\leq k} r_{i}$. Then, as $m, n \to \infty$ with $\frac{n}{n+m}\to\lambda \in (0,1)$,   there exits a non-degenerete law $\mu$, such that:
$$(n+m)^{\frac 1 {2r_0}}(\hat \gamma_{n,m}-\gamma(F,G)) \convw \mu.$$
\end{Theo}

The fact that $\mu$ is in general unknown, suggests to use bootstrap to apply this theorem. With this aim, 
for samples $x_1,\dots, x_n$ and $y_1,\dots, y_m$, we compute $\hat \gamma_{n,m}$. 
Bootstrapping with resampling sizes $\tilde n, \tilde m$ (sampling with replacement) from the original samples, 
we  obtain $B$ bootstrap d.f.'s $\tilde F_{\tilde n}, \tilde G_{\tilde m}$, and compute $\tilde \gamma_{\tilde n,\tilde m}:=\gamma(\tilde F_{\tilde n}, \tilde G_{\tilde m})$. If the bootstrap works for the index $\gamma$, we could use $\tilde z_\alpha,$ the $\alpha-$quantile of the bootstrap estimates $\tilde \gamma_{\tilde n,\tilde m}-\hat \gamma_{n,m}$, to obtain  upper and lower $(1-\alpha)$ confidence bounds for $\gamma(F,G)$.  
{
Taking into account the possible different sizes of the original  and the bootstrap samples, considering $C_{n,m}:=\left(\frac {\tilde n+\tilde m}{n+m}\right)^{1/(2r_0)}$, the  upper and lower $(1-\alpha)$ confidence bounds for $\gamma(F,G)$, would be: 
\begin{equation}
 U_{n,m,\tilde n,\tilde m}^\alpha:=\hat{\gamma}_{n,m}-\tilde z_\alpha C_{n,m}
\label{NPboundstrue}
\  \mbox{ and } \ 
 L_{n,m,\tilde n,\tilde m}^\alpha:=\hat{\gamma}_{n,m}-\tilde z_{1-\alpha}C_{n,m}.
\end{equation}
}


The disappointing fact is that the naif bootstrap ($\tilde n=n$ and $\tilde m=m$) seems to be useless (see Remark \ref{noboots} below) despite this having been taken for granted in \cite{Alv2017}.
However, to circumvent this problem we can use the bootstrap with lower resampling sizes according with the following new theorem.

\begin{Theo}\label{Bootstrap}
Let $\tilde n=\tilde n(n), \tilde m=\tilde m(m)$ such that $\tilde n=o(\frac n {\log \log n}), \tilde m=o(\frac m {\log \log m})$ and $\tilde F_{\tilde n}, \tilde G_{\tilde m}$  be the bootstrap d.f.'s based on the samples $X_1,\dots,X_n$ and $Y_1,\dots,Y_m$, with resampling sizes $\tilde n,\tilde m$ and with replacement. Under the hypotheses of Theorem \ref{TCL.principal.App}, if $\{t_1,\ldots t_k\} \subset (0,1)$, we have that, when $n,m,\tilde n,\tilde m \to \infty,$  with  $\lim \frac {\tilde n}{\tilde n+\tilde m}=\lim \frac  n{n+ m} = \lambda\in (0,1)$, for almost every realization of the samples  \suc \ and \sucY , the conditional laws  of
\begin{equation}\label{bootslimit}
(\tilde n+\tilde m)^{1/(2r_0)}\left(\gamma(\tilde F_{\tilde n},\tilde G_{\tilde m})-\gamma(F_n,G_m)\right)
\end{equation}
given the samples \suc \ and \sucY \
weakly converge,  to the same law as
$$ \mathcal L\left[(n+m)^{1/(2r_0)}\left(\gamma(F_n,G_m)-\gamma(F,G)\right)\right].$$
\end{Theo}
{
The proof easily follows from Lemma \ref{lemmabootstrap} below, which allows us to replicate the proof of Theorem 4.9 in \cite{del Barrio2021}. It suffices to replace the role of Lemma 3.3, there, by the new lemma involving the bootstrap distributions.

To obtain the pertinent results, we briefly introduce some theory involving useful representations for sample quantile functions:} without loss of generality, we can  
  assume that our samples, $\{X_n\}$ and $\{Y_m\}$, have been obtained from independent $U(0,1)$ samples  $\{U_n\}$ and $\{V_m\}$ through the transformations $X_i=F^{-1}(U_i), Y_j=G^{-1}(V_j).$  We will denote the empirical quantile functions of the uniform samples by $\mathbb U_n$ and $\mathbb V_m$. We have the obvious relations 
\begin{equation}\label{Eq.represent_2}
F_n^{-1}=F^{-1}(\mathbb U_n)\ \mbox{ and } \ G_m^{-1}=G^{-1}(\mathbb V_m).
\end{equation}

 With the usual notation $u_n$ and $v_m$ for the quantile processes based on the $U_i$'s and the $V_j$'s, respectively, ($u_n(t)=\sqrt n (\mathbb U_n(t)-t)$ and similarly for $v_m$) we have that

\begin{equation}\label{Eq.represent}
F_n^{-1}(t)=F^{-1}{\textstyle \Big(t+\frac{u_n(t)}{\sqrt{n}}\Big)} \ \mbox{ and } \ G_m^{-1}(t)=G^{-1}{\textstyle \Big(t+\frac{v_m(t)}{\sqrt{m}}\Big).}
\end{equation}

In an analogous way, we can obtain appropriate representations of the bootstrap distribution and quantile functions. By resorting to additional independent $U(0,1)$ samples $\{\tilde U_n\}$ and $\{\tilde V_m\}$  (and independent of the sequences above), and using the obvious modifications in the notation,  \eqref {Eq.represent_2} gives that
\begin{eqnarray*}\label{qf_bootstrap}
\tilde F_{\tilde n}^{-1}(t):=F^{-1}(\mathbb U_n(\tilde{\mathbb U}_{\tilde n}(t))), 
\ \ \tilde G_{\tilde m}^{-1}(t):=G^{-1}(\mathbb V_n(\tilde{\mathbb V}_{\tilde m}(t))), \ \ t\in (0,1),
\end{eqnarray*}
which are versions of the bootstrap quantile functions associated to independent bootstrap samples, with resampling sizes $\tilde n, \tilde m$, obtained without replacement of the original samples $X_1,\dots,X_n$ and $Y_1,\dots,Y_m$. 

Let us now focus on $\tilde F_{\tilde n}^{-1}$. The iterated use of (\ref{Eq.represent}) leads to
\begin{equation}\label{qf_bootstrap2}
\tilde F_{\tilde n}^{-1}(t)=F^{-1}\left(t+\frac {\tilde u_{\tilde n}(t)}{\sqrt{\tilde n}}+\frac {u_n\left(t+\frac {\tilde u_{\tilde n}(t)}{\sqrt{\tilde n}}\right)}{\sqrt n}\right),
\end{equation}
where $\tilde u_{\tilde n}$ is the quantile process associated to a sample with size $\tilde n$ taken from $\{\tilde U_n\}$.

The following result  is a consequence of a refined version of the Komlos-Major-Tusnady construction (the Hungarian construction) for the quantile process (see, e.g., Theorem 3.2.1, p. 152
in \cite{CsorgoHorvath}),
 from which we know that there exists a sequence of Brownian bridges on $(0,1)$, $\{B_{n}\}$, versions of $u_n$  and positive constants, $C_1,C_2$ and $C_3$, such that, for any $x>0$,
$$
P\Big\{\sup_{0\leq t\leq 1} |u_n(t)-B_{n}(t)| >{\textstyle \frac{x+C_1 \log n}{\sqrt{n}}} \Big\}\leq C_2 e^{-C_3 x}.
$$

Making use of this construction for the quantile processes $\tilde u_{\tilde n}$ and $\tilde v_{\tilde m}$, and taking $x=\frac a {C_3}\log n$  with $a>1$ and $K=\frac a {C_3}+C_1>0$, we obtain useful independent sequences of Brownian bridges  $\{B_{\tilde n}^1\}$,  $\{B_{\tilde m}^2\}$  and versions of $\tilde u_{\tilde n}$ and $\tilde v_{\tilde m}$, independent of the sequences of processes $u_n$ and $v_m$, as stated in the next theorem.
 
\begin{Theo}\label{aproxBridge}
With the previous notation, in a probability one set, the sequences $\{B_{\tilde n}^1\}$,  $\{B^2_{\tilde m}\}$,  $\tilde u_{\tilde n}$ and $\tilde v_{\tilde m}$
eventually satisfy
\begin{equation}\label{eventualbound2}
\sup_{0\leq t\leq 1} |\tilde u_{\tilde n}(t)-B_{\tilde n}^1(t)| \leq {\textstyle K\frac{\log \tilde n}{\sqrt{\tilde n}}}\ \ \ \ \ \
\sup_{0\leq t\leq 1} |\tilde v_{\tilde m}(t)-B_{\tilde m}^2(t)| \leq {\textstyle K\frac{\log \tilde m}{\sqrt{\tilde m}}}.
\end{equation}
\end{Theo}

Another consequence of the Hungarian construction and the functional law of iterated logarithm for the Kiefer process  is the following law of iterated logarithm for the uniform quantile  process (see, e.g., (3.3.2) in Theorem 3.1, p. 164 in  \cite{CsorgoHorvath}):
$$\lim \sup \sqrt{\frac 2 {\log \log n}}\sup_{t\in (0,1)}|u_n(t)|=1 \mbox{ a.s.}$$

Thus, there exists a constant $C$ and a probability one set such that 
\begin{equation}\label{bound}
 \frac{\sup_{t\in (0,1)}|u_n(t)|}{\sqrt n} \leq C\sqrt{\frac {\log \log n}n} \mbox{ eventually.}
\end{equation}

 If  now, $f:(0,1)\to \mathbb R$ is  any Lipschitzian function,  relations (\ref{bound}) and (\ref{eventualbound2}) imply that, for a suitable $L$, the following inequality holds eventually in a probability one set 
 \begin{eqnarray} \label{equiv}
\sqrt{\tilde n} \left|
f\left(t+\frac {\tilde u_{\tilde n}(t)}{\sqrt{\tilde n}}+\frac {u_n\left(t+\frac {\tilde u_{\tilde n}(t)}{\sqrt{\tilde n}}\right)}{\sqrt n}\right)-f\left(t+\frac {B^1_{\tilde n}(t)}{\sqrt{\tilde n}}\right)\right|
\leq
 L\left(\sqrt{\frac {\tilde n\log \log n}n}   +   \frac{\log \tilde n}{\sqrt{\tilde n}}\right)
 \end{eqnarray}
 
If the resampling size $\tilde n \to \infty$ and it satisfies $\tilde n=o(\frac n {\log \log n})$ as the sample size $n\to \infty$, relations \eqref{qf_bootstrap2} and (\ref{equiv}) show that, if $F^{-1}$ is Lipschitz, then, the asymptotic behaviour of the processes $\sqrt{\tilde n}\left(\tilde F_{\tilde n}^{-1}(t)-F_n^{-1}(t)\right)$ and $\sqrt{\tilde n}\left(F^{-1}\left(t+\frac {B^1_{\tilde n}(t)}{\sqrt{\tilde n}}\right)-F_n^{-1}(t)\right)$ conditionally to $\{X_n\}$ are a.s. the same. In fact, since $\sqrt n(F_n^{-1}(t)-F^{-1}(t))$ converges in distribution, $\sqrt{\tilde n}(F_n^{-1}(t)-F^{-1}(t))\to 0$ in probability, and that behaviour is also the same as that  $\sqrt{\tilde n}\left(F^{-1}\left(t+\frac {B^1_{\tilde n}(t)}{\sqrt{\tilde n}}\right)-F^{-1}(t)\right)$. 

We can also elaborating with a view on the estimation of the parameter $\gamma$ in the following terms. Given two real functions $f$ and $g$ and versions of independent sequences of Brownian bridges $\{B_{\tilde n}^1\}$,  $\{B_{\tilde m}^2\}$  and of independent uniform quantile processes, $u_n$, $\tilde u_n$, $v_m$ and $\tilde v_m$, as in  Theorem \ref{aproxBridge},  we set 
\begin{eqnarray*}\label{Eq.Las Tildes}
\begin{array}{rcl}
{f}_{\tilde n}(t)&:=&f\left(t+\frac {\tilde u_{\tilde n}(t)}{\sqrt{\tilde n}}+\frac {u_n\left(t+\frac {\tilde u_{\tilde n}(t)}{\sqrt{\tilde n}}\right)}{\sqrt n}\right),
 {g}_{\tilde m} :=g\left(t+\frac {\tilde v_{\tilde m}(t)}{\sqrt{\tilde m}}+\frac {v_n\left(t+\frac {\tilde v_{\tilde m}(t)}{\sqrt{\tilde m}}\right)}{\sqrt m}\right),
\\
[4mm]
\tilde{f}_{\tilde n}(t)&:=&f\left({\textstyle t +\frac{B_{\tilde n}^1(t)}{\sqrt{\tilde n}} }\right),
\tilde{g}_{\tilde m} :=g\left({\textstyle t +\frac{B_{\tilde m}^2(t)}{\sqrt{\tilde m}} }\right).
\end{array}
\end{eqnarray*}

{
Finally, the announced Lemma \ref{lemmabootstrap} is merely the adaptation of Lemma 3.3 in \cite{del Barrio2021} to the present setup. Its proof follows the same arguments, with the additional already presented ingredients.}

\begin{Lemm} \label{lemmabootstrap}
Consider $A\subset (0,1)$ such that $\ell(A)>0$. With the notation and construction above, let the resampling sizes  satisfy $\tilde n=o(\frac n {\log \log n})$, $\tilde m=o(\frac m {\log \log m})$. If we assume that $f,g$ are two real Lipschitz functions, then there exists $L>0$ such that, if $C_{\tilde n,\tilde m}:= L \left({\textstyle \frac {\log{\tilde n}} {\tilde n} + \frac {\log{\tilde m}} {\tilde m}+\sqrt{\frac {\log \log n}n}+\sqrt{\frac {\log \log m}m}}\right)$,  then whenever $n,m \to \infty$, on a probability one set eventually,
\begin{eqnarray*}
\ell\big\{t\in A:\, \tilde{f}_{\tilde n}(t)>\tilde{g}_{\tilde m}(t)+C_{\tilde n,\tilde m}\big\}
&\leq& \ell\big\{t\in A:\, {f}_{\tilde n}(t)>{g}_{\tilde m}(t)\big\}\\
\label{Eq.TgEnCero_2}
 &\leq& \ell\big\{t\in A:\, \tilde{f}_{\tilde n}(t) > \tilde{g}_{\tilde m}(t)-C_{\tilde n,\tilde m}\big\}.
\end{eqnarray*}
\end{Lemm}


{
 \begin{Nota}\label{noboots} It is not clear when the naif bootstrap works for the gamma index. We are unaware of any positive or negative results.  
 The instability shown in the simulations performed, even in very simple models and with moderate and large sample sizes, effectively invalidates it in practice, so we have not insisted on a proof.
 
   It seems that the role of the terms $\frac {u_n\left(t+\frac {\tilde u_{\tilde n}(t)}{\sqrt{\tilde n}}\right)}{\sqrt n}$ and $\frac {v_n\left(t+\frac {\tilde v_{\tilde m}(t)}{\sqrt{\tilde m}}\right)}{\sqrt m}$ in ${f}_{\tilde n}$ and ${g}_{\tilde m}$, depending on the original samples, cannot be ignored when taking the resampling sizes $\tilde n=n$ and $\tilde m=m$.  This fact leads to noticeable variations in the bootstrap estimates of the index $\gamma(F,G)$, which can be observed empirically. For example, take $F$ and $G$ uniform distributions on $(0,1)$ and on $(0.1,0.9)$ respectively, and compare the Monte Carlo and the bootstrap distributions for, say, $\tilde n=n=\tilde m=m=50000$. The former is obtained by sampling directly from the distributions $F$ and $G$, and the (conditional) bootstrap distribution is obtained by resampling from a given sample of each of the distributions $F$ and $G$.  \end{Nota}
}

\section{Computational details}\label{Subs.Technicalities}
\subsection{$\gamma$-index}\label{gamma_index_Technicalities}

{
Inference on this index} requires  the estimation of  the exponent $1/(2r_0)$ in \eqref{bootslimit}.
 To do
this we have used  the expression in the right hand side of  relation (17) in the main paper 
with the following choices:

\begin{itemize}

\item[-]
We  use  three subsample sizes  $(n+m)^{x}$, with  $x=0.75, 0.85, 0.95$, making  the three possible comparisons between them. We  take  $\tilde n^x$ and $\tilde m^x$as explained in Remark 3.3 in the main paper.

\item[-]
{
In order to largely avoid the appearance of $0/0$  inside} the logarithm in the numerator of  (17) in the main paper 
(this happened 1-2\% of times at most and those case were deleted), we have  taken  $\alpha$ in the set  $\{0.05, 0.10, \ldots, 0.40\}$ and we have fixed $\beta=\alpha+0.55$.

\end{itemize}

With such parameters, we have obtained   24 estimates of  $2r_0$. These estimates   can be disperse. For instance,  we have run again just once  the INE dataset with  2005 as initial year,  and  the 24 obtained values go from   0.495 to 3.467 with median at  0.964. The range was  from 1.641 to 6.837, with median at   2.246, for the new repetition of the NHANES dataset at age 13.

We have used the percentile    95 of those values as the  {
estimation} of $2r_0$. This  choice gives a safe estimate of the real speed, because according to \eqref{NPboundstrue}, the higher the used $2r_0$,  the larger the obtained confidence interval.

{
 As shown in Table \ref{Tabla.Exponentes}, the choice of $2r_0$ has little impact on the resulting confidence interval, especially with large samples.} Table \ref{Tabla.Exponentes} shows the confidence intervals obtained for the above cases, whose sample sizes are $N=$ 25,728 and $N=$ 1,485 respectively, {
for several confidence levels}.

\begin{table}[h] 
\centering
\begin{tabular}{c|cc|cc}
 \hline
 Chosen& \multicolumn{2}{c|}{INE data. 2005 vs 2009} & \multicolumn{2}{c}{NHANES data. Age=13}
\\
exponent &  Estimated $2r_0$ & Conf. interval  &  Estimated $2r_0$  & Conf. interval 

\\
\hline
Minimum & .405 & $(.035,.053)$ & 1.641 & $(.060,.206)$
\\
Median& .966 & $(.030,060)$ & 2.246 & $(.055,.209)$
\\
Maximum & 3.468& $(.024,.067)$  & 6.837 & $(.044,.219)$
\\
\hline
\end{tabular}
 \caption{Order statistic used to estimate  $2r_0$,  obtained estimates and $.95$ confidence intervals for $\gamma$ in the INE data  with 2005 as initial year and in the NHANES data for age 13. }

\label{Tabla.Exponentes}
 \end{table}

We also needed to use {
low resampling bootstrap }
to compute the confidence interval. Here,  as explained, in Section 3 in the main paper
, the subsampling rate {
was set to} $(n+m)^{.95}$.

\subsection{ $\pi$-index} 
Upper and lower bounds have been computed separately, each to the level $0.025$, as follows:

\vspace{2mm}

\noindent
{\it Lower bound:} {In \cite{Alv2014}, it is shown that  the quantiles of the limit law in Theorem 3.1 in the main paper 
can be suitably bounded 
above 
by  the quantiles of the law of:  
\begin{eqnarray*}
\bar{B}({\pi (F,G)},\lambda):=
\sup_{t \in [{\pi (F,G)},1]}\left(\sqrt{ \lambda} \ 
B_1(t)- \sqrt{1-\lambda} \ B_2(t-{\pi (F,G)}\right).
\end{eqnarray*}
To find the {lower} bound, we need to obtain the .975 quantile of the random variable  $\overline B$ defined here. We have replaced $\overline B$  by its 
plug-in estimate: $\overline B_{n,m}:=\overline B(\pi(F_n,G_m),n/(n+m))$.
}

The  .975 quantile of $\overline B_{n,m}$ can be obtained through numerical computation, {alternatively we runned} 5000 simulations of $\overline B_{n,m}$ and take the  .975 quantile of the sample obtained. The involved Brownian Bridges have been approximated using the values $t=.0001, .0002, \ldots, .9999, 1$.

\vspace{2mm}

 \noindent
{\it Upper bound:} It was computed using the procedure described in Section 3.3 in \cite{Alv2015}  using the worst possible variance.

\subsection{ $\rho$-index} 
{
 We used the fact (see Theorem 5.3 in \cite{Kotz}) that  $(\hat \rho_{n,m} -\rho)/S_{n,m}$ converges in distribution to the standard normal distribution. Here $S_{n,m}^2$ is a sequence of estimators of the variance of $\hat\rho_{n,m}$. There are several available posibilites for $S_{n,m}$;  we have chosen that one given in  (5.43) in \cite{Kotz}.
 }

\section{Tables of Subsections 4.1.1 
and 4.1.2
}\label{Tablas_Simul}

{The complete results of the simulations for the three indices $\gamma, \pi$ and $\rho$ appear in Tables \ref{TablaSimulNormal_gamma}, \ref{TablaSimulNormal_pi} and  \ref{TablaSimulNormal_rho} for pairs of Gaussian distributions  and  in Tables \ref{TablaSimulUnif_gamma}, \ref{TablaSimulUnif_pi} and \ref{TablaSimulUnif_rho} for the uniform case.
}

\begin{table}[H]
\centering
 {\tiny
\begin{tabular}{ccccccccc}
  \hline
$\gamma$ &  $\mu$ &  $\sigma$ &   size & $\widehat{2r_0}$ & $\hat \gamma_L$ & $\hat \gamma$ & $\hat \gamma_U$ & Coverage  \\ 
  \hline
  .01 & .233 & 1.10 & 250 & 1.648 & .0052 & .0830 & .1469 & .8240 \\ 
 &  & &  & (.971) & (.0353) & (.1026) & (.1822) & (.3812) \\ 
  &    &   & 1000 & 1.291 & .0019  &  .0340  &  .0581  &  .7140 \\ 
 &  &  &  & (.411) & (.0106) & (.0403) & (.0697) & (.4523) \\ 
  &  & & 5000 & 1.414 & .0012  &  .0160  &  .0270  &  .6260 \\ 
 &  &  &  & (.413) & (.0050) & (.0156) & (.0267) & (.4843) \\ 
  \cline{2-9} 
  & 1.163 & 1.50 & 250 & 1.145 & .0007 & .0175 & .0307 & .6380 \\ 
 &  & &  & (.351) & (.0042) & (.0176) & (.0308) & (.4811) \\ 
&  &  & 1000 & 1.407 & .0011  &  .0126  &  .0217  &  .7060 \\ 
 &  &  &  & (.434) & (.0033) & (.0093) & (.0159) & (.4560) \\
  &  & & 5000 & 1.860 & .0021  &  .0097  &  .0159  &  .7160 \\ 
 &  &  &  & (.441) & (.0032) & (.0047) & (.0076) & (.4514) \\ 
 \cline{2-9}  
& 2.326 & 2.00 & 250 & 1.085 & .0007 & .0122 & .0214 & .5972 \\ 
 &  & &  & (.279) & (.0029) & (.0119) & (.0209) & (.4910) \\ 
 &  &  & 1000 & 1.240 & .0015  &  .0103  &  .0171  &  .7060 \\ 
 &  &  &  & (.314) & (.0031) & (.0066) & (.0107) & (.4560) \\ 
 &  & & 5000 & 1.720 & .0041  &  .0100  &  .0151  &  .8600 \\ 
 &  &  &  & (.367) & (.0030) & (.0030) & (.0042) & (.3473) \\ 
  \hline 
   .05 & .164 & 1.10 & 250 & 2.159 & .0229 & .1522 & .2634 & .6700 \\ 
 &  & &  & (1.295) & (.0892) & (.1611) & (.2658) & (.4707) \\ 
   &    &   & 1000 & 1.569 & .0045  &  .0762  &  .1311  &  .6380 \\ 
 &  &  &  & (.605) & (.0231) & (.0721) & (.1260) & (.4811) \\ 
 &  & & 5000 & 1.714 & .0059  &  .0548  &  .0916  &  .6640 \\ 
 &  &  &  & (.513) & (.0171) & (.0380) & (.0645) & (.4728) \\ 
 \cline{2-9} 
&  .822&  1.50& 250 & 1.461 & .0062 & .0562 & .0970 & .6760 \\ 
 &  & &  & (.426) & (.0186) & (.0417) & (.0710) & (.4685) \\ 
 &    &   & 1000 & 1.788 & .0122  &  .0509  &  .0830  &  .7820 \\ 
 &  &  &  & (.443) & (.0167) & (.0235) & (.0375) & (.4133) \\ 
 &  & & 5000 & 1.788 & .0280  &  .0497  &  .0690  &  .8740 \\ 
 &  &  &  & (.378) & (.0129) & (.0108) & (.0143) & (.3322) \\ 
 \cline{2-9} 
  & 1.645 & 2.00 & 250 & 1.483 & .0094 & .0521 & .0880 & .7480 \\ 
 &  & &  & (.377) & (.0162) & (.0288) & (.0467) & (.4346) \\ 
  &  &  & 1000 & 1.649 & .0202  &  .0491  &  .0737  &  .8520 \\ 
 &  &  &  & (.374) & (.0150) & (.0147) & (.0207) & (.3555) \\ 
 &  & & 5000 & 1.711 & .0362  &  .0494  &  .0617  &  .8900 \\ 
 &  &  &  & (.241) & (.0075) & (.0069) & (.0084) & (.3132) \\ 
 \hline 
 .10 & .128 & 1.10 & 250 & 2.667 & .0449 & .2212 & .3772 & .6680 \\ 
 &  & &  & (1.415) & (.1354) & (.1949) & (.3030) & (.4714)\\
   &   &   & 1000 & 1.908 & .0141  &  .1316  &  .2281  &  .6480 \\ 
 &  &  &  & (.759) & (.0498) & (.1115) & (.1938) & (.4781) \\ 
 &  & & 5000 & 1.889 & .0201  &  .1057  &  .1738  &  .7020 \\ 
 &  &  &  & (.543) & (.0386) & (.0596) & (.0979) & (.4578) \\ 
 \cline{2-9} 
 & .641 & 1.50 & 250 & 1.638 & .0179 & .1022 & .1728 & .7060 \\ 
 &  & &  & (.441) & (.0342) & (.0586) & (.0963) & (.4560) \\ 
 &    &   & 1000 & 1.842 & .0404  &  .1031  &  .1586  &  .8200 \\ 
 &  &  &  & (.466) & (.0351) & (.0350) & (.0516) & (.3846) \\ 
 &  & & 5000 & 1.703 & .0701  &  .0994  &  .1267  &  .8700 \\ 
 &  &  &  & (.279) & (.0171) & (.0148) & (.0187) & (.3366) \\ 
 \cline{2-9}
 & 1.282 & 2.00 & 250 & 1.631 & .0298 & .1016 & .1628 & .8160 \\ 
 &  & &  & (.356) & (.0329) & (.0399) & (.0592) & (.3879) \\ 
  &  &  & 1000 & 1.692 & .0595  &  .1005  &  .1374  &  .8960 \\ 
 &  &  &  & (.312) & (.0224) & (.0202) & (.0260) & (.3056) \\ 
 &  & & 5000 & 1.706 & .0816  &  .0997  &  .1170  &  .9260 \\ 
 &  &  &  & (.205) & (.0100) & (.0093) & (.0105) & (.2620) \\ 
   \hline
\end{tabular}
 }
\caption{Results from 500  simulations of $N(0,1)$ vs. $N(\mu,\sigma^2)$. Means of the estimations of $2r_0$, of $\gamma$, and its confidence intervals at   level 0.05. Coverage is the proportion of times that $\gamma\in (\hat \gamma_L,\hat\gamma_U)$. Between parenthesis are the standard deviations of the  estimations. } 
\label{TablaSimulNormal_gamma}
\end{table}

\begin{table}[H]
\centering
{\tiny
\begin{tabular}{ccccccccc}
  \hline
  $\gamma$ & $a$ & $H$ & size & $\widehat{2r_0}$ & $\hat \gamma_L$ & $\hat \gamma$ & $\hat \gamma_U$ & Coverage \\ 
  \hline
  .01 & -.051 & 6.00 & 250 & 1.001 & .0005 & .0114 & .0194 & .7065 \\ 
 &  && & (.141) & (.0019) & (.0079) & (.0131) & (.4558) \\ 
 &   &   & 1000 & 1.048 &   .0029 &  .0100 &  .0151 &  .8300 \\ 
 &  &  &  &  (.132) &  (.0027) &  (.0038) &  (.0053) &  (.3760) \\ 
 &  &  & 5000 & 1.431 &   .0068 &  .0101 &  .0129 &  .9040 \\ 
 &  &  &  &  (.212) &  (.0014) &  (.0017) &  (.0020) &  (.2949) \\  
 \cline{2-9} 
 & -.101 & 11.00 & 250 & .987 & .0007 & .0108 & .0183 & .6989 \\ 
 &  & & & (.096) & (.0022) & (.0072) & (.0118) & (.4592) \\ 
 &   &   & 1000 & 1.026 &   .0038 &  .0101 &  .0149 &  .8300 \\ 
 &  &  &  &  (.096) &  (.0027) &  (.0037) &  (.0049) &  (.3760) \\  
 &  &  & 5000 & 1.393 &   .0071 &  .0100 &  .0125 &  .8920 \\ 
 &  &  &  &  (.227) &  (.0014) &  (.0016) &  (.0019) &  (.3107) \\  
 \cline{2-9} 
 & -.202 & 21.00 & 250 & .982 & .0006 & .0106 & .0180 & .7228 \\ 
 &   & & & (.077) & (.0017) & (.0063) & (.0103) & (.4481) \\ 
 &   &   & 1000 & 1.011 &   .0041 &  .0099 &  .0144 &  .8580 \\ 
 &  &  &  &  (.048) &  (.0023) &  (.0032) &  (.0041) &  (.3494) \\  
 &  &  & 5000 & 1.378 &   .0073 &  .0101 &  .0125 &  .8880 \\ 
 &  &  &  &  (.233) &  (.0013) &  (.0015) &  (.0018) &  (.3157) \\  
 \hline
 .05 & -.050 & 1.95 & 250 & 1.353 & .0059 & .0568 & .0920 & .7820 \\ 
 &  & & & (.397) & (.0134) & (.0308) & (.0499) & (.4133) \\  
  &   &   & 1000 & 1.508 &   .0168 &  .0514 &  .0724 &  .8080 \\ 
 &  &  &  &  (.304) &  (.0141) &  (.0164) &  (.0237) &  (.3943) \\  
 &  &  & 5000 & 1.705 &   .0358 &  .0505 &  .0619 &  .8860 \\ 
 &  &  &  &  (.220) &  (.0072) &  (.0069) &  (.0088) &  (.3181) \\  
 \cline{2-9} 
 & -.100 & 2.90 & 250 &1.331 & .0095 & .0520 & .0808 & .8020 \\ 
 &  & & & (.280) & (.0137) & (.0234) & (.0351) & (.3989) \\ 
 &   &   & 1000 & 1.507 &   .0273 &  .0504 &  .0671 &  .8380 \\ 
 &  &  &  &  (.259) &  (.0108) &  (.0113) &  (.0150) &  (.3688) \\  
 &  &  & 5000 & 1.683 &   .0402 &  .0501 &  .0585 &  .9000 \\ 
 &  &  &  &  (.181) &  (.0049) &  (.0050 &  (.0060) &  (.3003) \\  
 \cline{2-9}
 & -.200 & 4.80 & 250 & 1.245 & .0164 & .0519 & .0777 & .8600 \\ 
 & &  & & (.231) & (.0135) & (.0174) & (.0247) & (.3473) \\ 
 &   &   & 1000 & 1.421 &   .0321 &  .0494 &  .0636 &  .8820 \\ 
 &  &  &  &  (.234) &  (.0078) &  (.0085) &  (.0106) &  (.3229) \\  
 &  &  & 5000 & 1.650 &   .0426 &  .0503 &  .0573 &  .9640 \\ 
 &  &  &  &  (.153) &  (.0035) &  (.0036) &  (.0040) &  (.1865) \\  
 \hline
  .10 & -.050 & 1.45 & 250 & 1.591 & .0158 & .1165 & .1893 & .7160 \\ 
 & &  & & (.555) & (.0377) & (.0683) & (.1154) & (.4514) \\  
  &   &   & 1000 & 1.672 &   .0316 &  .1054 &  .1516 &  .8140 \\ 
 &  &  &  &  (.402) &  (.0315) &  (.0334) &  (.0512) &  (.3895) \\  
 &  &  & 5000 & 1.682 &   .0681 &  .1007 &  .1248 &  .8480 \\ 
 &  &  &  &  (.267) &  (.0165) &  (.0154) &  (.0208) &  (.3594) \\  
 \cline{2-9} 
 & -.100 & 1.90 & 250 & 1.509 & .0229 & .1062 & .1616 & .7940 \\ 
 &  & & & (.369) & (.0309) & (.0430) & (.0667) & (.4048) \\
 &   &   & 1000 & 1.611 &   .0570 &  .1024 &  .1353 &  .8720 \\ 
 &  &  &  &  (.285) &  (.0212) &  (.0218) &  (.0298) &  (.3344) \\  
 &  &  & 5000 & 1.726 &   .0817 &  .1010 &  .1177 &  .8820 \\ 
 &  &  &  &  (.213) &  (.0103) &  (.0101) &  (.0123) &  (.3229) \\  
 \cline{2-9} 
 & -.200 & 2.80 & 250 & 1.439 & .0391 & .1018 & .1461 & .8520 \\ 
 &  & & & (.241) & (.0275) & (.0295) & (.0411) & (.3555) \\
 &   &   & 1000 & 1.592 &   .0698 &  .0999 &  .1250 &  .9060 \\ 
 &  &  &  &  (.228) &  (.0143) &  (.0145) &  (.0183) &  (.2921) \\  
 &  &  & 5000 & 1.705 &   .0872 &  .1004 &  .1124 &  .9520 \\ 
 &  &  &  &  (.160) &  (.0063) &  (.0065) &  (.0073) &  (.2140) \\
   \hline
\end{tabular}
}
\caption{Results from 500 simulations of $U(0,1)$ vs. $U(a,H)$. Means of estimations of the inverse of the convergence rate, of $\gamma$, and of confidence intervals for $\gamma$, at the .05 level. Coverage is the proportion of times that $\gamma\in (\hat \gamma_L,\hat\gamma_U)$. In parenthesis are the standard deviations of the above estimations.  } 
\label{TablaSimulUnif_gamma}
\end{table}

\begin{table}[H]
\centering
{\tiny
\begin{tabular}{cccccccc}
  \hline
 $\pi$ & $\mu$ & $\sigma$ & size & $\hat \pi_L$ & $\hat \pi$ & $\hat \pi_U$ & Coverage \\ 
  \hline
  .0004    & .233   &  1.10    & 250   &  0 & .0172    &  .0712    &  1 \\ 
   &    &    &     &  (0) & (.0149)    &  (.0231)     &  (0) \\ 
   &    &    & 1000   &  0 & .0062    &  .0254    &  1 \\ 
   &    &    &     &  (0) & (.0051)    &  (.0087)     &  (0) \\ 
   &    &    & 5000   &  0 & .0021    &  .0076    &  1 \\ 
   &    &    &     &  (0) & (.0015)    &  (.0027)     &  (0) \\ 
   \hline
 .0019    &  .164    &  1.10    & 250   &  0 & .0270    &  .0843    &  1 \\ 
   &    &    &     &  (0) & (.0202)    &  (.0268)     &  (0) \\ 
   &    &    & 1000   &  0 & .0104    &  .0349    &  1 \\ 
   &    &    &     &  (0) & (.0077)    &  (.0107)     &  (0) \\ 
   &    &    & 5000   &  0 & .0051    &  .0137    &  1 \\ 
   &    &    &     &  (0) & (.0029)    &  (.0039)     &  (0) \\ 
   \hline
 .0039    &  .128    &  1.10    & 250   &  .0001    &  .0321    &  .0938    &  1 \\ 
   &    &    &    &  (.0016)    &  (.0233)    &  (.0286)     &  (0) \\ 
   &    &    & 1000   &  0 & .0140    &  .0400    &  1 \\ 
   &    &    &     &  (0) & (.0091)    &  (.0115)     &  (0) \\ 
   &    &    & 5000   &  0 & .0077    &  .0181    &  1 \\ 
   &    &    &     &  (0) & (.0037)    &  (.0044)     &  (0) \\ 
   \hline 
 .0016    & 1.163   &  1.50    & 250   &  0 & .0082    &  .0310    &  .992 \\ 
   &    &    &     &  (0) & (.0072)    &  (.0120)     &  (.089) \\ 
   &    &    & 1000   &  0 & .0045    &  .0135    &  1 \\ 
   &    &    &     &  (0) & (.0031)    &  (.0046)     &  (0) \\ 
   &    &    & 5000   &  0 & .0025    &  .0058    &  1 \\ 
   &    &    &     &  (0) & (.0011)    &  (.0015)     &  (0) \\ 
   \hline
 .0081    &  .822    &  1.50    & 250   &  0 & .0199    &  .0521    &  1 \\ 
   &    &    &     &  (0) & (.0122)    &  (.0162)     &  (0) \\ 
   &    &    & 1000   &  0 & .0131    &  .0268    &  1 \\ 
   &    &    &     &  (0) & (.0056)    &  (.0058)     &  (0) \\ 
   &    &    & 5000   &  0 & .0098    &  .0155    &  .998 \\ 
   &    &    &     &  (0) & (.0026)    &  (.0023)     &  (.045) \\ 
   \hline 
 .0165    &  .641    &  1.50    & 250   &  0 & .0319    &  .0659    &  1 \\ 
   &    &    &     &  (0) & (.0160)    &  (.0192)     &  (0) \\ 
   &    &    & 1000   &  0 & .0229    &  .0375    &  1 \\ 
   &    &    &     &  (0) & (.0079)    &  (.0092)     &  (0) \\ 
   &    &    & 5000   &  0 & .0190    &  .0249    &  .986 \\ 
   &    &    &    &  (.0003)    &  (.0034)    &  (.0041)     &  (.118) \\ 
   \hline 
 .0026    & 2.326   &  2.00    & 250   &  0 & .0081    &  .0262    &  .982 \\ 
   &    &    &     &  (0) & (.0066)    &  (.0103)     &  (.133) \\ 
   &    &    & 1000   &  0 & .0050    &  .0130    &  1 \\ 
   &    &    &     &  (0) & (.0027)    &  (.0038)     &  (0) \\ 
   &    &    & 5000   &  0 & .0034    &  .0064    &  .992 \\ 
   &    &    &     &  (0) & (.0011)    &  (.0014)     &  (.089) \\ 
   \hline 
 .0136    & 1.645   &  2.00    & 250   &  0 & .0230    &  .0483    &  1 \\ 
   &    &    &     &  (0) & (.0113)    &  (.0142)     &  (0) \\ 
   &    &    & 1000   &  0 & .0175    &  .0285    &  1 \\ 
   &    &    &     &  (0) & (.0058)    &  (.0065)     &  (0) \\ 
   &    &    & 5000   &  0 & .0151    &  .0196    &  .994 \\ 
   &    &    &     &  (0) & (.0025)    &  (.0028)     &  (.077) \\ 
   \hline
 .0277    & 1.282   &  2.00    & 250   &  0 & .0393    &  .0683    &  .994 \\ 
   &    &    &     &  (0) & (.0166)    &  (.0212)     &  (.077) \\ 
   &    &    & 1000   &  0 & .0327    &  .0462    &  .980 \\ 
   &    &    &     &  (0) & (.0078)    &  (.0096)     &  (.140) \\ 
   &    &    & 5000   &  .0024    &  .0297    &  .0358    &  .984 \\ 
   &    &    &    &  (.0029)    &  (.0037)    &  (.0041)     &  (.126) \\ 
   \hline
\end{tabular}
   }
\caption{Results from 500  simulations of $N(0,1)$ vs. $N(\mu,\sigma^2)$. Means of the estimations of $\pi$, and its confidence intervals at   level 0.05. Coverage is the proportion of times that $\pi\in (\hat \pi_L,\hat\pi_U)$. Between parenthesis are the standard deviations of the  estimations. } 
\label{TablaSimulNormal_pi}
\end{table}

 \begin{table}[H]
\centering
 {\tiny
\begin{tabular}{cccccccc}
  \hline
 $\pi$ & $a$ & $H$& size & $\hat \pi_L$ & $\hat \pi$ & $\hat \pi_U$ & Coverage \\ 
  \hline
.0083 & -.051 & 6 & 250 & 0 & .0088 & .0221 & .9300 \\ 
 &  &  &  & (0) & (.0058) & (.0091) & (.2554) \\ 
 &  &  & 1000 & 0 & .0085 & .0147 & .9520 \\ 
 &  &  &  & (0) & (.0030) & (.0037) & (.2140)  \\ 
 &  &  & 5000 & 0 & .0083 & .0110 & .9620  \\ 
 &  &  &  & (0) & (.0013) & (.0014) & (.1914)  \\ 
 \hline 
.0091 & -.101 & 11 & 250 & 0 & .0093 & .0214 & .9200 \\ 
 &  &  &  & (0) & (.0060) & (.0100) & (.2716)  \\ 
 &  &  & 1000 & 0 & .0093 & .0155 & .9620 \\ 
 &  &  &  & (0) & (.0030) & (.0037) & (.1914)  \\ 
 &  &  & 5000 & 0 & .0091 & .0118 & .9660 \\ 
 &  &  &  & (0) & (.0014) & (.0015) & (.1814)  \\ 
 \hline 
.0095 & -.202 & 21 & 250 & 0 & .0099 & .0220 & .9220 \\ 
 &  &  &  & (0) & (.0062) & (.0102) & (.2684)  \\ 
 &  &  & 1000 & 0 & .0096 & .0158 & .9560 \\ 
 &  &  &  & (0) & (.0031) & (.0040) & (.2053)  \\ 
 &  &  & 5000 & 0 & .0096 & .0123 & .9720 \\ 
 &  &  &  & (0) & (.0013) & (.0014) & (.1651)  \\ 
 \hline 
.0250 & -.050 & 1.95 & 250 & 0 & .0302 & .0514 & .9840 \\ 
 &  &  &  & (0) & (.0113) & (.0154) & (.1256)  \\ 
 &  &  & 1000 & 0 & .0264 & .0357 & .9740 \\ 
 &  &  &  & (0) & (.0052) & (.0062) & (.1593)  \\ 
 &  &  & 5000 & .0002 & .0253 & .0294 & .9660 \\ 
 &  &  &  & (.0007) & (.0023) & (.0025) & (.1814)  \\ 
 \hline 
.0333 & -.100 & 2.90 & 250 & 0 & .0353 & .0572 & .9500 \\ 
 &  &  &  & (0) & (.0115) & (.0154) & (.2182)  \\ 
 &  &  & 1000 & 0 & .0334 & .0440 & .9460 \\ 
 &  &  &  & (0) & (.0058) & (.0068) & (.2262)  \\ 
 &  &  & 5000 & .0063 & .0334 & .0383 & .9600 \\ 
 &  &  &  & (.0027) & (.0027) & (.0029) & (.1962)  \\ 
 \hline 
.0400 & -.200 & 4.80 & 250 & 0 & .0405 & .0640 & .9540 \\ 
 &  &  &  & (0) & (.0124) & (.0161) & (.2097)  \\ 
 &  &  & 1000 & 0 & .0411 & .0532 & .9720 \\ 
 &  &  &  & (0) & (.0061) & (.0070) & (.1651)  \\ 
 &  &  & 5000 & .0129 & .0400 & .0454 & .9840 \\ 
 &  &  &  & (.0028) & (.0028) & (.0030) & (.1256)  \\ 
 \hline 
.0333 & -.050 & 1.45 & 250 & 0 & .0398 & .0650 & .9720 \\ 
 &  &  &  & (0) & (.0138) & (.0201) & (.1651)  \\ 
 &  &  & 1000 & 0 & .0347 & .0453 & .9560 \\ 
 &  &  &  & (0) & (.0060) & (.0073) & (.2053)  \\ 
 &  &  & 5000 & .0065 & .0339 & .0386 & .9760 \\ 
 &  &  &  & (.0027) & (.0026) & (.0028) & (.1532)  \\ 
\hline 
 0500 & -.100 & 1.90 & 250 & 0 & .0530 & .0796 & .9520 \\ 
 &  &  &  & (0) & (.0147) & (.0188) & (.2140)  \\ 
 &  &  & 1000 & .0003 & .0509 & .0639 & .9660 \\ 
 &  &  &  & (.0013) & (.0069) & (.0077) & (.1814)  \\ 
 &  &  & 5000 & .0230 & .0502 & .0561 & .9780 \\ 
 &  &  &  & (.0031) & (.0030) & (.0032) & (.1468)  \\ 
 \hline 
.0667 & -.200 & 2.80 & 250 & 0 & .0695 & .0996 & .9560 \\ 
 &  &  &  & (.0001) & (.0159) & (.0194) & (.2053)  \\ 
 &  &  & 1000 & .0072 & .0669 & .0819 & .9480 \\ 
 &  &  &  & (.0069) & (.0084) & (.0093) & (.2222)  \\ 
 &  &  & 5000 & .0395 & .0666 & .0735 & .9640 \\ 
 &  &  &  & (.0036) & (.0036) & (.0038) & (.1865)  \\ 
   \hline
\end{tabular}
 }
\caption{Results from 500  simulations of  $U(0,1)$ vs. $U(a,H)$. Means of the estimations of $\pi$ and of its confidence intervals at the .05 level. Coverage is the proportion of times that $\pi\in (\hat \pi_L,\hat\pi_U)$. In parenthesis are the standard deviations of the above estimations.} 
\label{TablaSimulUnif_pi}
\end{table}

\begin{table}[H]
\centering
{\tiny
\begin{tabular}{cccccccc}
  \hline
  	 $\rho$  & 	  $\mu$  & 	  $\sigma$  & 	  size  & 	  $\hat \rho_L$  & 	  $\hat \rho$  & 	  $\hat \rho_U$  & 	  Coverage \\ 
  \hline
 .5623 & .233 & 1.10 &250& .5121 & .5624 & .6126 & .9520 \\ 
&&&& (.0257) & (.0253) & (.0249) & (.2140) \\ 
&&&1000& .5372 & .5623 & .5874 & .9480 \\ 
&&&& (.0126) & (.0125) & (.0124) & (.2222) \\ 
&&&5000& .5511 & .5623 & .5735 & .9600 \\ 
&&&& (.0059) & (.0058) & (.0058) & (.1962) \\ 
\hline
 .5439 & .164 & 1.10 &250& .4927 & .5432 & .5937 & .9540 \\ 
&&&& (.0266) & (.0263) & (.0260) & (.2097) \\ 
&&&1000& .5187 & .5439 & .5691 & .9440 \\ 
&&&& (.0133) & (.0132) & (.0131) & (.2302) \\ 
&&&5000& .5326 & .5439& .5551 & .9620 \\ 
&&&& (.0056) & (.0056) & (.0056) & (.1914) \\ 
\hline
 .5343 & .128 & 1.10 &250& .4850 & .5356 & .5862 & .9560 \\ 
&&&& (.0255) & (.0253) & (.0251) & (.2053) \\ 
&&&1000& .5089 & .5341 & .5594 & .9620 \\ 
&&&& (.0125) & (.0124) & (.0124) & (.1914) \\ 
&&&5000& .5231 & .5344 & .5457 & .9640 \\ 
&&&& (.0055) & (.0055) & (.0055) & (.1865) \\ 
\hline
 .7406 &1.163& 1.50 &250& .6956 & .7394 & .7832 & .9500 \\ 
&&&& (.0237) & (.0222) & (.0207) & (.2182) \\ 
&&&1000& .7186 & .7405 & .7623 & .9460 \\ 
&&&& (.0113) & (.0109) & (.0106) & (.2262) \\ 
&&&5000& .7310 & .7408 & .7505 & .9340 \\ 
&&&& (.0052) & (.0051) & (.0050) & (.2485) \\ 
\hline
 .6758 & .822 & 1.50 &250& .6284 & .6758 & .7232 & .9480 \\ 
&&&& (.0253) & (.0241) & (.0230) & (.2222) \\ 
&&&1000& .6516 & .6753 & .6989 & .9620 \\ 
&&&& (.0115) & (.0113) & (.0110) & (.1914) \\ 
&&&5000& .6646 & .6752 & .6858 & .9580 \\ 
&&&& (.0053) & (.0053) & (.0052) & (.2008) \\ 
\hline
 .6389 & .641 & 1.50 &250& .5884 & .6374 & .6864 & .9460 \\ 
&&&& (.0255) & (.0246) & (.0238) & (.2262) \\ 
&&&1000& .6147 & .6391 & .6635 & .9520 \\ 
&&&& (.0130) & (.0128) & (.0125) & (.2140) \\ 
&&&5000& .6283 & .6392 & .6502 & .9700 \\ 
&&&& (.0053) & (.0053) & (.0052) & (.1708) \\ 
\hline
 .8509 &2.326& 2.00 &250& .8145 & .8497 & .8850 & .9440 \\ 
&&&& (.0204) & (.0182) & (.0160) & (.2302) \\ 
&&&1000& .8339 & .8514 & .8690 & .9440 \\ 
&&&& (.0092) & (.0086) & (.0081) & (.2302) \\ 
&&&5000& .8428 & .8507 & .8586 & .9400 \\ 
&&&& (.0043) & (.0042) & (.0041) & (.2377 \\ 
\hline
 .7690 &1.645& 2.00 &250& .7271 & .7699 & .8128 & .9600 \\ 
&&&& (.0222) & (.0206) & (.0189) & (.1962) \\ 
&&&1000& .7475 & .7690 & .7904 & .9660 \\ 
&&&& (.0107) & (.0103) & (.0099) & (.1814) \\ 
&&&5000& .7593 & .7689 & .7785 & .9480 \\ 
&&&& (.0050) & (.0050) & (.0049) & (.2222) \\ 
\hline
 .7168 &1.282& 2.00 &250& .6710 & .7175 & .7639 & .9480 \\ 
&&&& (.0246) & (.0232) & (.0218) & (.2222) \\ 
&&&1000& .6943 & .7175 & .7408 & .9440 \\ 
&&&& (.0125) & (.0121) & (.0118) & (.2302) \\ 
&&&5000& .7061 & .7165 & .7269 & .9420 \\ 
&&&& (.0055) & (.0055) & (.0054) & (.2340) \\ 
   \hline
\end{tabular}
}
\caption{Results from 500  simulations of $N(0,1)$ vs. $N(\mu,\sigma^2)$. Means of the estimations of $\rho$, and its confidence intervals at   level 0.05. Coverage is the proportion of times that $\rho\in (\hat \rho_L,\hat\rho_U)$. Between parenthesis are the standard deviations of the  estimations. } 
\label{TablaSimulNormal_rho}
\end{table}

\begin{table}[H]
\centering
{\tiny
\begin{tabular}{cccccccc}
  \hline
 $\rho$ & $a$ & $H$ & size & $\hat \rho_L$ & $\hat \rho$ & $\hat \rho_U$ & Coverage \\ 
  \hline
 .9090 & -.051 & 6.00 &250& .8797 & .9094 & .9390 & .9520 \\ 
&&&& (.0171) & (.0146) & (.0122) & (.2140) \\ 
&&&1000& .8942 & .9091 & .9239 & .9580 \\ 
&&&& (.0077) & (.0071) & (.0065) & (.2008) \\ 
&&&5000& .9025 & .9091 & .9158 & .9460 \\ 
&&&& (.0035) & (.0034) & (.0032) & (.2262) \\ 
 \hline
.9459 & -.101 & 11.00 &250& .9216 & .9453 & .9691 & .9480 \\ 
&&&& (.0149) & (.0121) & (.0093) & (.2222) \\ 
&&&1000& .9337 & .9456 & .9576 & .9500 \\ 
&&&& (.0066) & (.0059) & (.0052) & (.2182) \\ 
&&&5000& .9406 & .9460 & .9513 & .9460 \\ 
&&&& (.0028) & (.0026) & (.0025) & (.2262) \\ 
\hline
 .9669 & -.202 & 21.00 &250& .9470 & .9663 & .9856 & .9160 \\ 
&&&& (.0134) & (.0101) & (.0069) & (.2777) \\ 
&&&1000& .9576 & .9672 & .9768 & .9460 \\ 
&&&& (.0054) & (.0047) & (.0039) & (.2262) \\ 
&&&5000& .9626 & .9669 & .9712 & .9460 \\ 
&&&& (.0025) & (.0023) & (.0021) & (.2262) \\ 
\hline
 .7250 & -.050 & 1.95 &250& .6800 & .7259 & .7718 & .9580 \\ 
&&&& (.0246) & (.0231) & (.0216) & (.2008) \\ 
&&&1000& .7028 & .7257 & .7486 & .9360 \\ 
&&&& (.0129) & (.0125) & (.0121) & (.2450) \\ 
&&&5000& .7149 & .7252 & .7354 & .9280 \\ 
&&&& (.0056) & (.0055) & (.0054) & (.2587) \\ 
\hline
 .8000 & -.100 & 2.90 &250& .7584 & .8002 & .8421 & .9500 \\ 
&&&& (.0228) & (.0209) & (.0189) & (.2182) \\ 
&&&1000& .7795 & .8004 & .8212 & .9580 \\ 
&&&& (.0108) & (.0104) & (.0099) & (.2008) \\ 
&&&5000& .7904 & .7997 & .8091 & .9440 \\ 
&&&& (.0050) & (.0049) & (.0048) & (.2302) \\ 
\hline
 .8600 & -.200 & 4.80 &250& .8225 & .8598 & .8970 & .9560 \\ 
&&&& (.0207) & (.0183) & (.0160) & (.2053) \\ 
&&&1000& .8414 & .8600 & .8786 & .9660 \\ 
&&&& (.0096) & (.0090) & (.0085) & (.1814) \\ 
&&&5000& .8514 & .8597 & .8681 & .9460) \\ 
&&&& (.0045) & (.0044) & (.0042) & (.2262 \\ 
\hline
 .6333 & -.050 & 1.45 &250& .5827 & .6321 & .6816 & .9460 \\ 
&&&& (.0260) & (.0251) & (.0241) & (.2262) \\ 
&&&1000& .6080 & .6327 & .6573 & .9500 \\ 
&&&& (.0125) & (.0123) & (.0121) & (.2182) \\ 
&&&5000& .6220 & .6330 & .6441 & .9320 \\ 
&&&& (.0060) & (.0059) & (.0059) & (.2520) \\ 
\hline
 .7000 & -.100 & 1.90 &250& .6508 & .6986 & .7464 & .9520 \\ 
&&&& (.0254) & (.0241) & (.0228) & (.2140) \\ 
&&&1000& .6755 & .6993 & .7231 & .9680 \\ 
&&&& (.0119) & (.0116) & (.0113) & (.1762) \\ 
&&&5000& .6891 & .6998 & .7104 & .9540 \\ 
&&&& (.0054) & (.0054) & (.0053) & (.2097) \\ 
\hline
 .7667 & -.200 & 2.80 &250& .7228 & .7678 & .8129 & .9460 \\ 
&&&& (.0239) & (.0222) & (.0205) & (.2262) \\ 
&&&1000& .7445 & .7671 & .7897 & .9680 \\ 
&&&& (.0112) & (.0108) & (.0103) & (.1762) \\ 
&&&5000& .7566 & .7667 & .7768 & .9460 \\ 
&&&& (.0052) & (.0051) & (.0050) & (.2262) \\
   \hline
\end{tabular}
 }
\caption{Results from 500 simulations of  $U(0,1)$ vs. $U(a,H)$. Means of the estimations of $\rho$, and its confidence intervals at   level 0.05. Coverage is the proportion of times that $\rho\in (\hat \rho_L,\hat\rho_U)$. Between parenthesis are the standard deviations of the  estimations. } 
\label{TablaSimulUnif_rho}
\end{table}

\section{Tables of Subsections 4.2.1 
and 4.2.2 
}\label{Tablas_Datos}

The complete set of estimates and auxiliary values related with the  indices $\gamma, \pi$ and $\rho$ for the INE-data  (resp. NHANES-data) appear in Tables \ref{Tabla.INE_gamma}, \ref{Tabla.INE_pi} and \ref{Tabla.INE_rho} (resp.  \ref{Tabla.NHANES_gamma}, \ref{Tabla.NHANES_pi} and \ref{Tabla.NHANES_rho}). 
{
Additionally, we include in Figures \ref{Fig.QFDF_INE} and \ref{Fig.QFDF_NHANES} the representation of the 
corresponding quantile and empirical distribution functions for the involved  data sets.
}

\begin{figure}[H]
\centering

\includegraphics[width=5.5cm,height=4.35cm]{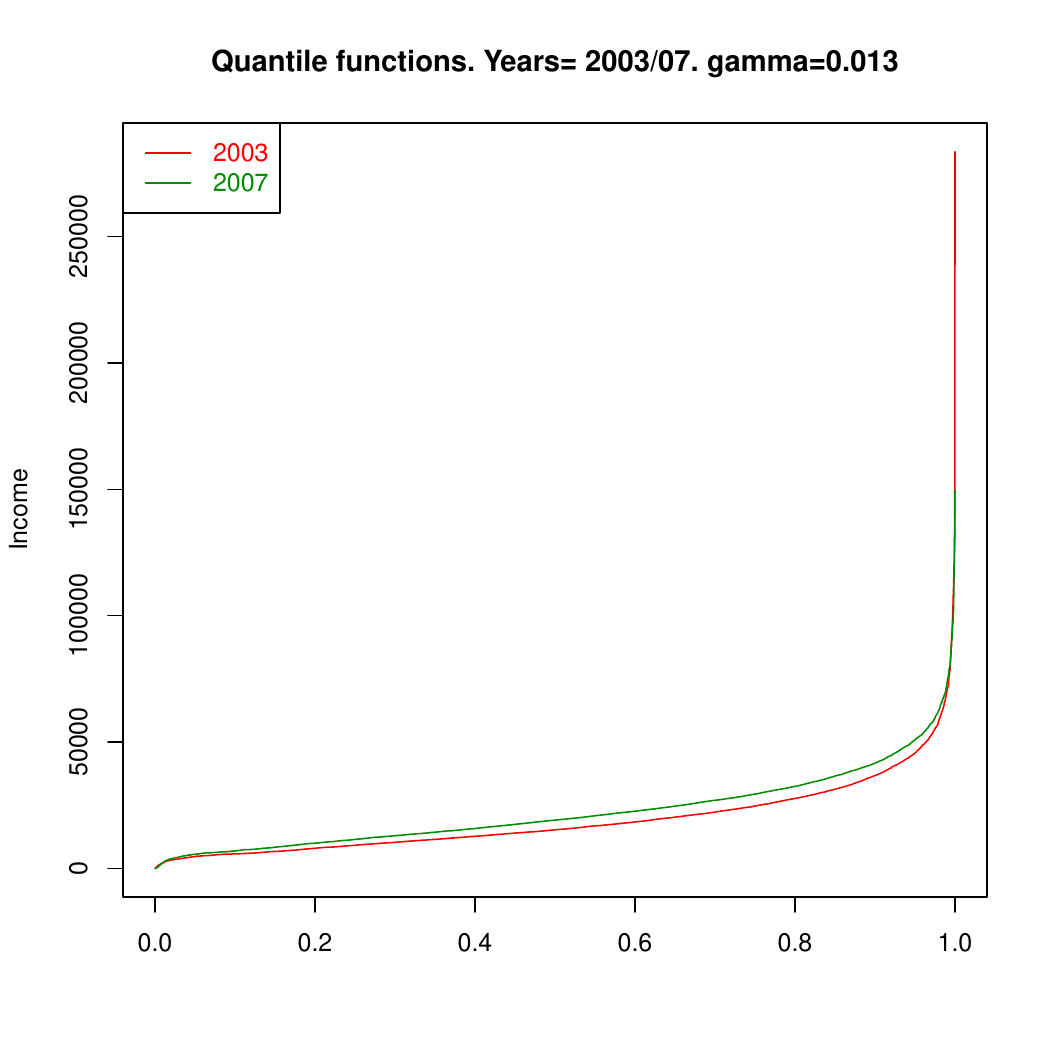}
\includegraphics[width=5.5cm,height=4.35cm]{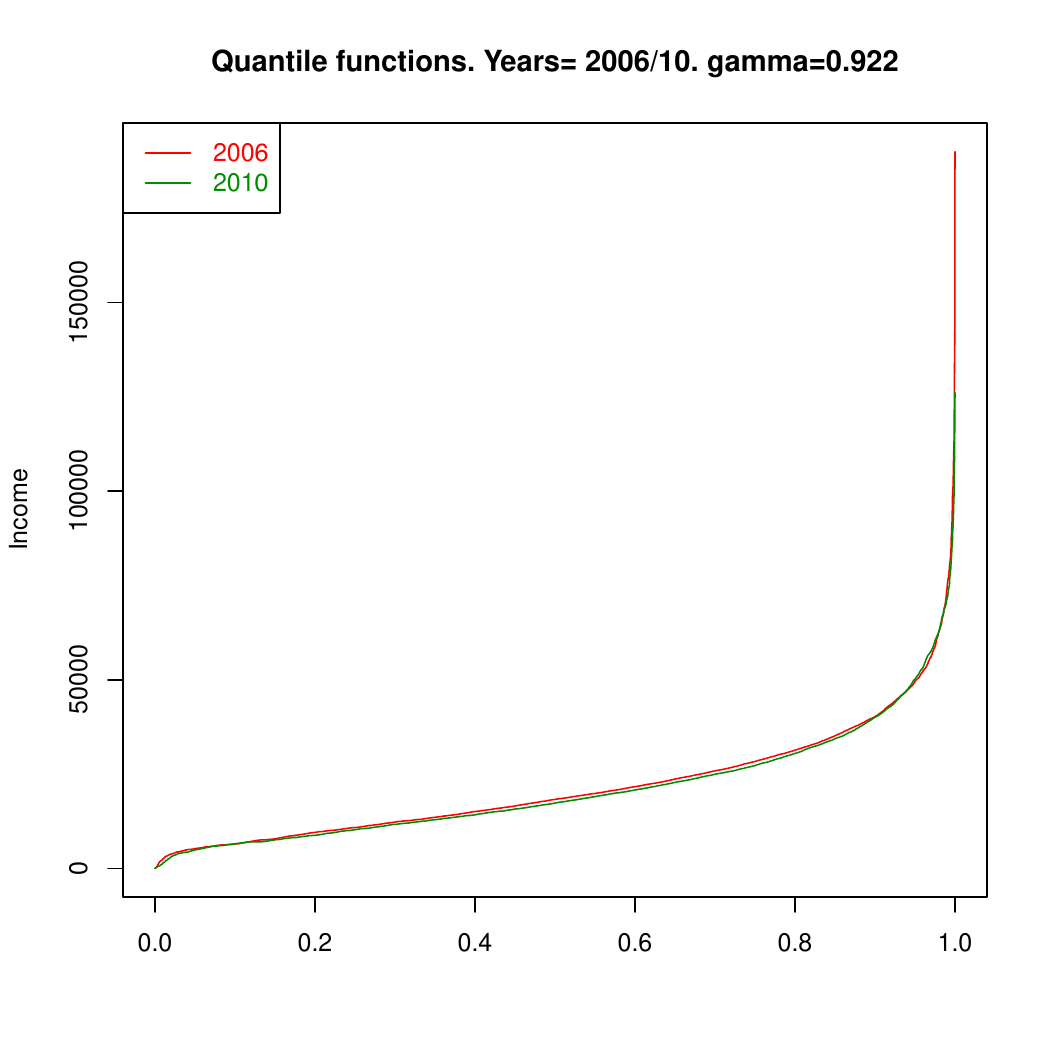}
\includegraphics[width=5.5cm,height=4.35cm]{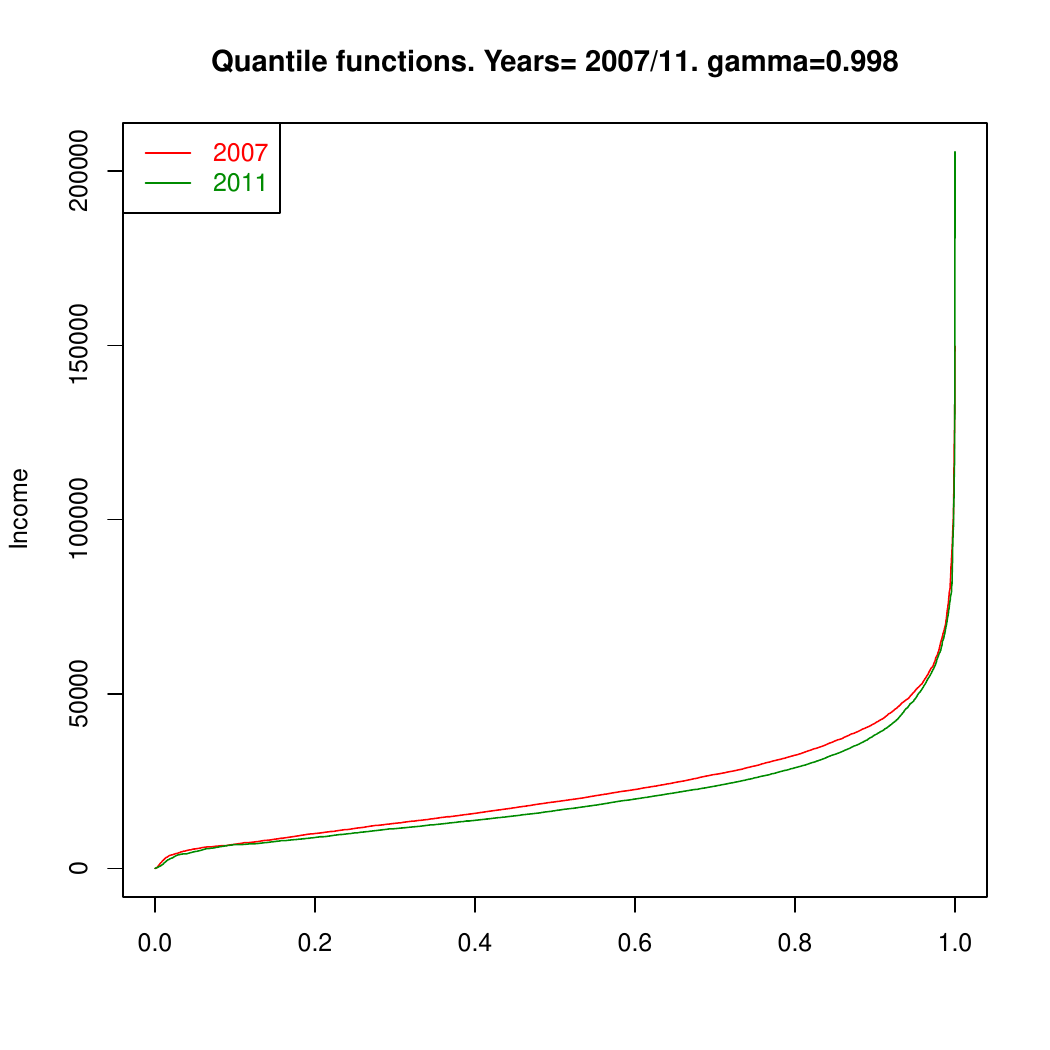}

\includegraphics[width=5.5cm,height=4.35cm]{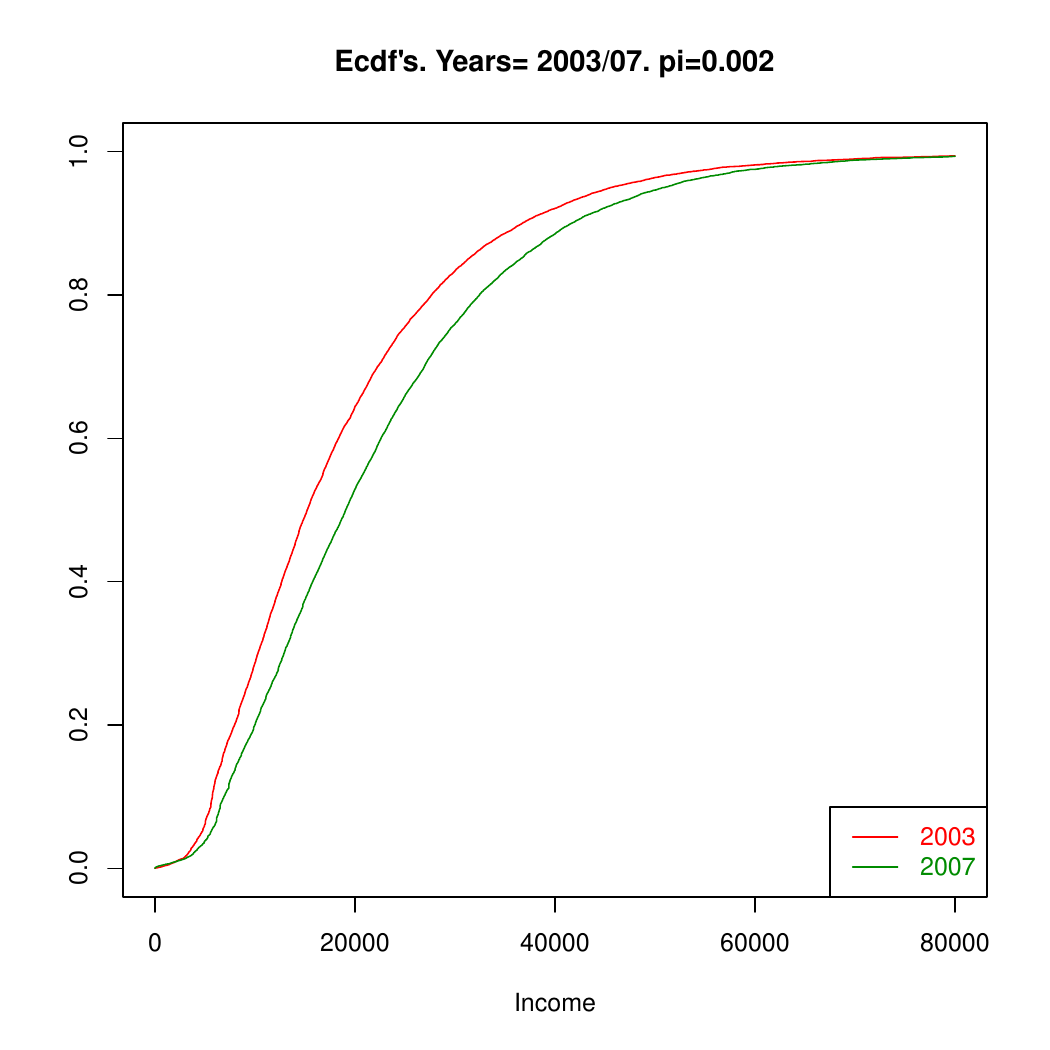}
\includegraphics[width=5.5cm,height=4.35cm]{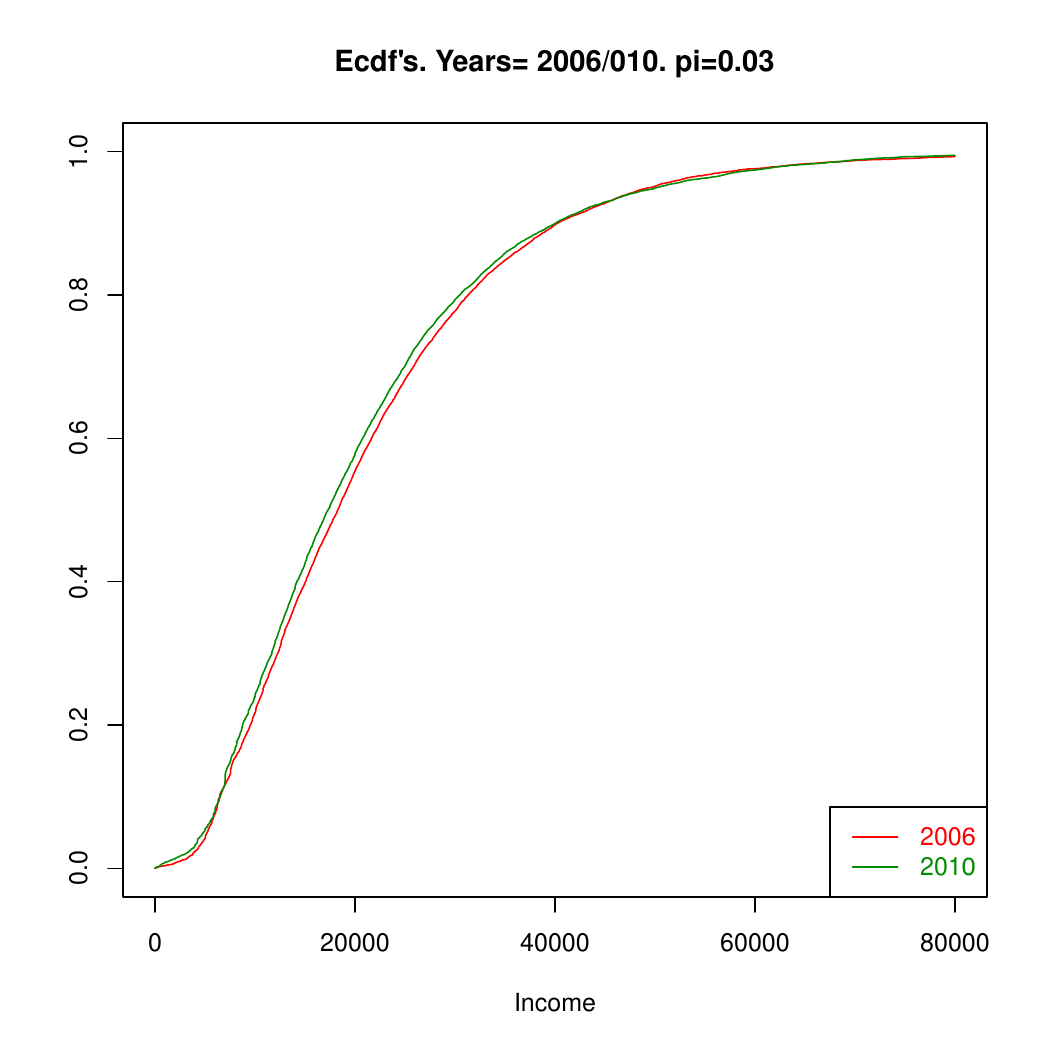}
\includegraphics[width=5.5cm,height=4.35cm]{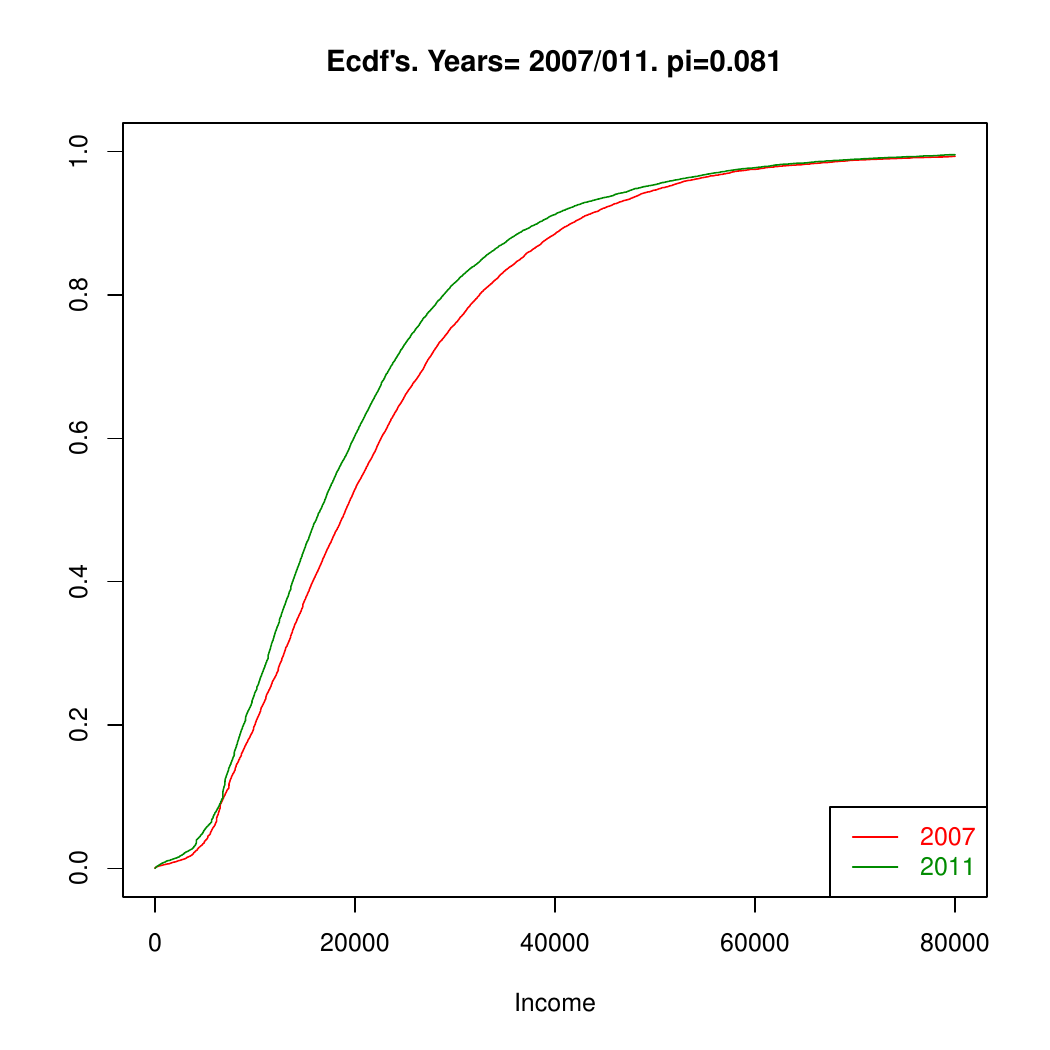}

\caption{Plots of the quantile (above) and empirical distribution (below) functions for the annual net disposable incomes of households at periods 2003/07, 2006/10 and 2007/11. We denote  ``gamma" (resp. ``pi") the value of $\hat \gamma$ (resp. $\pi$) of the final year over the initial one.
}
\label{Fig.QFDF_INE}
\end{figure}

\begin{figure}[H] 
\centering

\includegraphics[width=5.5cm,height=4.35cm]{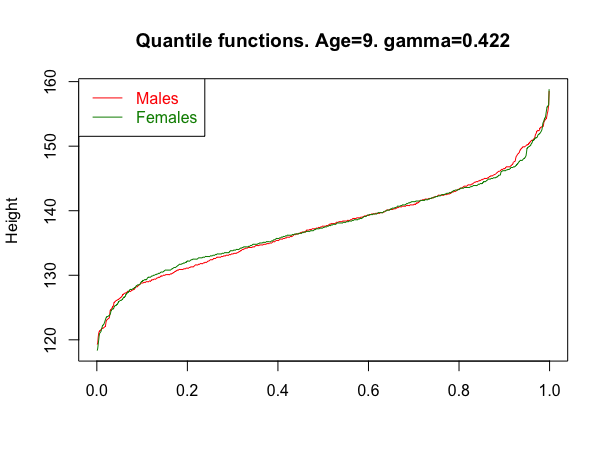}
\includegraphics[width=5.5cm,height=4.35cm]{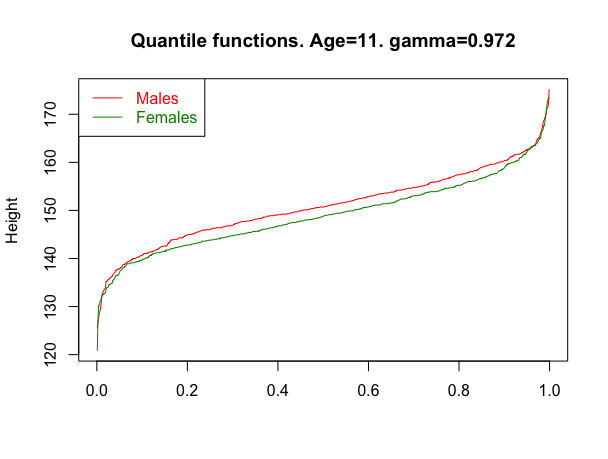}
\includegraphics[width=5.5cm,height=4.35cm]{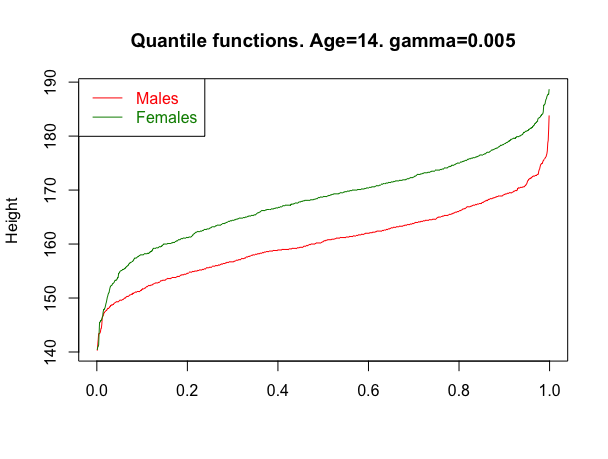}

\includegraphics[width=5.5cm,height=4.35cm]{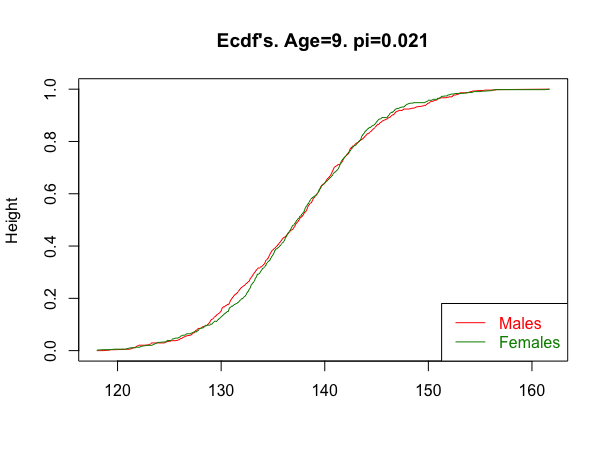}
\includegraphics[width=5.5cm,height=4.35cm]{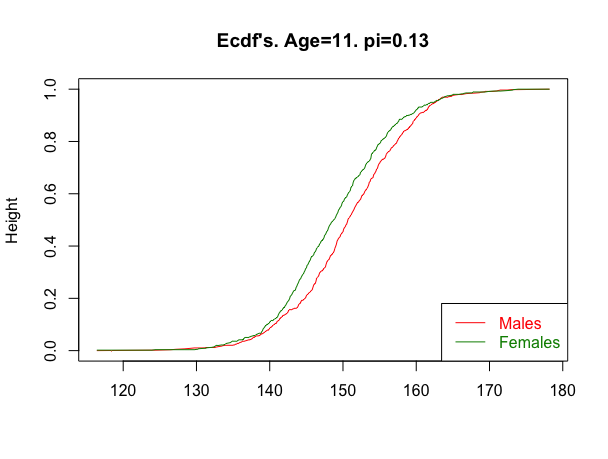}
\includegraphics[width=5.5cm,height=4.35cm]{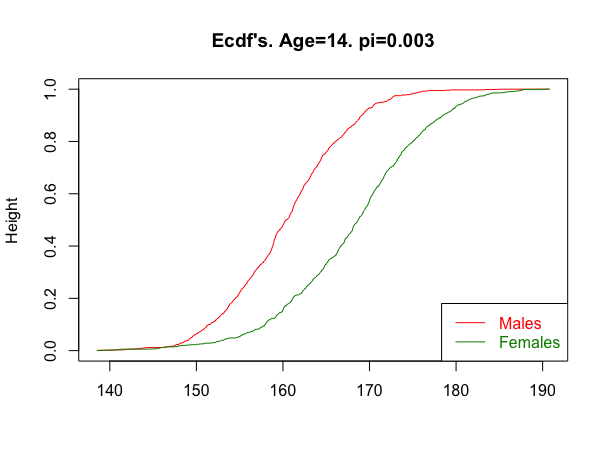}
\caption{Plots of the quantile (above) and empirical distribution (below) functions for the heights of males and females at ages 2, 3, 11 and 14. The plots include as ``gamma" (resp. ``pi") the value of $\hat \gamma$ (resp. $\pi$) {
of girls over boys at such age}.
}
\label{Fig.QFDF_NHANES}
\end{figure}

\begin{table}[H]
\centering
\begin{tabular}{rcccc}
  \hline
  Year & $\widehat{ 2r_0}$ & LowerBound & Estimation & UpperBound \\ 
  \hline
 2003 & 2.5679 & .0028 & .0133 & .0191 \\ 
 2004 & 1.8853 & 0 & .0287 & .0421 \\ 
   2005 & 2.6928 & .0257 & .0417 & .0685 \\ 
2006 & 2.1691 & .8693 & .9220 & 1 \\ 
 2007 & 1.8990 & .9965 & .9978 & 1 \\ 
   \hline
\end{tabular}
\caption{INE data. Estimations of $2r_0$, of $\gamma$ and confidence intervals for $\gamma$. Level = 0.05} 
\label{Tabla.INE_gamma}
\end{table}

\

\begin{table}[H]
\centering
\begin{tabular}{rccc}
  \hline
 Year & Lower Bound  & Estimation & Upper Bound  \\ 
  \hline
    2003 & 0 & .0021 & .0038 \\ 
2004 & 0 & .0024 & .0053 \\ 
2005 & 0 & .0033 & .0078 \\ 
2006 & .0130 & .0304 & .0399 \\ 
2007 & .0643 & .0807 & .0914 \\ 
   \hline
\end{tabular}
\caption{INE data. Estimations and confidence intervals for $\pi$. Level = 0.05} 
\label{Tabla.INE_pi}
\end{table}

\

\begin{table}[H]
\centering
\begin{tabular}{rccc}
  \hline
  Year & LowerBound & Estimation & UpperBound \\ 
  \hline
  2003 & .5704 & .5790 & .5876 \\ 
  2004 & .5256 & .5343 & .5429 \\ 
  2005 & .5122 & .5212 & .5301 \\ 
  2006 & .4743 & .4832 & .4921 \\ 
  2007 & .4418 & .4506 & .4594 \\ 
   \hline
\end{tabular}
\caption{INE data. Estimations and confidence intervals for $\rho$; level = 0.05} 
\label{Tabla.INE_rho}
\end{table}

\begin{table}[H]
\centering
\begin{tabular}{rrccc}
  \hline
  Age &  inverse speed & LowerBound & Estimation & UpperBound \\ 
  \hline
 2 & 1.0582 & 0 & .0062 & .0105 \\ 
 3 & 5.0645 & 0 & .0346 & .0644 \\ 
 4 & 1.3416 & 0 & .0031 & .0055 \\ 
 5 & 1.2049 & 0 & .0140 & .0244 \\ 
  6 & 1.5472 & 0 & .0035 & .0063 \\ 
  7 & 3.1509 & 0 & .1176 & .2089 \\ 
  8 & 1.8509 & 0 & .0625 & .1071 \\ 
  9 & 13.7403 & 0 & .4220 & .7856 \\ 
 10 & 2.1305 & .9178 & .9549 & 1 \\ 
 11 & 2.0739 & .9514 & .9724 & 1 \\ 
 12 & 2.2004 & .5390 & .6999 & .7836 \\ 
 13 & 4.2655 & .0505 & .1374 & .2157 \\ 
 14 & 4.5763 & 0 & .0052 & .0100 \\     
\hline
\end{tabular}
\caption{NHANES data. Estimations of $2r_0$, of $\gamma$ and confidence intervals for $\gamma$. Level = 0.05} 
    \label{Tabla.NHANES_gamma}
\end{table}

\begin{table}[H]
\centering
\begin{tabular}{rccc}
  \hline
 Age &  Lower Bound  & Estimation & Upper Bound  \\ 
  \hline
     2 & 0 & 0.0013 & 0.0091 \\ 
     3 & 0 & 0.0066 & 0.0568 \\ 
    4 & 0 & 0.0016 & 0.0270\\ 
   5 & 0 & 0.0036 & 0.0233\\ 
    6 & 0 & 0.0014 & 0.0343 \\ 
    7 & 0 & 0.0145 & 0.0612\\ 
     8 & 0 & 0.0110 & 0.0416 \\ 
     9 & 0 & 0.0211 & 0.0858 \\ 
  10 & 0.0439 & 0.1284 & 0.1705 \\ 
    11 & 0.0479 & 0.1297 & 0.1698 \\ 
   12 & 0.0491 & 0.1229 & 0.1601\\ 
    13 & 0 & 0.0320 & 0.0508\\ 
    14 & 0 & 0.0031 & 0.0135\\
   \hline
\end{tabular}
\caption{NHANES data. Estimations and confidence intervals for $\pi$. Level = 0.05} 
    \label{Tabla.NHANES_pi}
    \end{table}

\

\begin{table}[H]
\centering
\begin{tabular}{cccc}
  \hline
Age & Lower Bound  & Estimation & Upper Bound  \\ 
  \hline
    2 & 0.3626 & 0.3978 & 0.4330 \\ 
    3 & 0.4283 & 0.4696 & 0.5109 \\ 
    4 & 0.3900 & 0.4294 & 0.4687 \\ 
    5 & 0.3824 & 0.4237 & 0.4650 \\ 
    6 & 0.3897 & 0.4318 & 0.4738 \\ 
    7 & 0.4199 & 0.4611 & 0.5024 \\ 
    8 & 0.4048 & 0.4458 & 0.4867 \\ 
    9 & 0.4509 & 0.4929 & 0.5350 \\ 
    10 & 0.5279 & 0.5702 & 0.6125 \\ 
    11 & 0.5299 & 0.5715 & 0.6130 \\ 
    12 & 0.5135 & 0.5497 & 0.5859 \\ 
    13 & 0.3357 & 0.3720 & 0.4084 \\ 
    14 & 0.1871 & 0.2241 & 0.2610 \\
   \hline
\end{tabular}
\caption{NHANES data. Estimations and confidence intervals for $\rho$. Level = 0.05} 
\label{Tabla.NHANES_rho}  
\end{table}

\end{appendix}

\end{document}